%% file: paper.tex
\documentclass[aps, pre, a4paper, floatfix,  twocolumn, longbibliography, nofootinbib]{revtex4-1}

\usepackage[utf8]{inputenc}
\usepackage{epsfig}
\usepackage[T1]{fontenc}
\usepackage[english]{babel}
\usepackage[table]{xcolor}
\usepackage{t1enc}
\usepackage{graphicx}
\usepackage{amssymb}
\usepackage{amsmath}
\usepackage{relsize}
\usepackage[percent]{overpic}
\usepackage{bm}
\usepackage{hyperref}

\usepackage{siunitx}
\usepackage{accents}
\usepackage{mathtools}

\usepackage{grffile}

\usepackage[normalem]{ulem}

\usepackage{pgf}
\usepackage{tikz}

\usepackage{longtable}

\newcommand{\includegraphicslp}[4]{
  \begin{tikzpicture}
    \node[anchor=south west,inner sep=0] (image) at (0,0) {\includegraphics[#3]{#4}};
    \begin{scope}[x={(image.south east)},y={(image.north west)}]
      \node[anchor=south west,inner sep=0] at (0,#2) {#1};
    \end{scope}

  \end{tikzpicture}
}

\newcommand{\includegraphicsl}[3]{\includegraphicslp{#1}{.95}{#2}{#3}}

\newcommand{\avg}[1]{{\left<#1\right>}}

\usepackage{mathtools}
\def\multiset#1#2{\ensuremath{\left(\kern-.3em\left(\genfrac{}{}{0pt}{}{#1}{#2}\right)\kern-.3em\right)}}

\usepackage{amsmath}

\newcommand{\A}{\bm{A}}
\newcommand{\G}{\bm{G}}

\newcommand{\bb}{\bm{b}}

\newcommand{\ee}{\bm{e}}
\newcommand{\kk}{\bm{k}}

\newcolumntype{P}[1]{>{\centering\arraybackslash}p{#1}}

\usepackage{verbatim}
\usepackage{overpic}
\usepackage{booktabs}
\usepackage{placeins}
\usepackage{bbm}

\usepackage{siunitx}
\usepackage{multirow}

\usepackage{xcolor}  

\begin{document}

\title{Systematic assessment of the quality of fit of the stochastic block model for empirical networks}

\author{Felipe Vaca-Ram\'{i}rez}
\email{vaca_felipe@phd.ceu.edu}
\affiliation{Department of Network and Data Science, Central European University, 1100 Vienna, Austria}
\author{Tiago P. Peixoto}
\email{peixotot@ceu.edu}
\affiliation{Department of Network and Data Science, Central European University, 1100 Vienna, Austria}

\begin{abstract}
  We perform a systematic analysis of the quality of fit of the
  stochastic block model (SBM) for 275 empirical networks spanning a
  wide range of domains and orders of size magnitude. We employ
  posterior predictive model checking as a criterion to assess the
  quality of fit, which involves comparing networks generated by the
  inferred model with the empirical network, according to a set of
  network descriptors. We observe that the SBM is capable of providing
  an accurate description for the majority of networks considered, but
  falls short of saturating all modeling requirements. In particular,
  networks possessing a large diameter and slow-mixing random walks tend
  to be badly described by the SBM. However, contrary to what is often
  assumed, networks with a high abundance of triangles can be well
  described by the SBM in many cases. We demonstrate that simple network
  descriptors can be used to evaluate whether or not the SBM can provide
  a sufficiently accurate representation, potentially pointing to
  possible model extensions that can systematically improve the
  expressiveness of this class of models.
\end{abstract}

\maketitle

\section{Introduction}

The stochastic block model
(SBM)~\cite{holland_stochastic_1983,karrer_stochastic_2011} is an
important family of generative network models used primarily for
community detection~\cite{peixoto_bayesian_2017} and link
prediction~\cite{guimera_missing_2009}. In its simplest formulation, it
describes a network formation mechanism where the nodes are divided into
discrete groups, and the probability of an edge existing between two
nodes is given as a function of their group memberships. Many variations
of this idea exist, including mixed-membership
SBMs~\cite{airoldi_mixed_2008}, where nodes are allowed to belong to
multiple groups, the degree-corrected SBM
(DCSBM)~\cite{karrer_stochastic_2011}, where nodes are allowed to
possess arbitrary degrees, as well as several extensions to other
domains, such as dynamical
networks~\cite{xu_dynamic_2014,peixoto_inferring_2015,ghasemian_detectability_2016}
and multilayer
networks~\cite{peixoto_inferring_2015,stanley_clustering_2016}, to name
a few.

SBMs also serve as generalizations of more fundamental random network
models. The basic SBM has the Erd\H{o}s-Rényi
model~\cite{erdos_random_1959} as a special case when there is a single
group, and likewise the DCSBM recovers the configuration
model~\cite{fosdick_configuring_2018} in the same situation. However,
differently from these more fundamental models, the SBM possesses a set
of parameters --- the partition of the nodes and the affinities between
groups --- that is not trivially recoverable from observed
networks. These parameters are \emph{latent} information that need to be
obtained via inference algorithms, which form the basis of the community
detection methods that use this
approach~\cite{peixoto_bayesian_2017}. Furthermore, the SBM has a
controllable level of complexity: by increasing the number of groups, we
have the ability to express increasingly elaborate types of network
structures, via arbitrary mixing patterns between the latent groups. In
fact, despite its stylized nature, it can be shown that the SBM can
approximate a broad class of generative models that are different from
it~\cite{olhede_network_2014}, and its inference functions similarly to
fitting a histogram to numeric data in order to estimate the underlying
probability density --- with the node groups playing a similar role to
the histogram bins. However, the expressiveness of the SBM is not
absolute, specially when the networks are \emph{sparse}, i.e. when their
average degree is much smaller than the total number of nodes. In such a
situation, there is no guarantee that the SBM is capable of arbitrarily
approximating the true underlying model, regardless of how we infer
it: By increasing the model complexity we move from a situation where we
are \emph{underfitting}, i.e. extracting patterns that do not
sufficiently capture all the features of the true model, to a situation
where we are \emph{overfitting}, i.e. incorporating randomness into the
model description, which is also a deviation from the true model. When
we find the most adequate inference that balances statistical evidence
against model complexity to prevent overfitting, we might still be
missing important features of the true model, simply because it cannot
be sufficiently well captured under the SBM parametrization.

Here we are not interested in evaluating the SBM as a plausible
generative process of networks across all domains, since it does not
represent an ultimately credible mechanism for any of them. Instead, our
objective is to assess how capable it is of providing a general
\emph{effective} description of empirical networks, and in which aspects
and to what extent (and not \emph{whether}) it tends to be
misspecified. Understanding the limits of the SBM representation in
empirical settings is therefore a nuanced undertaking that is likely to
be affected by a variety of possible sources of deviations. Since the
SBM tends to yield very good comparative performance in link prediction
tasks~\cite{ghasemian_evaluating_2019,ghasemian_stacking_2020}, it is
therefore known that it tends to outperform alternative models in
capturing the structure of networks, but we still lack a more accurate
assessment of its qualities and shortcomings in absolute terms.

In this work, we evaluate the quality of fit of the SBM in empirical
contexts by performing \emph{model checking} on Bayesian
inferences. Based on a diverse collection of 275 networks spanning
various domains and several orders of size magnitude, we compare the
values of many network descriptors computed on the observed network with
what would be typically obtained with networks sampled from the inferred
SBM. In this way, any significant discrepancy can be interpreted as a
form of ``residual'' that points to a shortcoming of the SBM in
capturing that particular network property.

Overall we find that the SBM is capable of encapsulating the network
structure to a significant degree for a large fraction of the networks
studied, but falls short of completely exhausting the modelling
requirements in many cases. We find that for networks with very large
diameter or a very slow mixing random walk the SBM tends to provide a
poor description. This includes, for example, many transportation
networks --- which are typically embedded in a low dimensional space ---
as well as some economic networks. However, for other kinds of networks
the quality of fit tends to be good overall.

We proceed with describing in detail the model and inference procedure
(Sec.~\ref{sec:model}), our criteria to evaluate the quality of fit
(Sec.~\ref{sec:quality}), the network corpus used
(Sec.~\ref{sec:corpus}), and the results of our analysis for it
(Sec.~\ref{sec:fit}). We finalize in Sec.~\ref{sec:conclusion} with
a conclusion.

\section{Model and inference}\label{sec:model}

For our analysis we will use the microcanonical degree-corrected SBM
(DCSBM)~\cite{karrer_stochastic_2011,peixoto_nonparametric_2017}, which
combines arbitrary mixing patterns between groups together with
arbitrary degree sequences. It has as parameters the partition of the
nodes into groups, $\bb=\{b_i\}$, with $b_i\in[1,B]$ being the group
membership of node $i$, the degree sequence $\kk=\{k_i\}$, where $k_i$
is the degree of node $i$, and the edge counts between groups
$\ee=\{e_{rs}\}$ (or twice that number for $r=s$), given by
$e_{rs}=\sum_{ij}A_{ij}\delta_{b_i,r}\delta_{b_i,s}$. Given these
constraints, the network is generated with
probability~\cite{peixoto_nonparametric_2017}
\begin{equation}\label{eq:dcsbm}
  P(\A | \kk, \ee, \bb) = \frac{\prod_{r<s}e_{rs}!\prod_re_{rr}!!\prod_ik_i!}{\prod_{i<j}A_{ij}!\prod_iA_{ii}!!\prod_re_r!},
\end{equation}
where $\A=\{A_{ij}\}$ is the adjacency matrix of an undirected
multigraph with potential self-loops, and $e_r=\sum_se_{rs}$.

All the networks we will be studying are undirected simple graphs, for
which the above model can give only an approximation. As demonstrated in
Ref.~\cite{peixoto_latent_2020}, the use of multigraph models based on
the Poisson distribution (or equivalently, microcanonical models based
on the pairing of half-edges, as above) cannot ascribe probabilities to
simple edges (i.e. $A_{ij}=1$) that are larger than $1/\mathrm{e}
\approx 0.37$. This limits the applicability of such models on networks
with heterogeneous density, either due to broad degree distributions or
sufficiently dense communities, which are ubiquitous properties of
empirical networks. To address this limitation, we use the latent
multigraph model of Ref.~\cite{peixoto_latent_2020}, where we assume
that an underlying unobserved multigraph $\A$ is in fact responsible for
the observed simple graph $\G$ simply via the removal of the edge
multiplicities and self-loops, i.e.
\begin{equation}
  P(\G|\A) = \prod_{i<j}\left(1-\delta_{A_{ij},0}\right)^{G_{ij}}{\delta_{A_{ij},0}}^{1-G_{ij}}.
\end{equation}
Note that $P(\G|\A)$
can only take a value of $0$ or $1$, depending on whether $\G$ and $\A$
are compatible. Via this mathematical construction, the final model
\begin{equation}
  P(\G | \kk, \ee, \bb) = \sum_{\A}P(\G | \A)P(\A | \kk, \ee, \bb)
\end{equation}
can express both arbitrary mixing patterns between groups as well as
degree correction, without the limitations of the multigraph model for
networks with large local densities~\cite{peixoto_latent_2020}. The
inference of this model is performed by sampling from the posterior
distribution
\begin{equation}\label{eq:posterior}
  P(\A, \kk, \ee, \bb | \G) = \frac{P(\G|\A)P(\A | \kk, \ee, \bb)P(\kk, \ee, \bb)}{P(\G)},
\end{equation}
which remains tractable.
Here we use the merge-split Markov chain Monte Carlo (MCMC) algorithm
described in Ref.~\cite{peixoto_merge-split_2020} to efficiently sample
from this distribution.

Note that for $P(\kk, \ee, \bb)$ we use the nonparametric microcanonical
hierarchical priors and hyperpriors described in
Refs.~\cite{peixoto_hierarchical_2014,peixoto_nonparametric_2017}.
Importantly, this kind of approach determines the appropriate model
complexity (via the number of groups) according to the statistical
evidence available in the data. As has been shown in these previous
works, this choice guarantees that only compressive inferences are made
in a manner that prevents overfitting (finding a number of groups that
is too large), but also with a substantial protection against
underfitting (finding a number that is too small), which tends to happen
when noninformative priors are used instead.

In addition to the DCSBM we will also use the configuration model as a
comparison, obtained by reshuffling the edges of the obtained network
while preserving its degree sequence (here we use the edge-switching
MCMC algorithm~\cite{fosdick_configuring_2018}). We note that the
configuration model is an approximate special case of the DCSBM
considered above when there is only a single group.\footnote{This is
only approximately true since the configuration model and the latent
Poisson models are not identical, but sufficiently similar for the
purposes of this work~\cite{peixoto_latent_2020}.} Therefore, whenever
the Bayesian approach above identifies more than one group with a large
probability, this automatically implies a selection of the DCSBM in
favor of the configuration model. This happens for every network that we
consider in this work, meaning that the DCSBM is the favored model for
all of them. Nevertheless, the configuration model serves as a good
baseline to determine to what extent the quality of fit obtained with
the DCSBM can be ascribed to the degree sequence alone or to the
group-based mixing patterns uncovered.

\section{Assessing quality of fit}\label{sec:quality}

The approach we use to assess the quality of fit of the DCSBM is based
on obtaining the \emph{posterior predictive distribution} of certain
network descriptors. More precisely, for a scalar network descriptor
$f(\G)$, its posterior predictive distribution is given by
\begin{multline}\label{eq:ppred}
  P(y|\G) = \sum_{\substack{\G',\A',\A\\\kk,\ee,\bb}}\delta(y-f(\G'))P(\G'|\A')\\\times
  P(\A'|\ee,\kk,\bb)P(\A, \kk, \ee, \bb|\G).
\end{multline}
In other words, for each inferred parameter set $(\kk, \ee, \bb)$,
weighted according to its posterior probability, we sample a new network
$\G'$ from the model defined above (which can be done in time $O(E + N)$
where $E$ and $N$ are the total number of edges and nodes, respectively,
as we show in Appendix~\ref{app:generation}), and obtain the descriptor
value $y=f(\G')$.\footnote{The posterior predictive distribution for the
configuration model is analogous,
i.e. $P(y|\G)=\sum_{G'}\delta(y-f(\G'))P(\G'|\kk)$, where $\kk$ are the
observed degrees, and $P(\G|\kk)$ is the likelihood of the configuration
model.}

We can say that a model captures well the value of a descriptor if its
predictive posterior distribution ascribes high probability to values
that are close to what was observed in the original network. We can
obtain a compact summary of the level of agreement in two different
ways. The first measures the statistical significance of the deviation,
e.g. via the $z$-score
\begin{equation}
  z = \frac{f(\G) - \avg{y}}{\sigma_y},
\end{equation}
where $\avg{y}$ and $\sigma_y$ are the mean and standard deviation of
$P(y|\G)$. The second criterion is the relative deviation, which here we
compute in two different ways,
\begin{equation}
  \Delta_1 = \frac{f(\G) - \avg{y}}{f(\G)},\quad
  \Delta_2 = \frac{f(\G) - \avg{y}}{f_{\text{max}}-f_{\text{min}}},
\end{equation}
depending on whether the descriptor values are bounded in a well defined
interval $[f_{\text{min}},f_{\text{max}}]$ ($\Delta_2$) or not ($\Delta_1$).

The $z$-score and relative deviation measure complementary aspects of
the agreement between data and model, and represent different criteria
which should be used together. While a high value of the $z$-score can
be used to reject the inferred model as a plausible explanation for the
data, by itself it tells us nothing about how good an approximation it
is. Conversely, the relative deviation tells us how well the descriptor
is being reproduced by the model, but nothing about the statistical
significance of the comparison.

\begin{figure}
  \begin{tabular}{cc}
      \includegraphicsl{(a)}{width=.5\columnwidth}{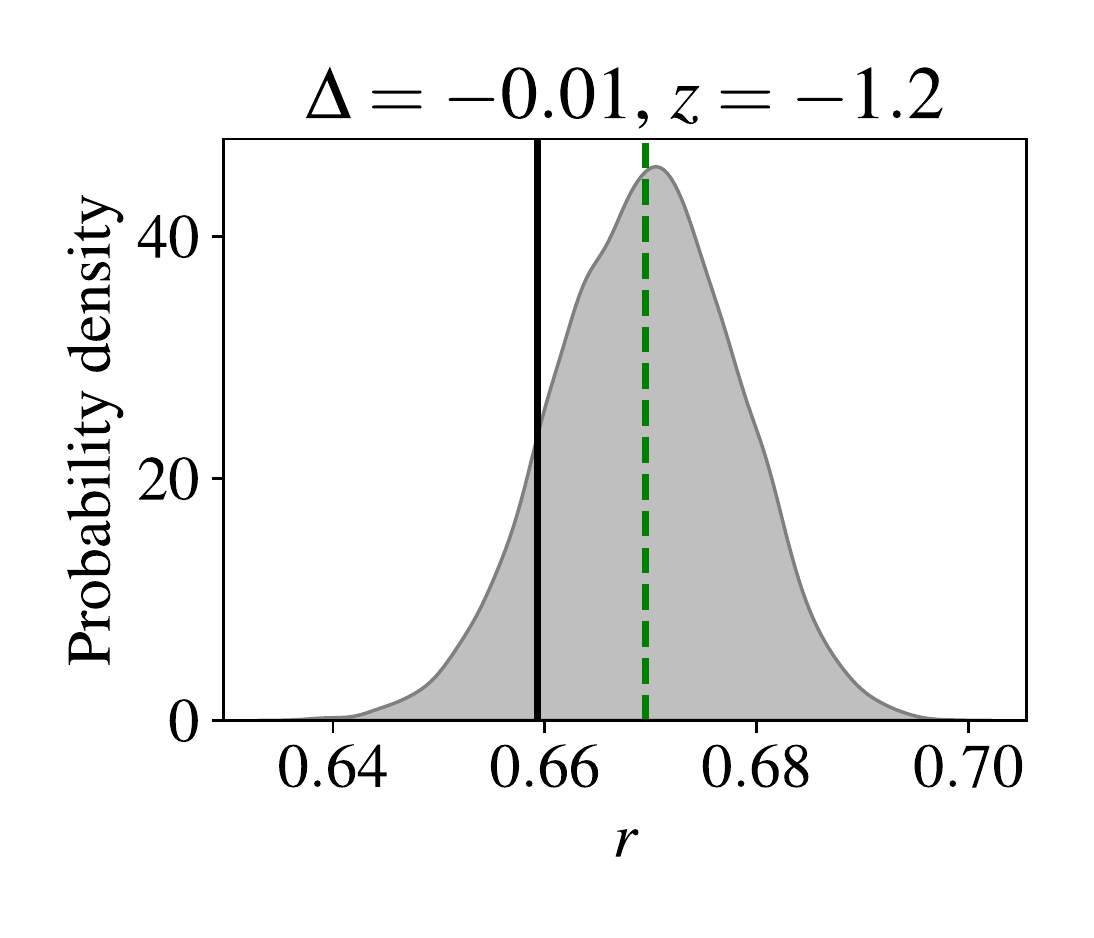} &
      \includegraphicsl{(b)}{width=.5\columnwidth}{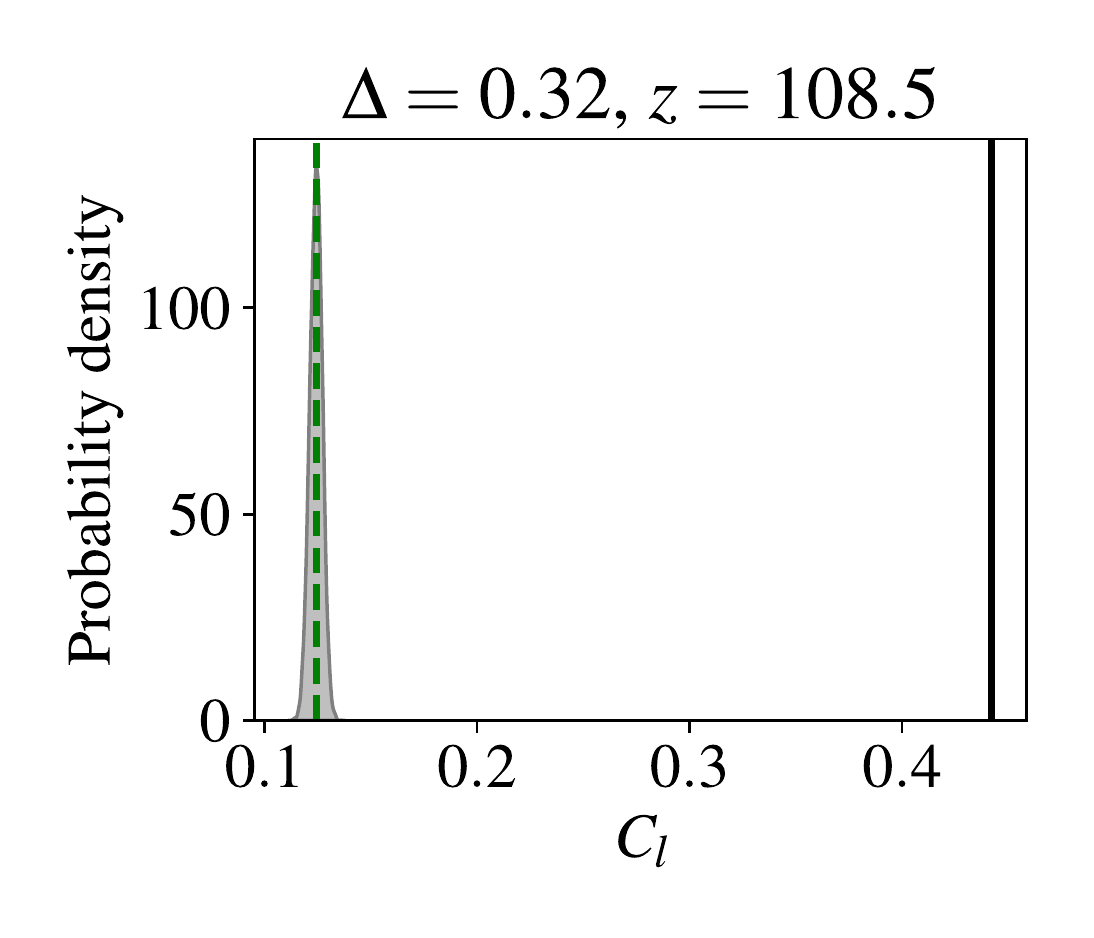} \\
      \includegraphicsl{(c)}{width=.5\columnwidth}{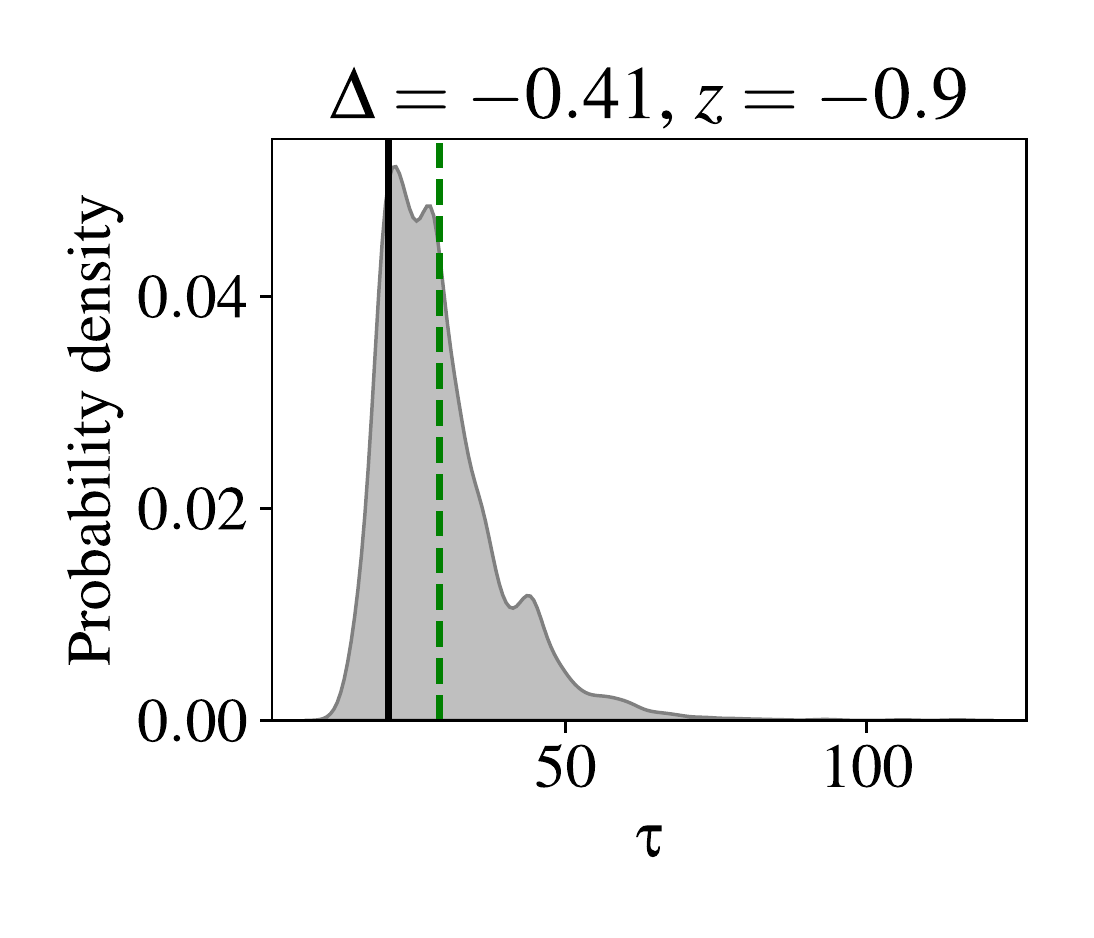} &
      \includegraphicsl{(d)}{width=.5\columnwidth}{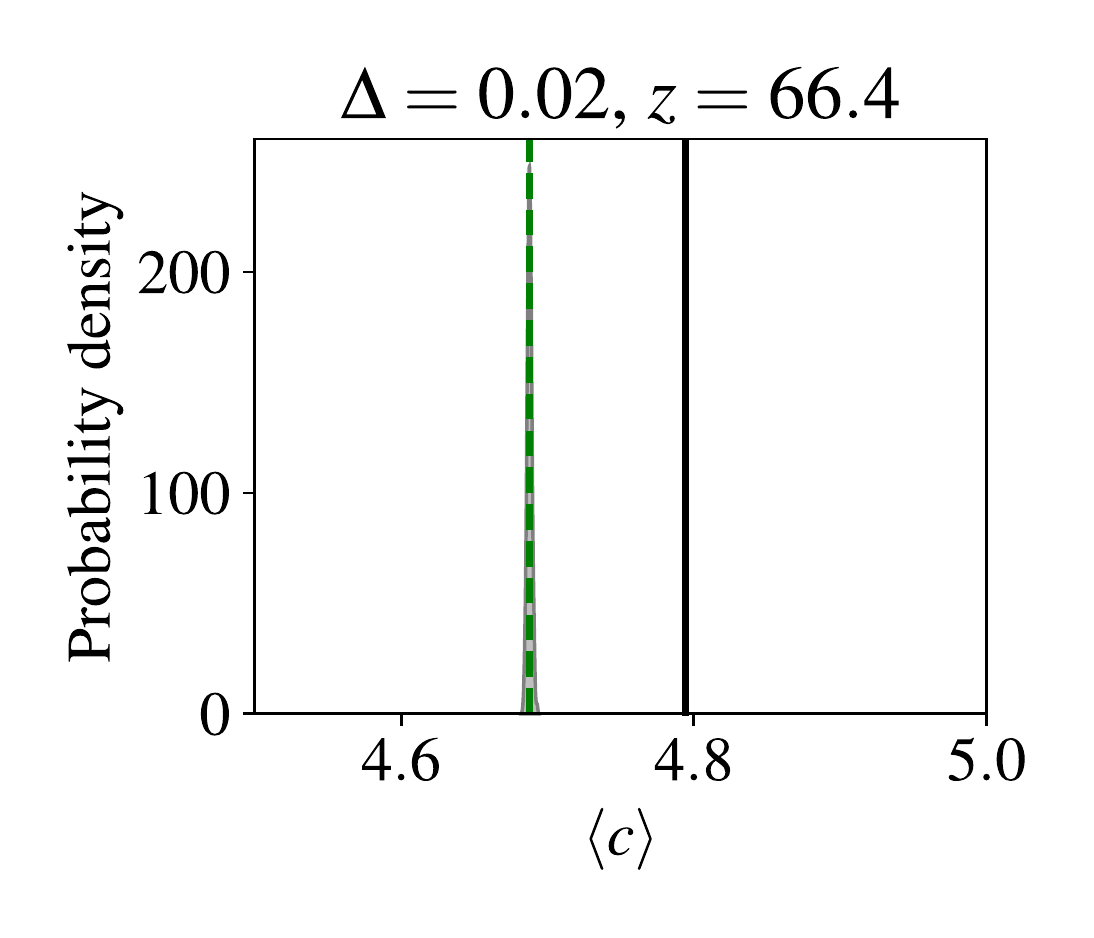}
  \end{tabular} \caption{Examples of posterior predictive distributions
  for some descriptors (see Table~\ref{tab:descriptors} for definitions)
  using the DCSBM, together with $z$-score and relative deviation. The
  solid black line shows the empirical value of the descriptor, and the
  dashed green line the mean of the predictive posterior distribution.
  In (a) and (b) we see examples where employing both criteria reveal
  unambiguously good and bad agreements, respectively, between data and
  model. However, in (c) we see a situation where despite a substantial
  disagreement with respect to the relative deviation, the $z$-score
  indicates that the model cannot be discarded as a plausible
  explanation for the data. In
  (d) we see a situation where the $z$-score points to decisive
  rejection of the model, but the small relative deviation allows us to
  accept it as an accurate
  approximation.\label{fig:discrepancy_examples}}
\end{figure}

In Fig.~\ref{fig:discrepancy_examples} we show examples that illustrate
how the different criteria operate. In
Fig.~\ref{fig:discrepancy_examples}(a) and (b) we see examples that show
good and bad agreements between model and data, respectively, according
to both criteria simultaneously. In these cases, the conclusion is
unambiguous: we either see no reason whatsoever to condemn the model, or
we see a definitive reason to do so. However, in
Fig.~\ref{fig:discrepancy_examples}(c) and (d) we reach mixed
conclusions. Fig.~\ref{fig:discrepancy_examples}(c) the model typically
yields different values than observed in the data, but it still ascribes
a large probability to it. We cannot condemn the model as an implausible
explanation for the data, but it is conceivable that the true generative
model would be more concentrated on the observed value. Conversely, in
Fig.~\ref{fig:discrepancy_examples} (d) we see a situation where the
model ascribes close to zero probability to the actual descriptor value
seen in the data, but, in absolute terms, the discrepancy is quite
small. Although we find evidence to condemn the plausibility of the
model, we could still claim that it is a good approximation.

\begin{table}
  \begin{tabular}{cp{17em}ll}
    Symbol & Descriptor & Range & $\Delta$\\\hline\hline\\[-.8em]
    $r$ & Degree assortativity & $[-1,1]$ & $\Delta_2$\\
    $\avg{c}$ & Mean $k$-core value& $[0,\infty]$ & $\Delta_1$\\
    $C_l$ & Mean local clustering coefficient & $[0,1]$ & $\Delta_2$\\
    $C_g$ & Global clustering coefficient & $[0,1]$ & $\Delta_2$\\
    $\lambda_1^A$ & Leading eigenvalue of the adjacency matrix & $[0,\infty]$ & $\Delta_1$\\
    $\lambda_1^H$ & Leading eigenvalue of the Hashimoto matrix & $[0,\infty]$ & $\Delta_1$\\
    $\tau$ & Characteristic time of a random walk & $[0,\infty]$ & $\Delta_1$\\
    $\varnothing$ & Pseudo-diameter & $[1,\infty]$ & $\Delta_1$\\
    $R_r$ & Node percolation profile (random removal) & 
    $[0,1/2]$ & $\Delta_2$\\
    $R_t$ & Node percolation profile (degree-targeted removal) & 
    $[0,1/2]$ & $\Delta_2$\\
    $S$ & Fraction of nodes in the largest component & $[0, 1]$ & $\Delta_1$\\\hline
  \end{tabular}

  \caption{List of network descriptors used in this work, with their
    respective symbol, range of values,
    and how the relative deviation was computed. More details on how the
    descriptors are computed are given in
    Appendix~\ref{app:descriptors}.\label{tab:descriptors}}
\end{table}

\begin{figure}[b!]
  \includegraphics[width=\columnwidth]{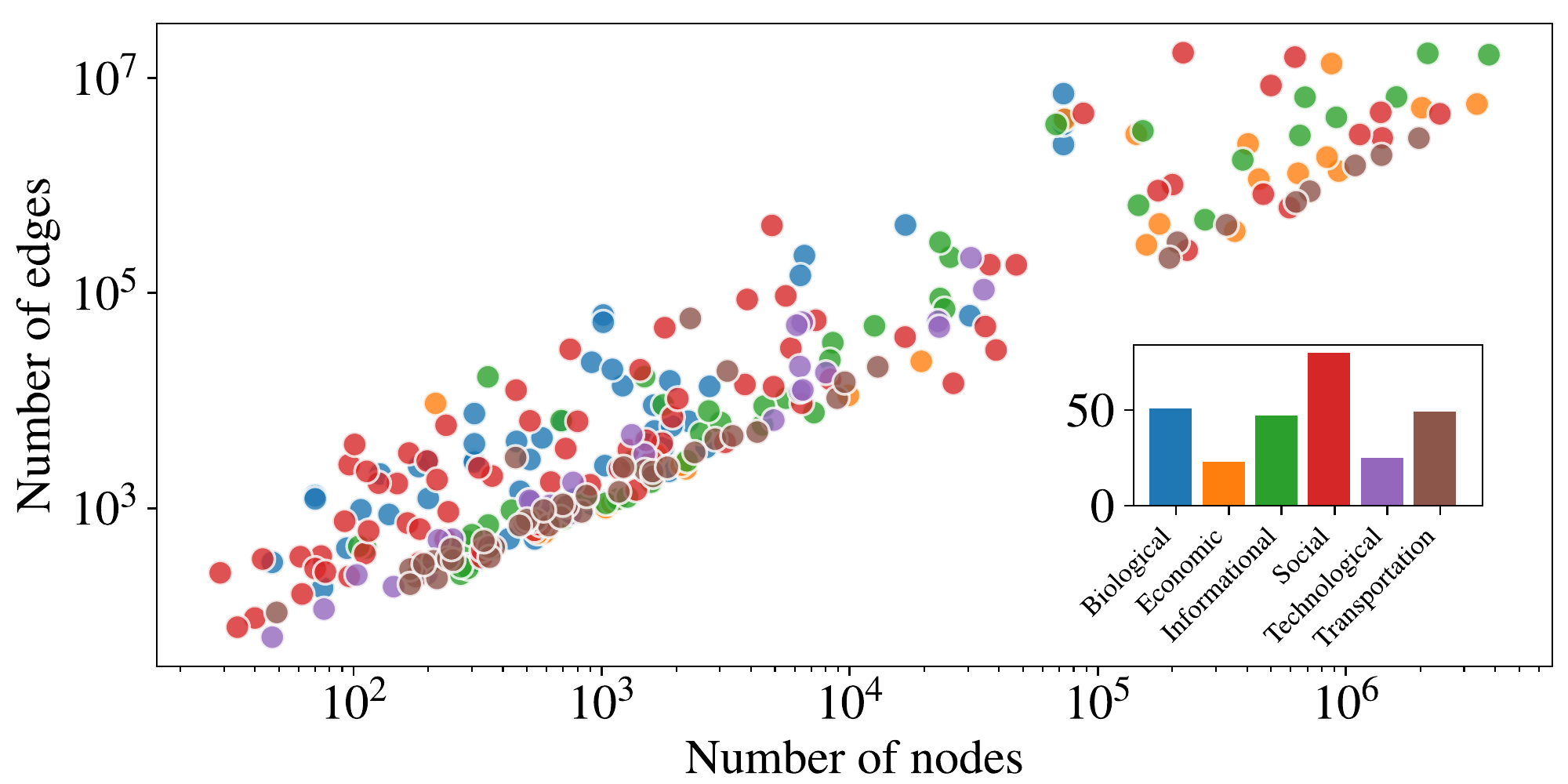}\\[1em]
  \includegraphics[width=\columnwidth]{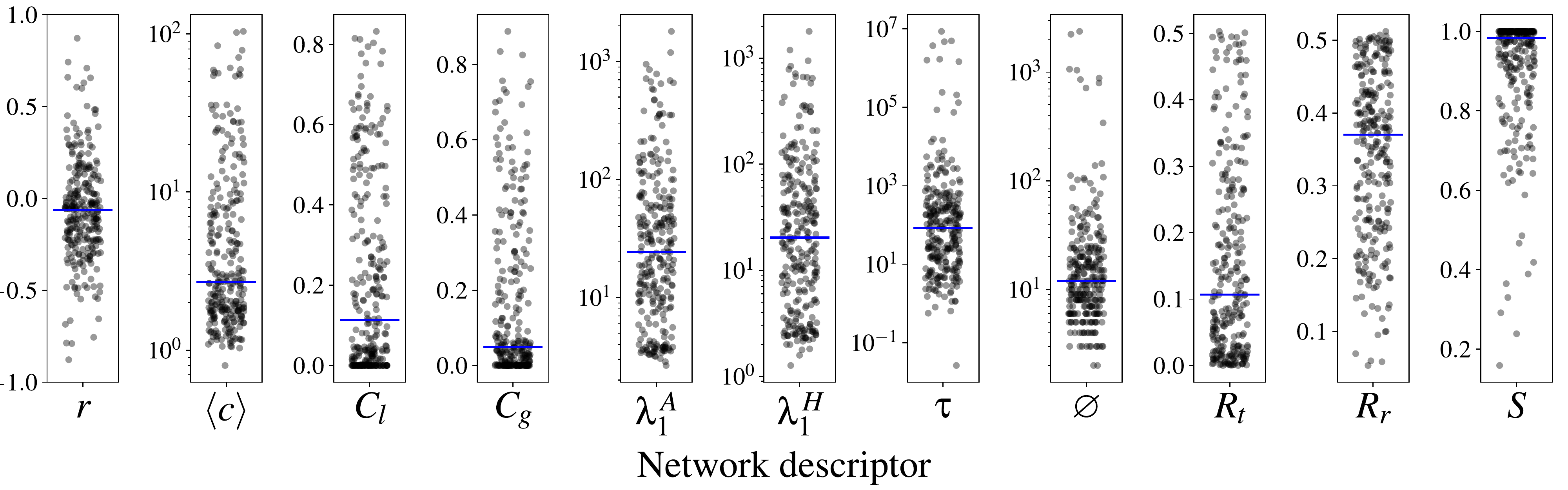}

  \caption{(Top) Number of nodes and edges for the networks in the
  corpus used in this work, and their domain composition
  (inset). (Bottom) Distribution of descriptor values for the networks
  in the corpus. The horizontal line marks the median
  values. \label{fig:corpus}}
\end{figure}

\begin{figure*}[t!]
  \begin{tabular}{cc}
    (a) Configuration model & (b) DCSBM \\
    \includegraphics[width=\columnwidth]{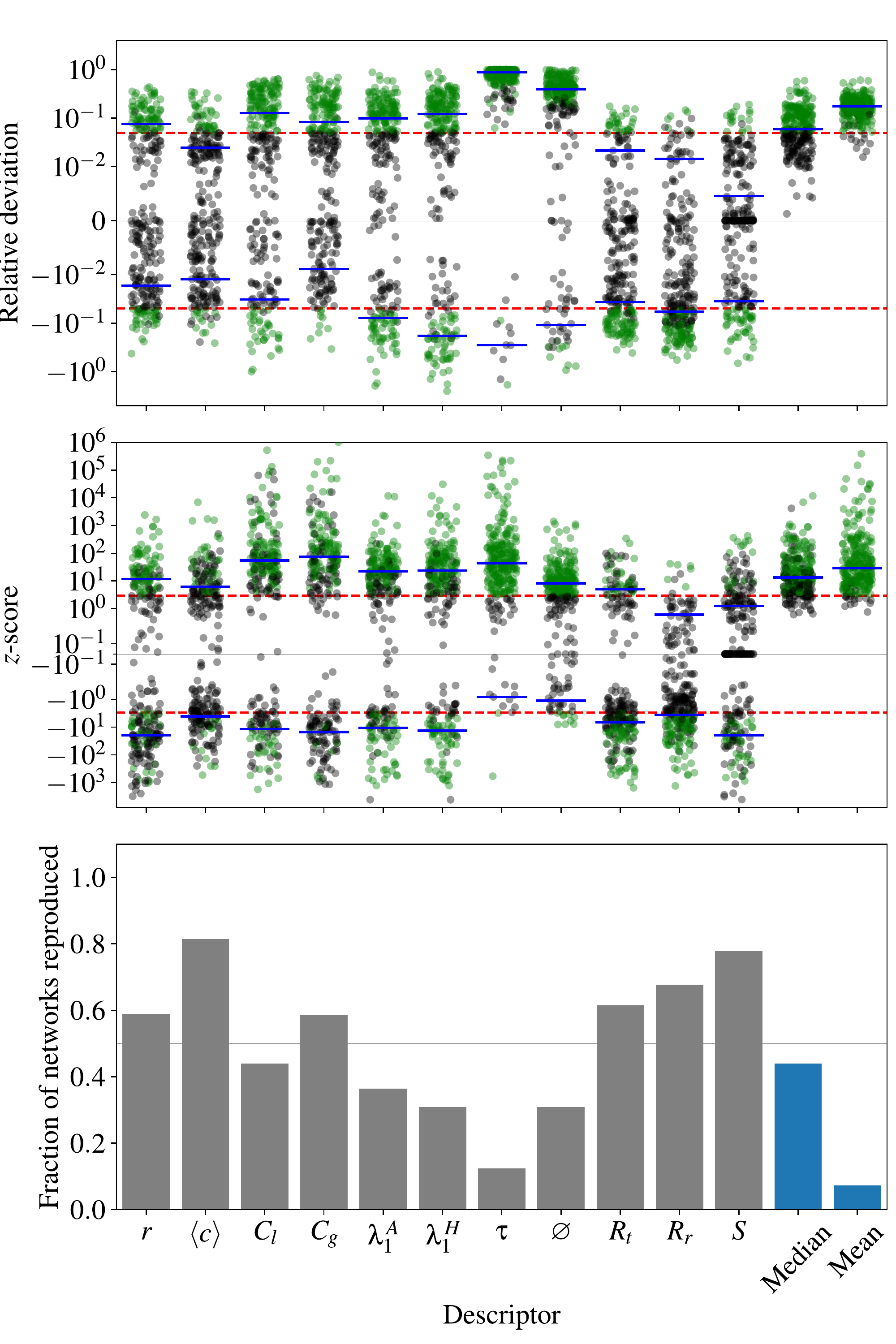} &
    \includegraphics[width=\columnwidth]{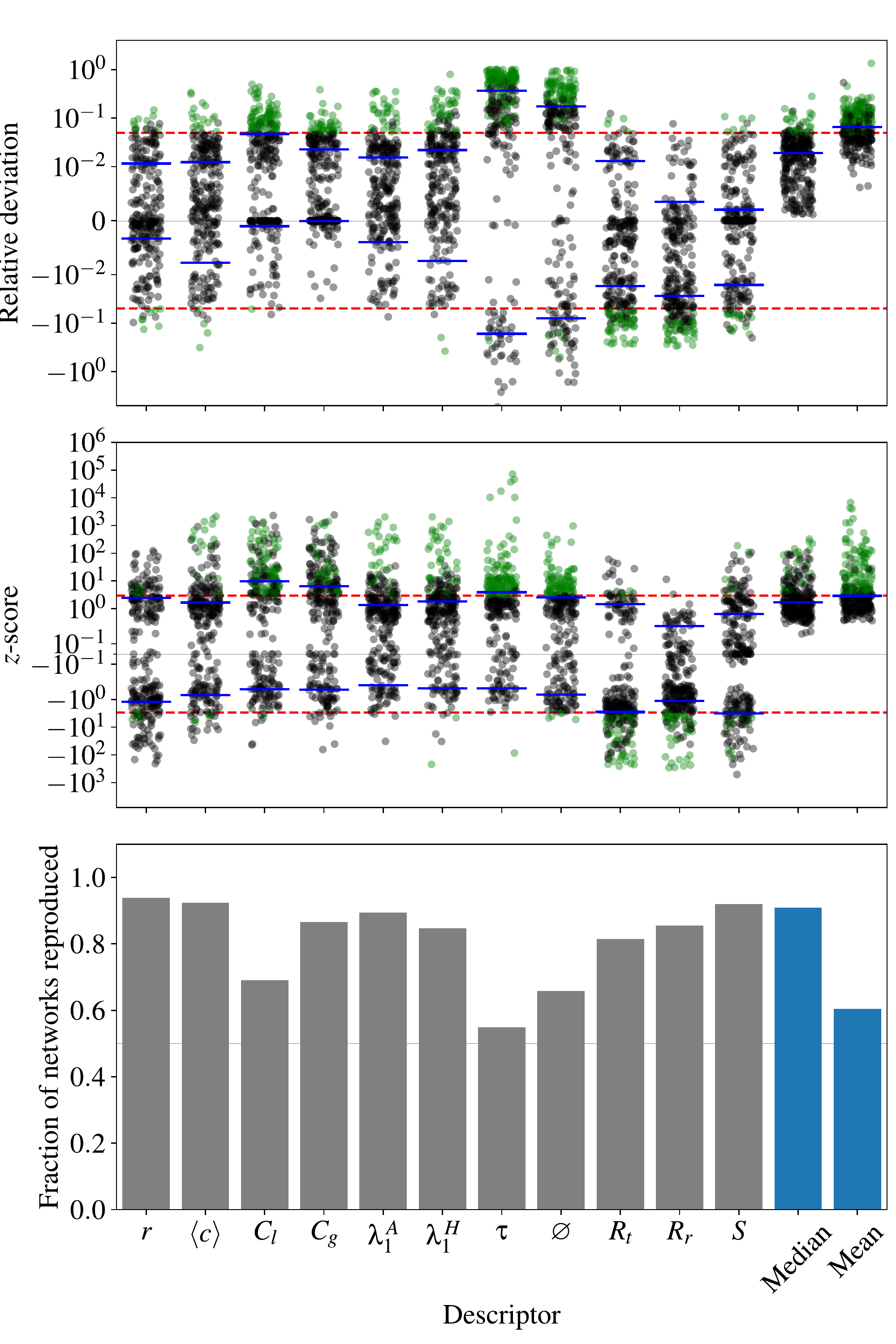}
  \end{tabular} \caption{Distribution of relative deviation (top),
  $z$-score (middle), and fraction of networks reproduced (bottom) for
  (a) the configuration model and (b) the DCSBM, according to their
  respective predictive posterior distributions for each descriptor. We
  also show the median and mean of the absolute values for all
  descriptors for each network.  The solid blue lines mark the negative
  and positive median values, and the dashed red line marks the values
  of $|\Delta|=0.05$ and $|z|=3$. The fraction of networks reproduced
  correspond to those that have the absolute value of either $\Delta$ or
  $z$ below these thresholds. The points in green color correspond to
  the networks that are not reproduced according to this combined
  criterion.\label{fig:desc-deviation}}
\end{figure*}
\begin{figure*}
  \begin{tabular}{cccc}
    \includegraphicsl{(a)}{width=.48\columnwidth}{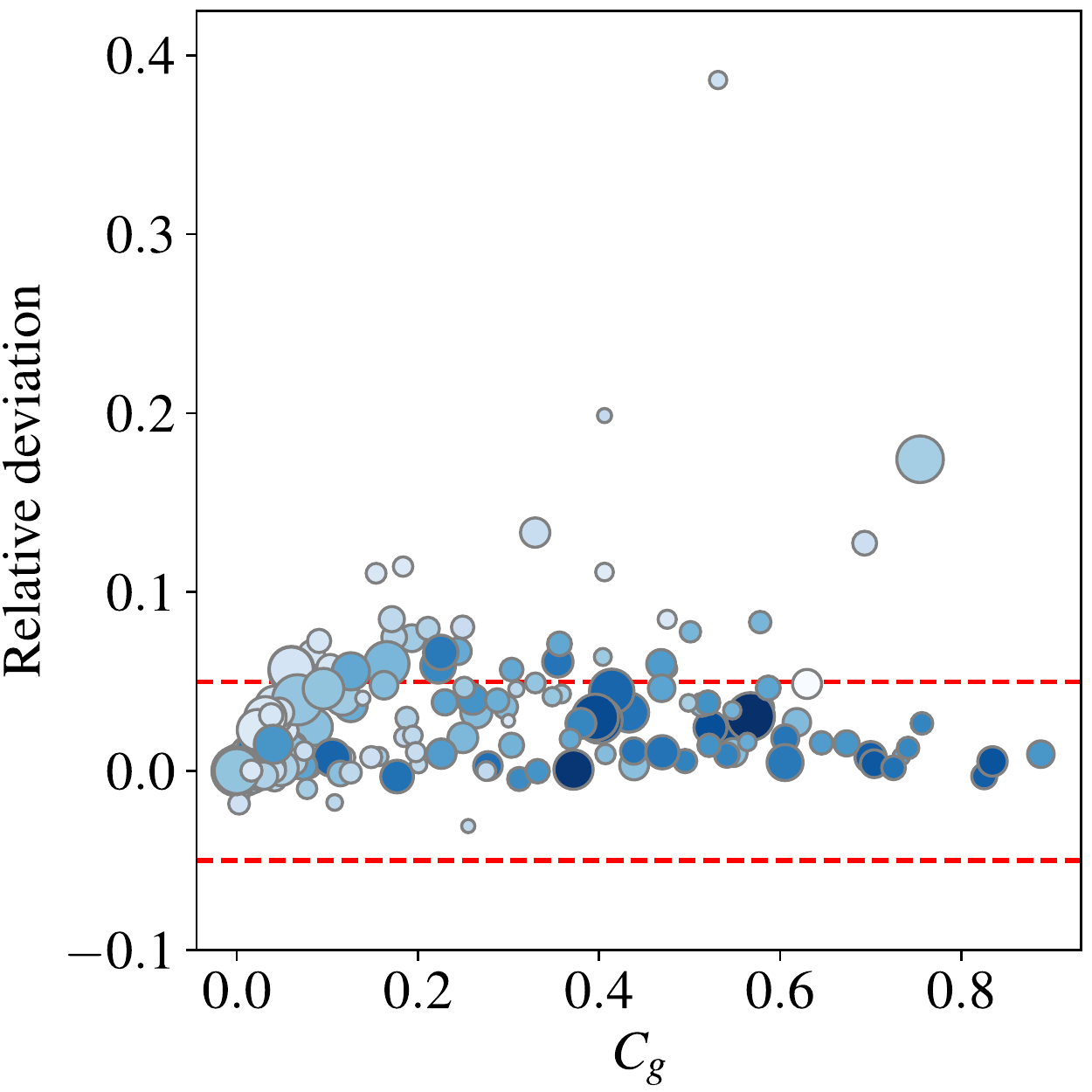} &
    \includegraphicsl{(b)}{width=.48\columnwidth}{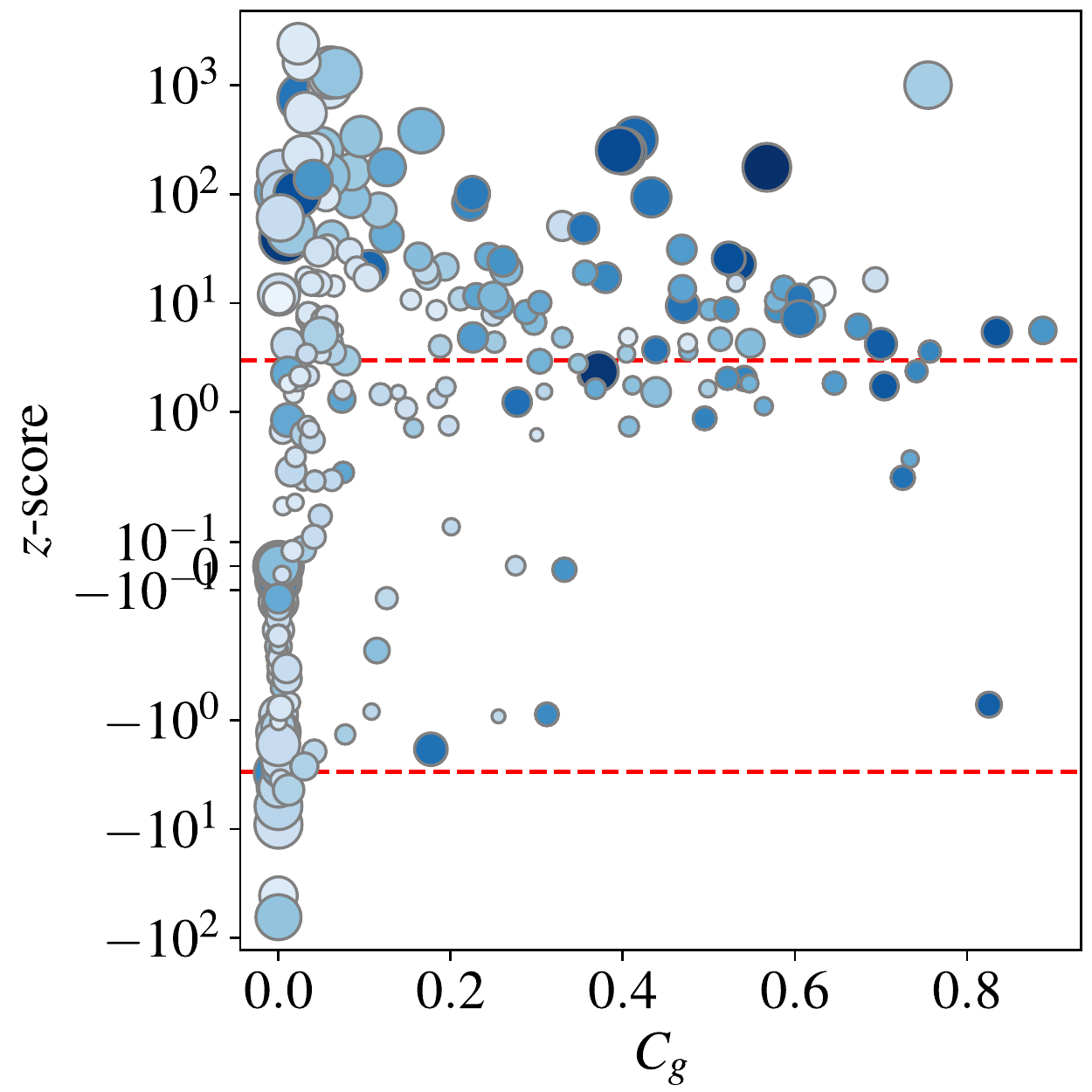} &
    \includegraphicsl{(c)}{width=.48\columnwidth}{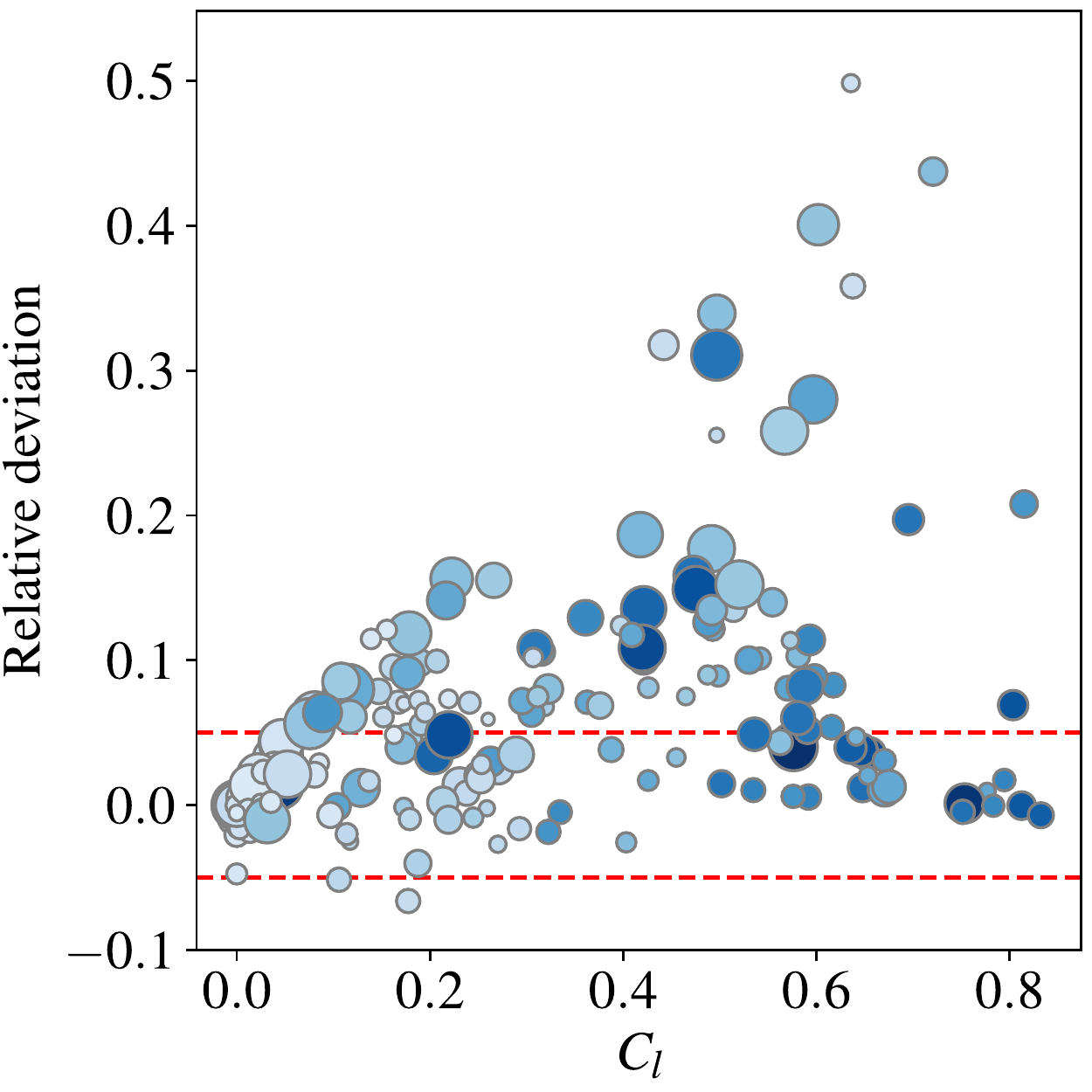}&
    \includegraphicsl{(d)}{width=.48\columnwidth}{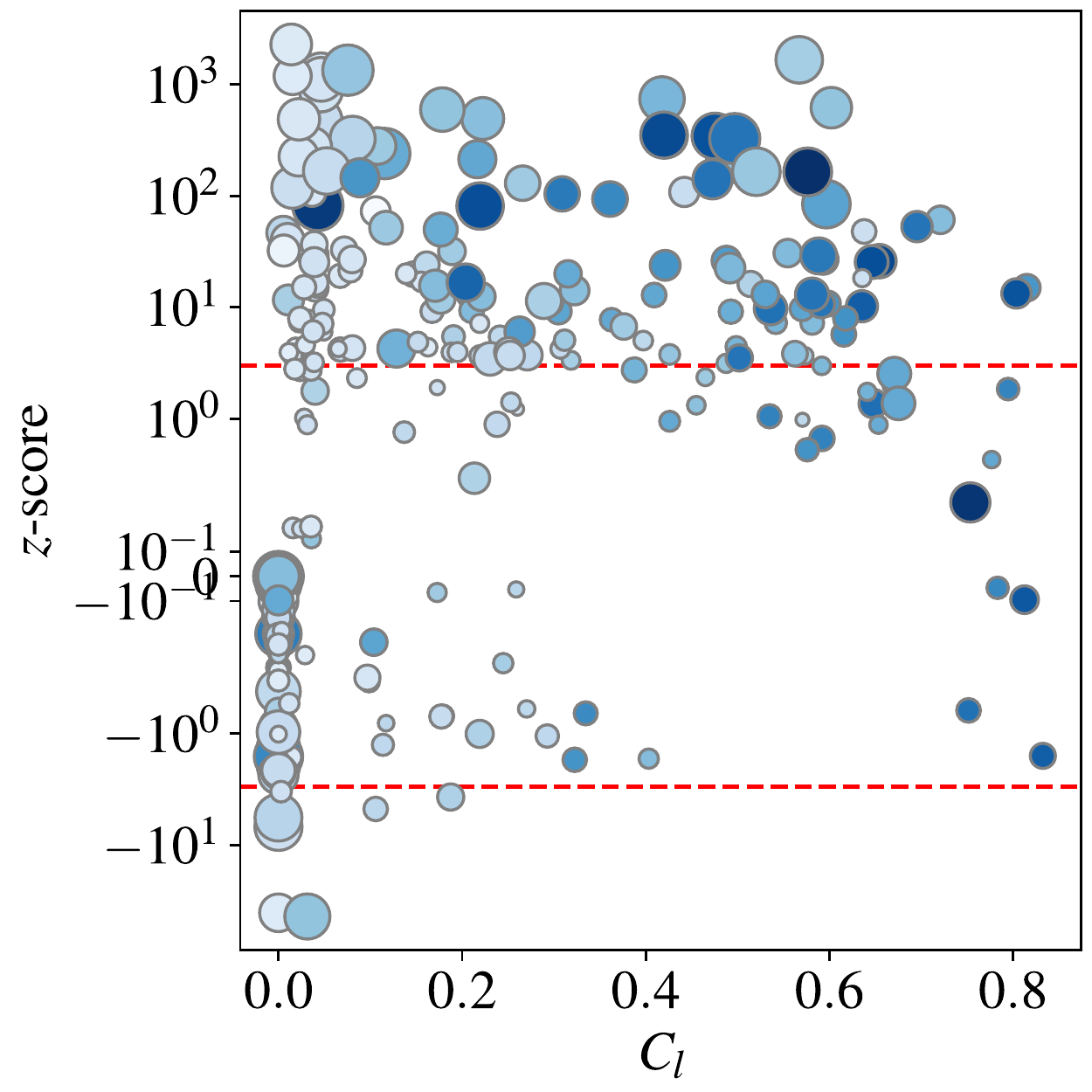}\\
    \includegraphicsl{(e)}{width=.48\columnwidth}{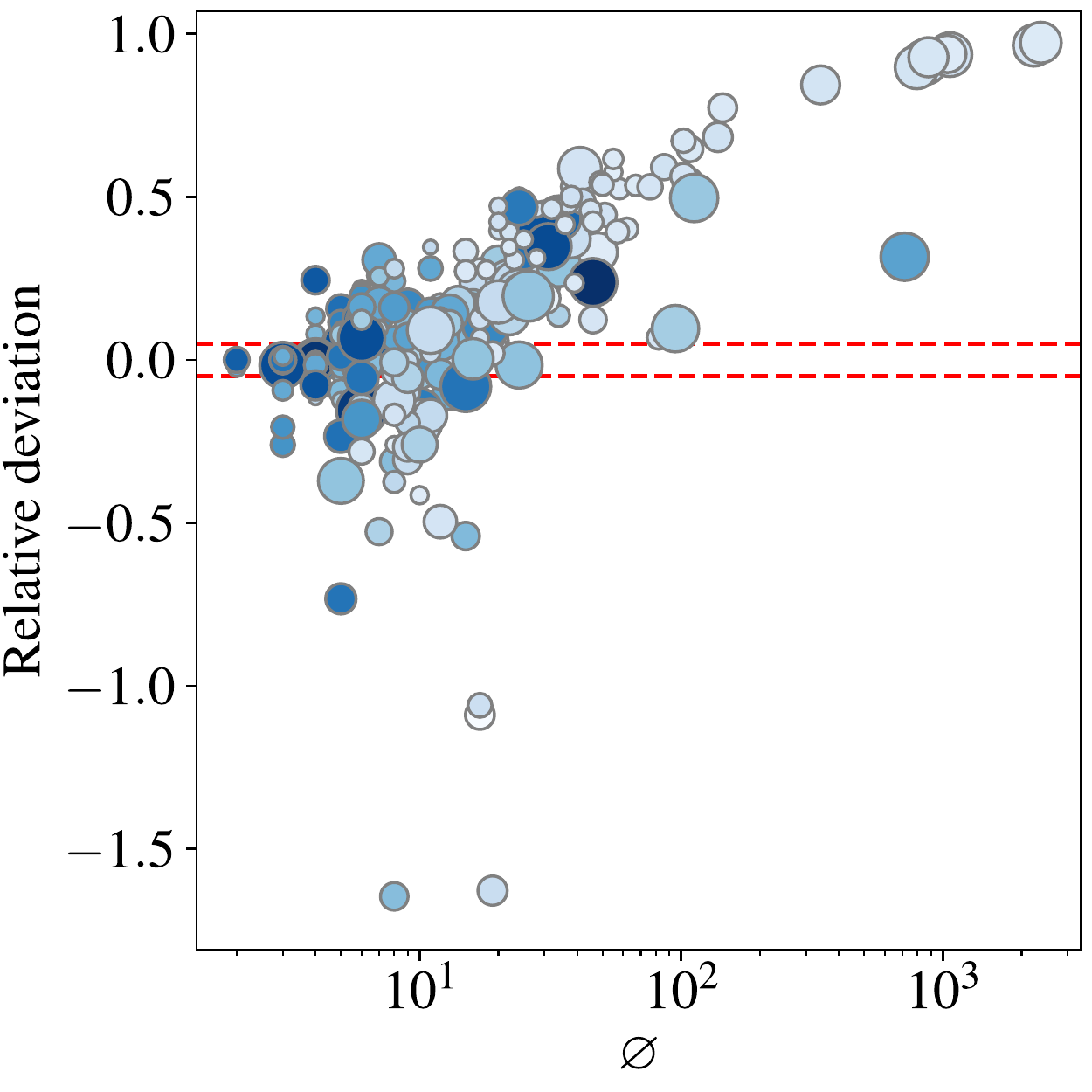} &
    \includegraphicsl{(f)}{width=.48\columnwidth}{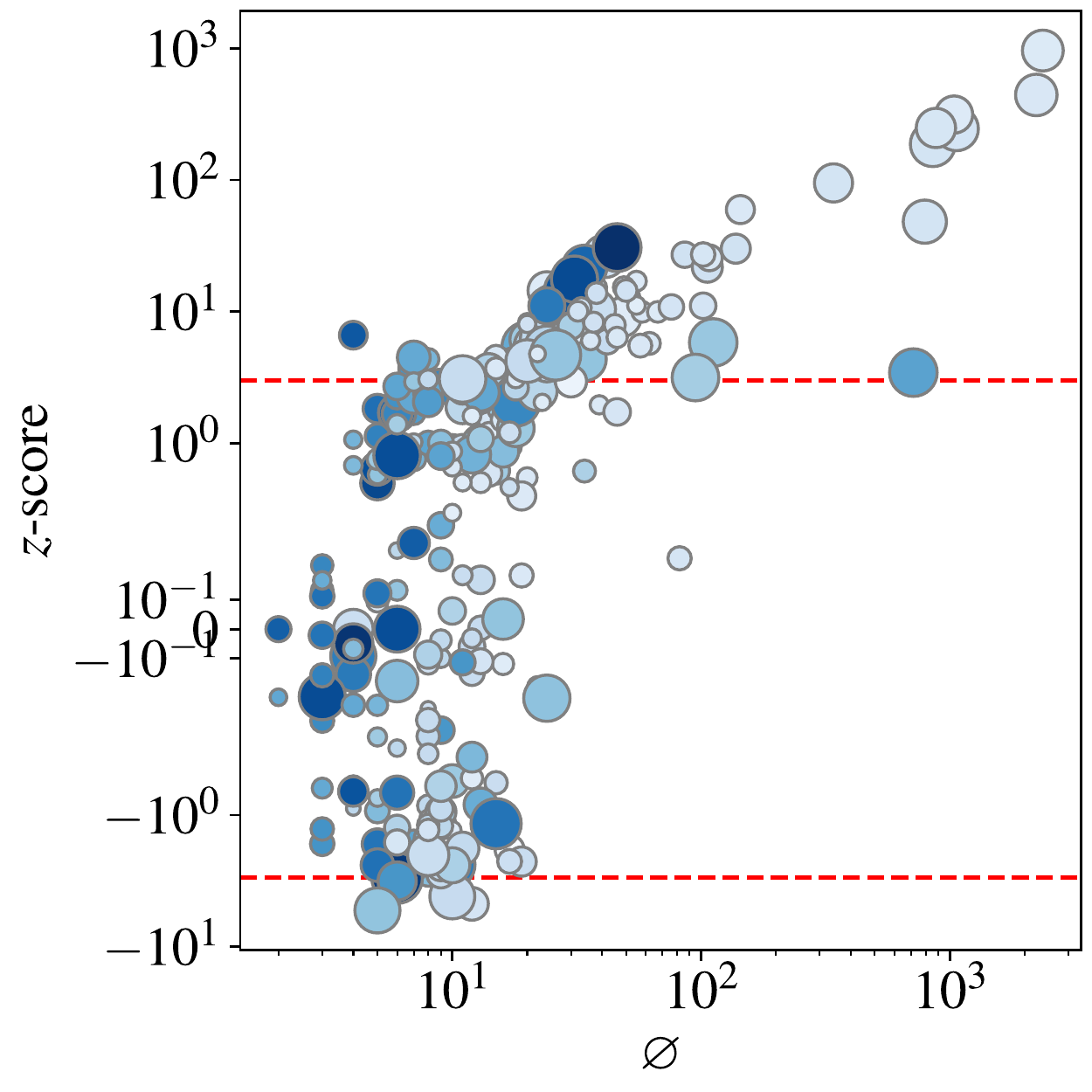} &
    \includegraphicsl{(g)}{width=.48\columnwidth}{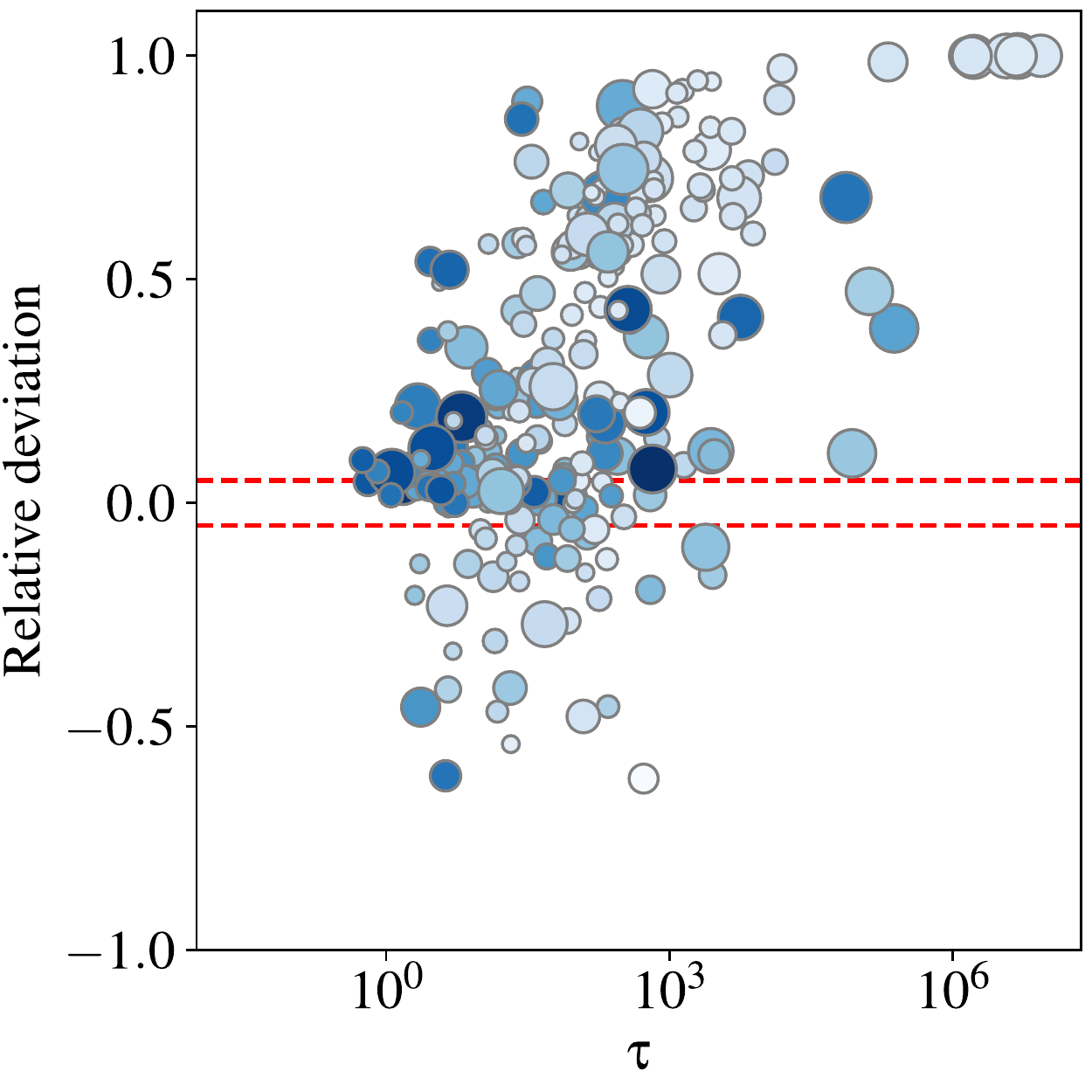}&
    \includegraphicsl{(h)}{width=.48\columnwidth}{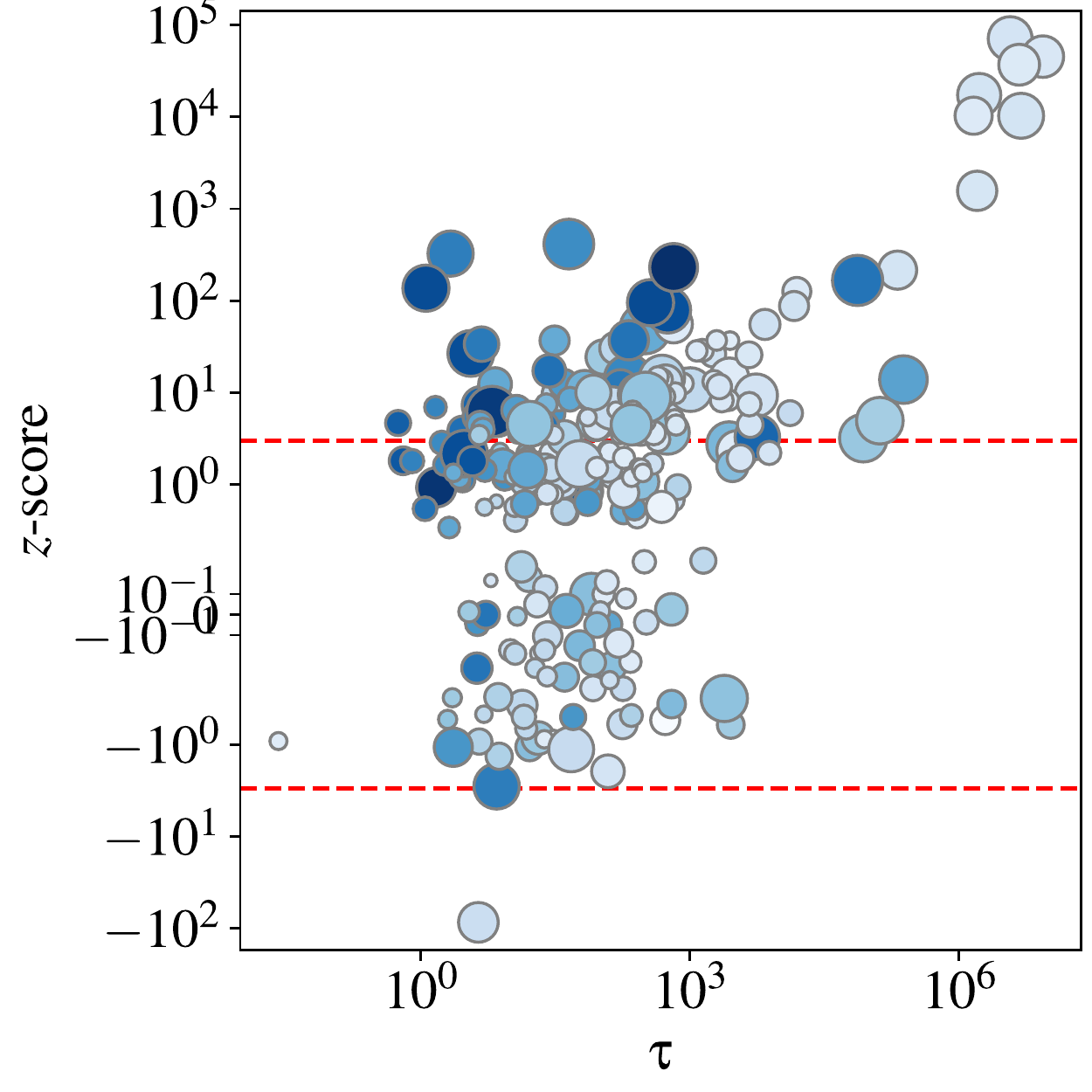}\\
  \end{tabular}

  \caption{Relative deviation and $z$-score values for the global and
    mean local clustering coefficients, $C_g$ and $C_l$, as well as
    diameter and characteristic time of a random walk, $\varnothing$ and
    $\tau$, as a function of
    their empirical values, for every network in the corpus, when using
    the DCSBM. The dashed red line marks the values of $|\Delta|=0.05$
    and $|z|=3$. The size of the symbol corresponds to the logarithm of
    the number of edges in the network, and the darkness to the mean
    degree.\label{fig:clust_diam}}
\end{figure*}

\begin{figure}
  \includegraphics[width=\columnwidth]{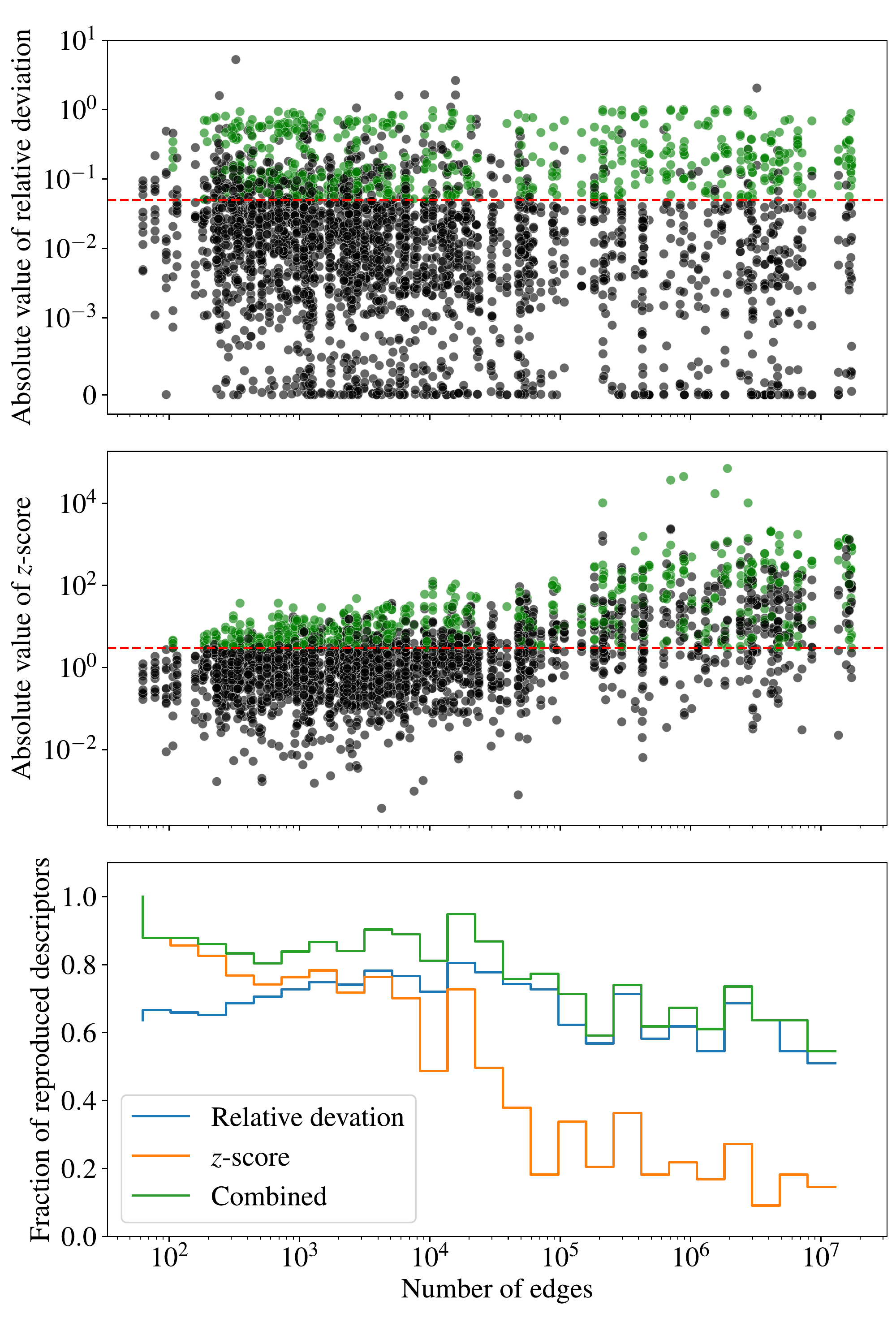} \caption{Absolute
  value of the relative deviation (top), $z$-score (middle) and fraction
  of reproduced descriptors (bottom), as a function of the number of
  edges, for every network in the corpus. The dashed red line marks the
  values of $|\Delta|=0.05$ and $|z|=3$. The fraction of descriptors
  reproduced correspond to those that have the value of either $\Delta$
  or $z$ below these thresholds. The points in green color correspond to
  the descriptors that are not reproduced according to this combined
  criterion.
  \label{fig:size}}
\end{figure}
\begin{figure}
  \begin{tabular}{ccc}
    \multicolumn{3}{c}{(a) Median}\\
    \smaller $z$-score &
    \smaller Relative deviation &
    \smaller Combined \\
    \includegraphics[width=.32\columnwidth]{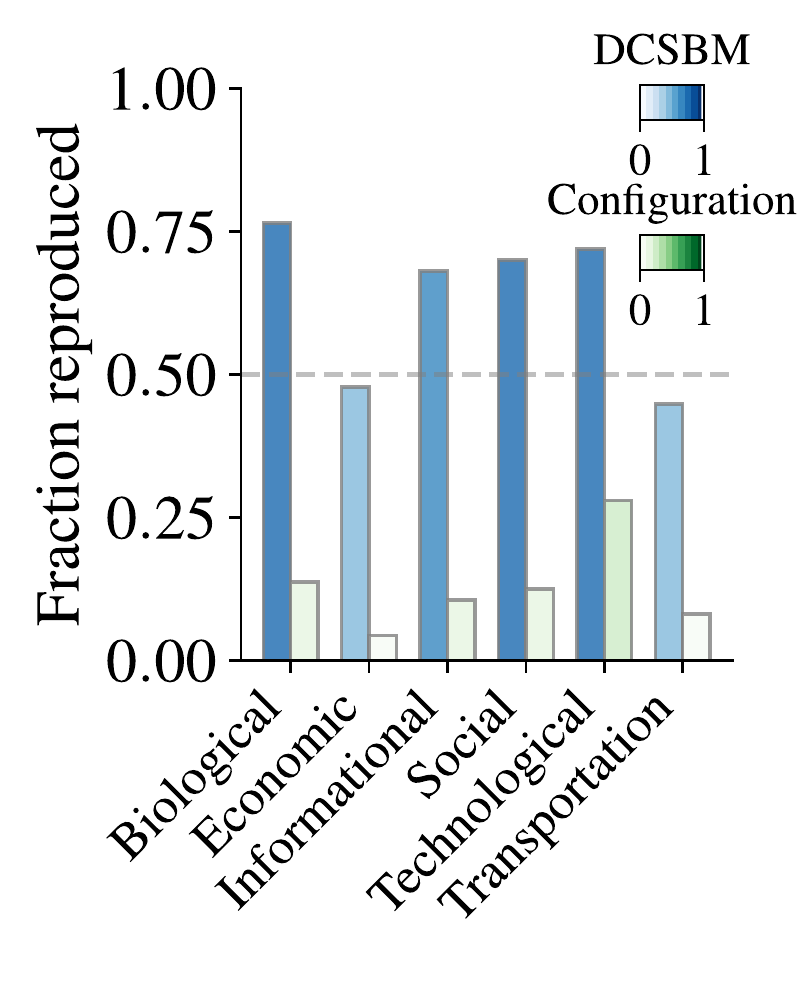} &
    \includegraphics[width=.32\columnwidth]{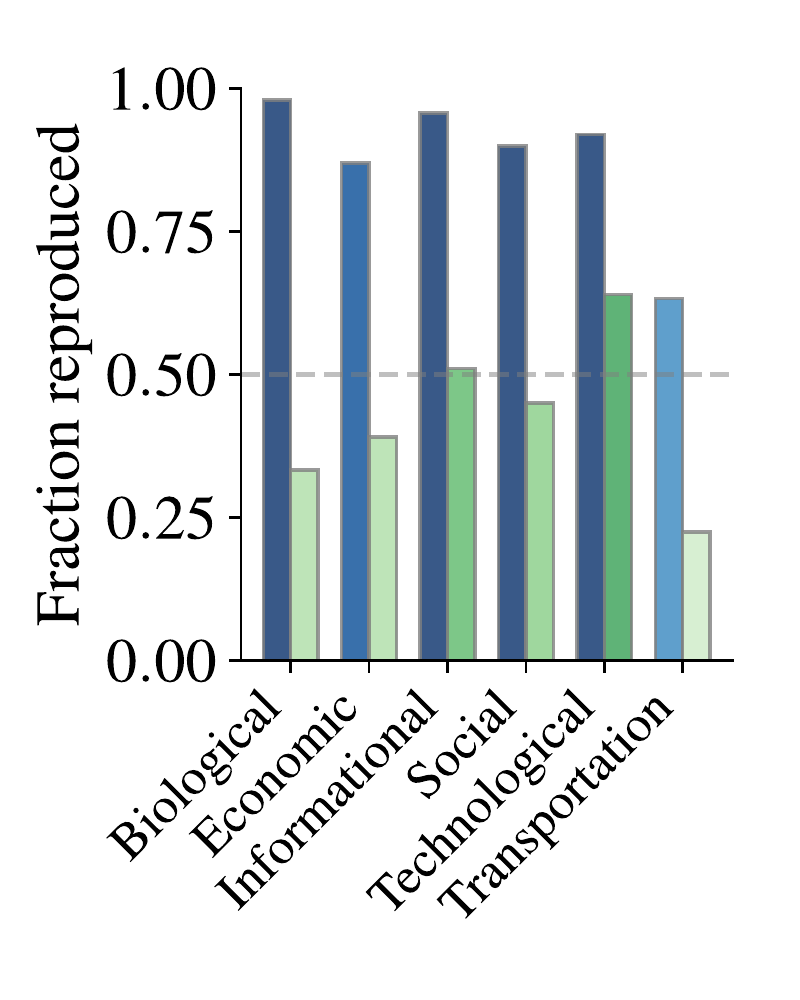} &
    \includegraphics[width=.32\columnwidth]{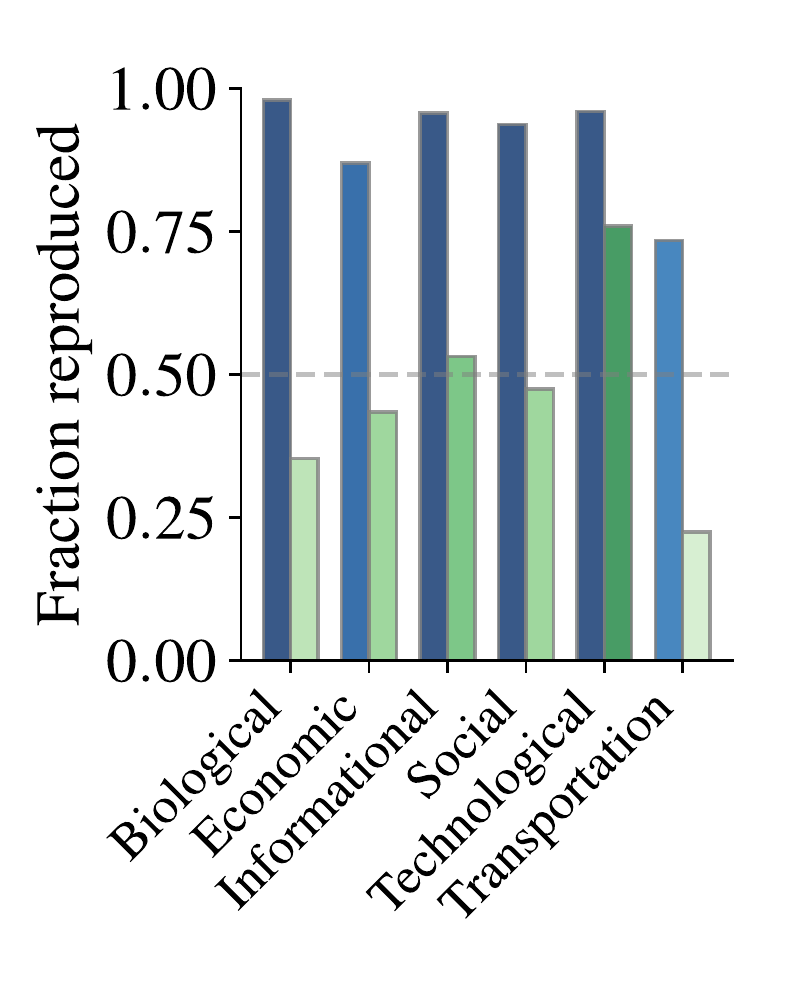}\\
    \multicolumn{3}{c}{(b) Mean}\\
    \smaller $z$-score &
    \smaller Relative deviation &
    \smaller Combined \\
    \includegraphics[width=.32\columnwidth]{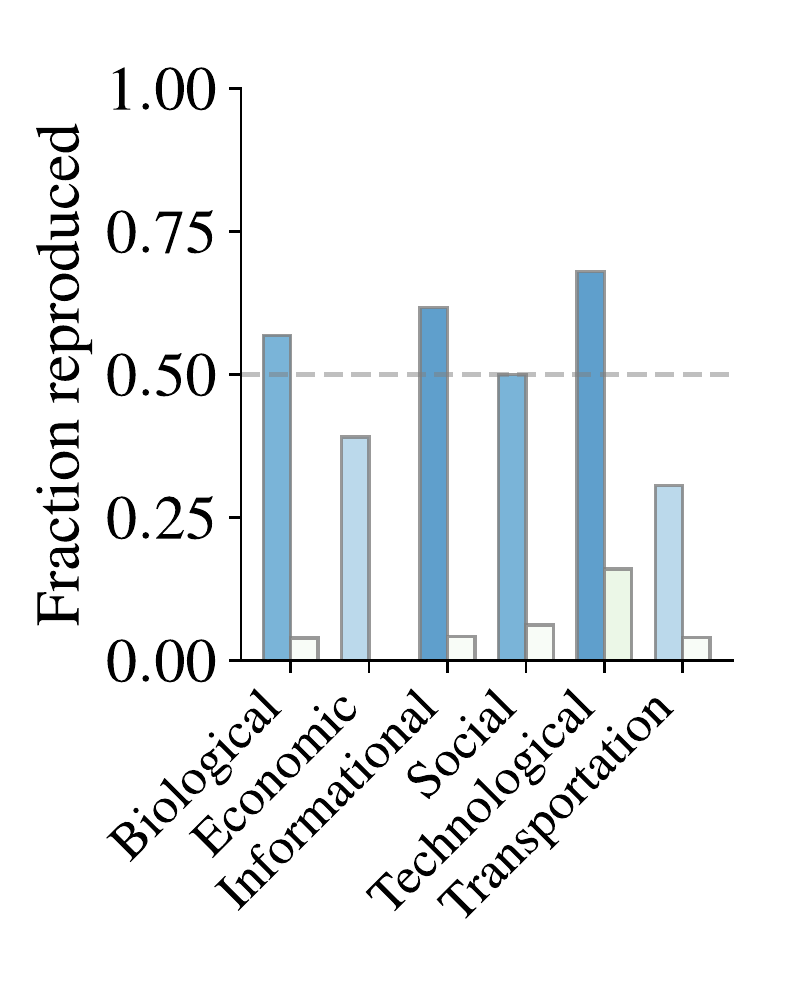} &
    \includegraphics[width=.32\columnwidth]{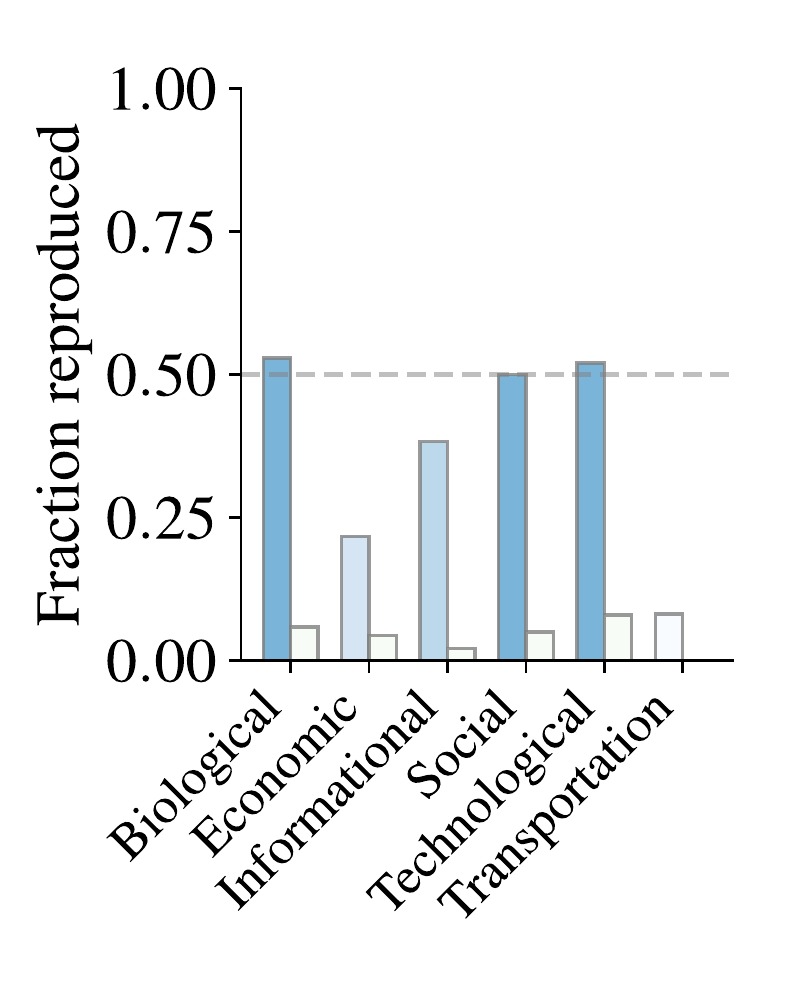} &
    \includegraphics[width=.32\columnwidth]{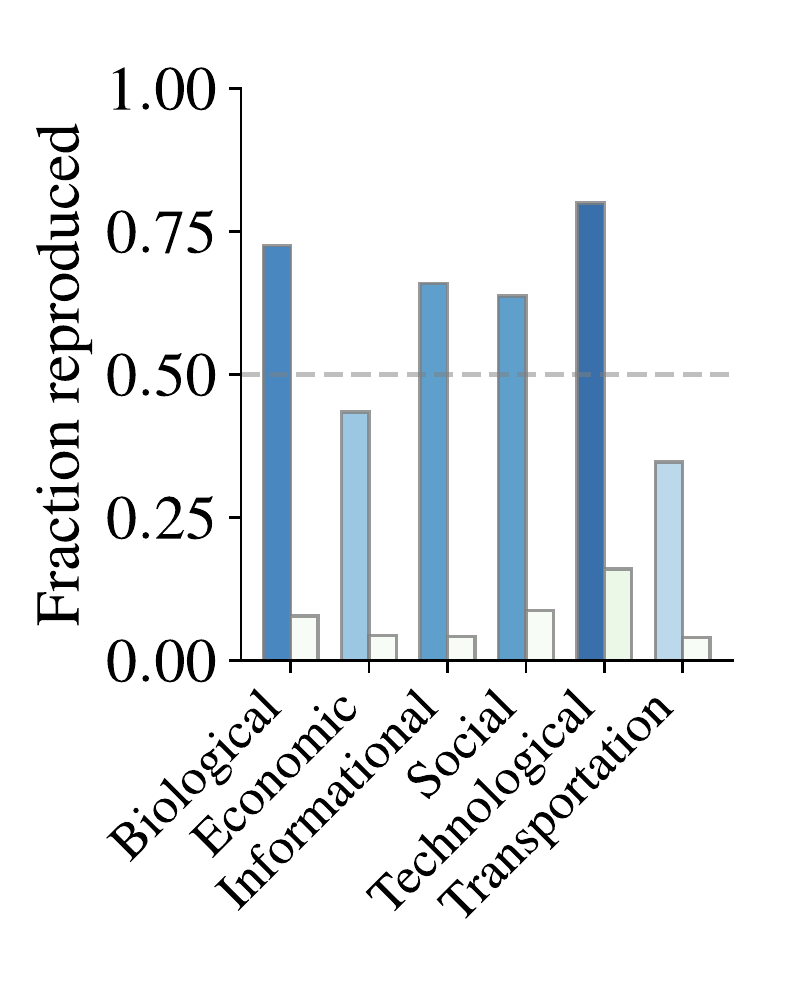}
  \end{tabular}
  \caption{Fraction of reproduced networks according to their domain,
    considering the (a) median and (b) mean values of either the
    $z$-score, the relative deviations, or their combined values, for
    both models (as shown in the legend). When the combined values are
    used, this means that a model is deemed compatible with a network
    when we obtain either $|\Delta|<0.05$ or $|z|<3$.
    \label{fig:desc-domain}}
\end{figure}

Overall, since we know that a model like the DCSBM cannot possibly
correspond to the true generative model of empirical networks, we should
expect that in situations where the network is sufficiently large, and
hence there is more abundant data, the values of the $z$-score will tend
to be high. Here we argue that since the objective of a model like the
DCSBM is to obtain a good approximation of the underlying model, not an
exact representation, the ultimate criterion is a combination of the
two, where we may deem the model compatible with the data when
\emph{either} the $z$-score \emph{or} the relative deviation has a
sufficiently low magnitude. For the purpose of clarity and simplicity of
our analysis, we will consider the thresholds $|z|=3$ and
$|\Delta|=0.05$ as reasonable choices to deem the model compatible with
data, although our results will not depend on these particular choices,
and we will always report the full range of values.

Before continuing, some important considerations regarding model
checking should be made. While an excellent model should fulfill both of
the above criteria simultaneously, we need to observe that a model that
maximally overfits, i.e. ascribes to the observed network a probability
of one, and to any other a probability of zero, will achieve the best
possible performance according to both relative deviation and
statistical significance. This occurs because we are using the same data
to perform both the model inference and evaluate its quality, which is
an invalid approach for \emph{model selection}. Therefore, it is
important to recognize the crucial difference between model checking and
model selection: the latter attempts to find the model alternative that
is better justified according to statistical evidence, while the former
simply finds systematic discrepancies between the inferred model and
data. In our analysis, protection against overfitting is obtained via
Bayesian inference, and we use model checking only to evaluate the
discrepancies (indeed, the fact we find discrepancies to begin with
shows that we cannot be massively overfitting). Another observation is
that when performing multiple comparison over many networks and
descriptors, some amount of ``statistically significant'' deviations are
always expected, even if the models inferred correspond to the true
ones, unless we incorporate the fact that we are doing multiple
comparisons in our criterion of statistical significance, which would be
the methodologically correct approach. We will not perform such a
correction in our analysis, because we do not seek to demonstrate the
absolute quality of DCSBM as a ultimately plausible hypothesis for
network formation. As we will see from our results, such a correction
would gain us very little.

Finally, in Table~\ref{tab:descriptors} we list the network descriptors
that are used in this work. Our approach requires scalar values, so we
constrained ourselves to this category, and furthermore we chose
quantities that can be computed quickly, so that robust statistics from
the predictive posterior distributions can be obtained. Given these
restrictions, we then chose descriptors that measure different aspects
of the network structure, both at a local and global levels. Further
details on the network descriptors are given in
Appendix~\ref{app:descriptors}.

\section{Network corpus}\label{sec:corpus}

We base our analysis on a corpus containing 275 networks spanning
various domains and several orders of size magnitude, as shown in
Fig.~\ref{fig:corpus}. We have not collected every network at our
disposal, but instead chosen networks that are as diverse as possible,
both in size and domain, and avoided many networks that are closely
related by belonging to the same subset. In Appendix~\ref{app:data} we
give more details about the datasets used.

\section{Results}\label{sec:fit}

In Fig.~\ref{fig:desc-deviation} we show the summaries of the posterior
predictive checks for each descriptor and network, for both models
considered. We observe a wide variety of deviation magnitudes, both for
the same descriptors across networks, and across descriptors. As
expected, the DCSBM results show systematically better agreement with
the data when compared with the configuration model. Overall, the
descriptors that show the worst agreement are the characteristic time of
a random walk ($\tau$) and the diameter ($\varnothing$), both of which
are particularly high for networks that are embedded in two dimensions,
and for which the DCSBM is an inaccurate approximation (more on this
below). Nevertheless, there is no single descriptor that the DCSBM does
not capture for fewer than 50\% of the networks. For descriptors like
$S$, $R_r$, $R_t$ and $\avg{c}$, the difference between the DCSBM and
the configuration model are relatively minor, indicating that those can
be captured to a substantial degree by the degree sequence alone.

When considering all descriptors simultaneously for each network, either
by the median or mean of the absolute values of the $z$-score and
relative deviation, we observe that a substantial majority of the
networks considered show good agreement with the DCSBM, as opposed to
the small minority that agree with the configuration model. The
difference between the median and the mean indicates that there a
sizeable fraction of the networks where the agreement is spoiled by a
few outlier descriptors --- typically $\tau$ and $\varnothing$.

The results obtained by the clustering coefficients are particularly
interesting, since it is often the case that they are well reproduced by
the DCSBM. This contrasts with what is commonly assumed, namely that the
DCSBM should not be able to capture the abundance of triangles often
seen in empirical networks, because in the limit where the number of
groups is much smaller than the total number of nodes, the DCSBM becomes
locally tree-like~\cite{decelle_asymptotic_2011}, with a vanishing
probability of forming triangles. Therefore, we may imagine that the
situations where there is an agreement with the DCSBM are those where
the clustering values are low. However, as we see in
Fig.~\ref{fig:clust_diam}(a) to (d), this is not quite true, and we
observe good agreements even when the clustering values are high. This
illustrates a point made in Ref.~\cite{peixoto_disentangling_2021}, that
it is possible to obtain an abundance of triangles with the SBM simply
by increasing the number of groups, in which case it can be explained as
a byproduct of homophily. Indeed this is a situation we see in
Fig.~\ref{fig:clust_diam}(a) to~(d), where both the relative deviation
and $z$-score values can be quite small even for extremal values of
clustering. However, we do notice a substantial variability between
agreements, and a fair amount of instances where the DCSBM cannot
capture the observed clustering values, even when they are moderate or
even small. This seems to indicate that there are a variety of processes
capable of resulting in high clustering values, with homophily being
only one of them~\cite{peixoto_disentangling_2021}. Overall, the mean
local clustering values tend to be harder to reproduce than the global
clustering values. In both cases, the $z$-scores are systematically
high, indicating that the clustering values are in general a good
criterion to reject the DCSBM as a statistically plausible model,
although the relative deviation values tend to be lower than what one
would naively expect, meaning that the model can still serve as a
reasonably accurate approximation for clustered networks in many cases.

The behavior seen for the clustering coefficient is different for the
diameter and characteristic time of a random walk, which are the least
well reproduced descriptors, as shown Fig.~\ref{fig:clust_diam}(e)
to~(h). For both these descriptors --- which are closely related, since
a network with a large diameter will also tend to result in a slow
mixing random walk --- it is rare to find a network with very high
empirical values which the DCSBM is able to accurately
describe. Therefore it seems indeed that the DCSBM offers an inadequate
ansatz to describe the structure of these networks, even by optimally
adjusting its complexity.

In Fig.~\ref{fig:size} we show how the model assessment depends on the
size of the network. As one could expect, the $z$-score values tend to
increase for larger networks, as more evidence becomes available against
the plausibility of the DCSBM as the true generative model. However, the
values of the relative deviation do not change appreciably for larger
networks, indicating that it remains a good approximation regardless of
the size of the system.\footnote{Sampling issues with MCMC could also
contribute to the elevated $z$-scores for larger networks, as we discuss
in Appendix~\ref{app:generation}.}

In Fig.~\ref{fig:desc-domain} we show a summary of the fraction of all
networks for which we obtain good agreement with either model, according
to the network domains. Overall, we see that most domains show similar
levels of agreements, except transportation and economic
networks. Transportation networks are often embedded in two-dimensional
spaces, resulting in large diameters and slow-mixing random walks. The
economic networks considered also tend to show large values of these
quantities, so the explanation for their discrepancy is the same.

\subsection{Predicting quality of fit}

\begin{figure}
  \begin{tabular}{cc}
    \multirow{2}{*}[5.5em]{\includegraphicsl{(a)}{width=.5\columnwidth}{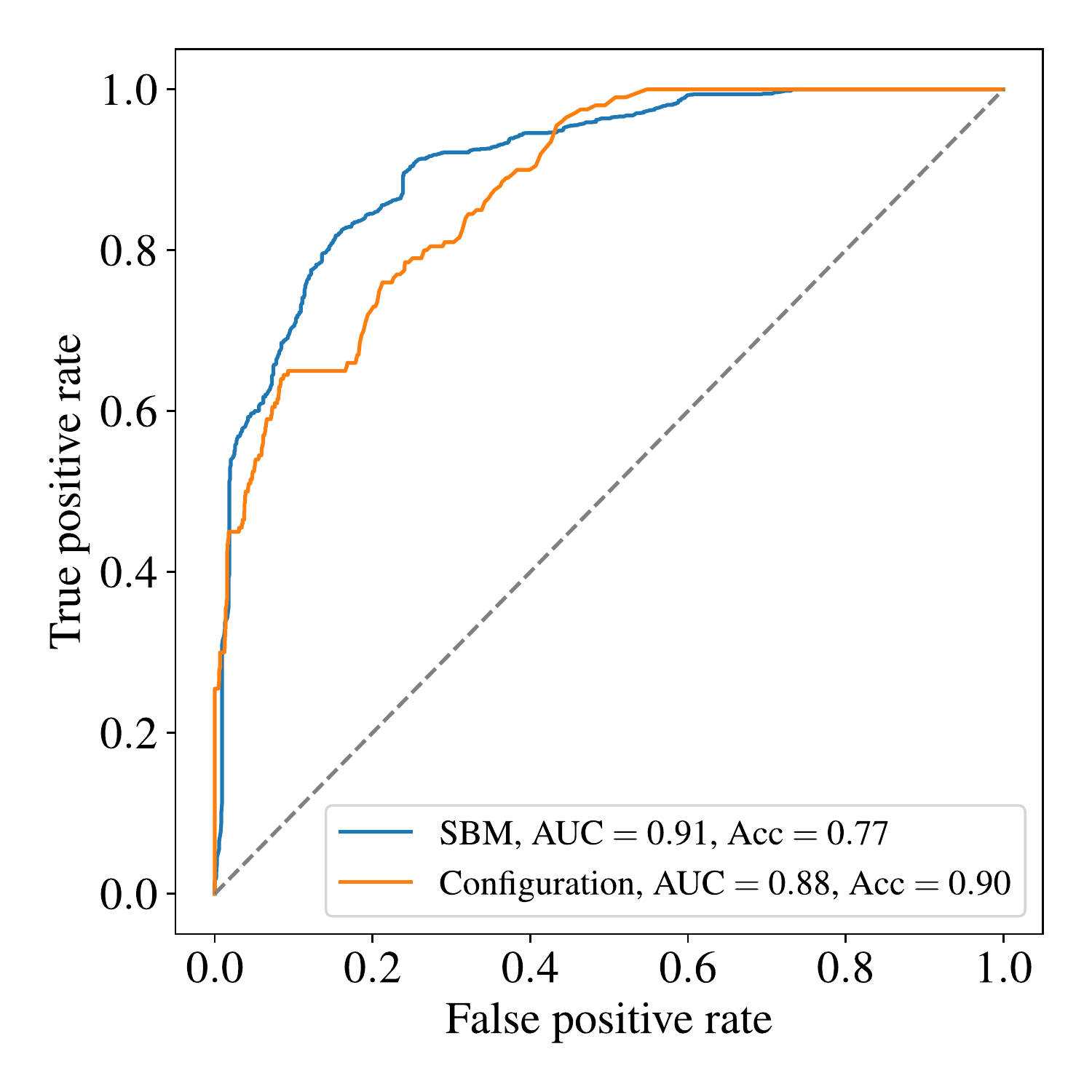}}
    &\includegraphicsl{(b)}{width=.5\columnwidth}{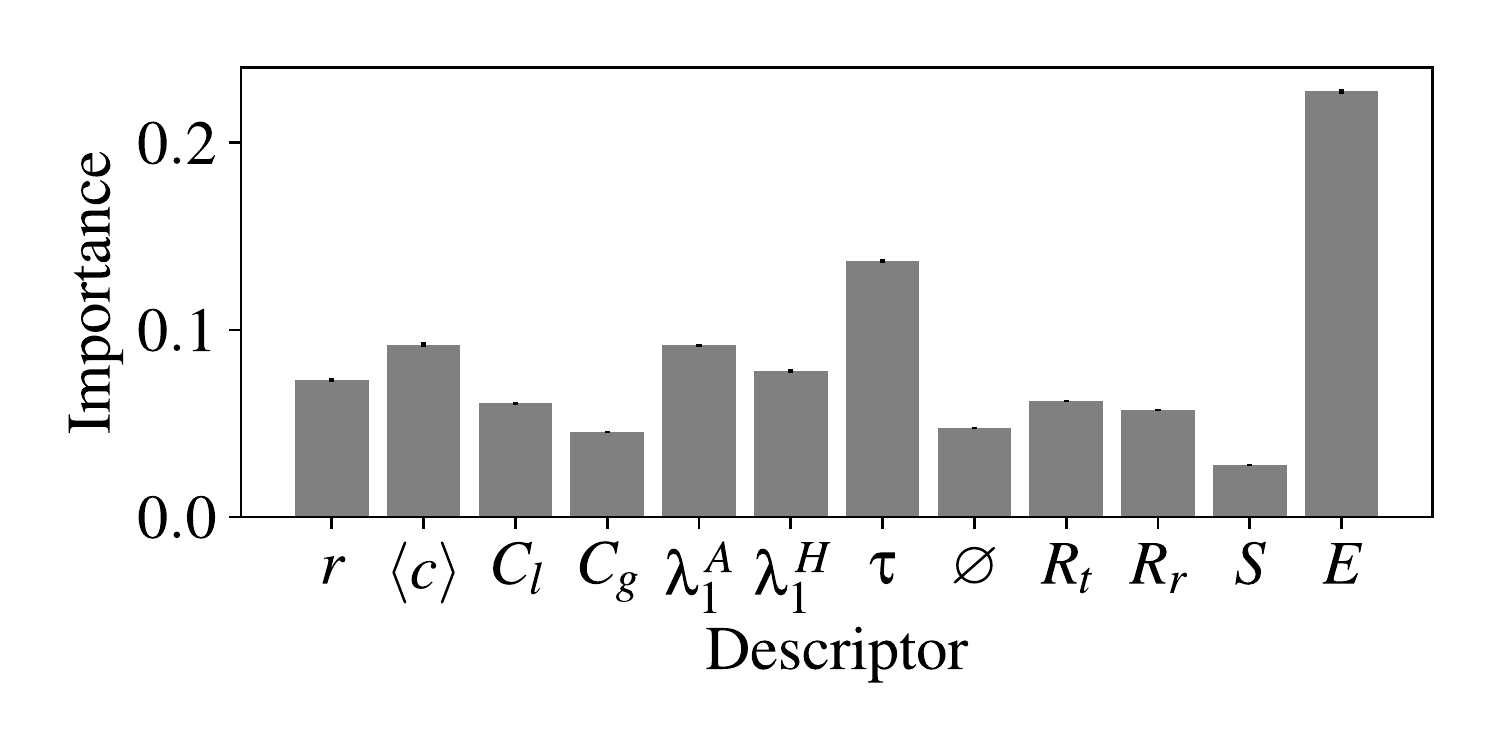}\\
    & \includegraphicsl{(c)}{width=.5\columnwidth}{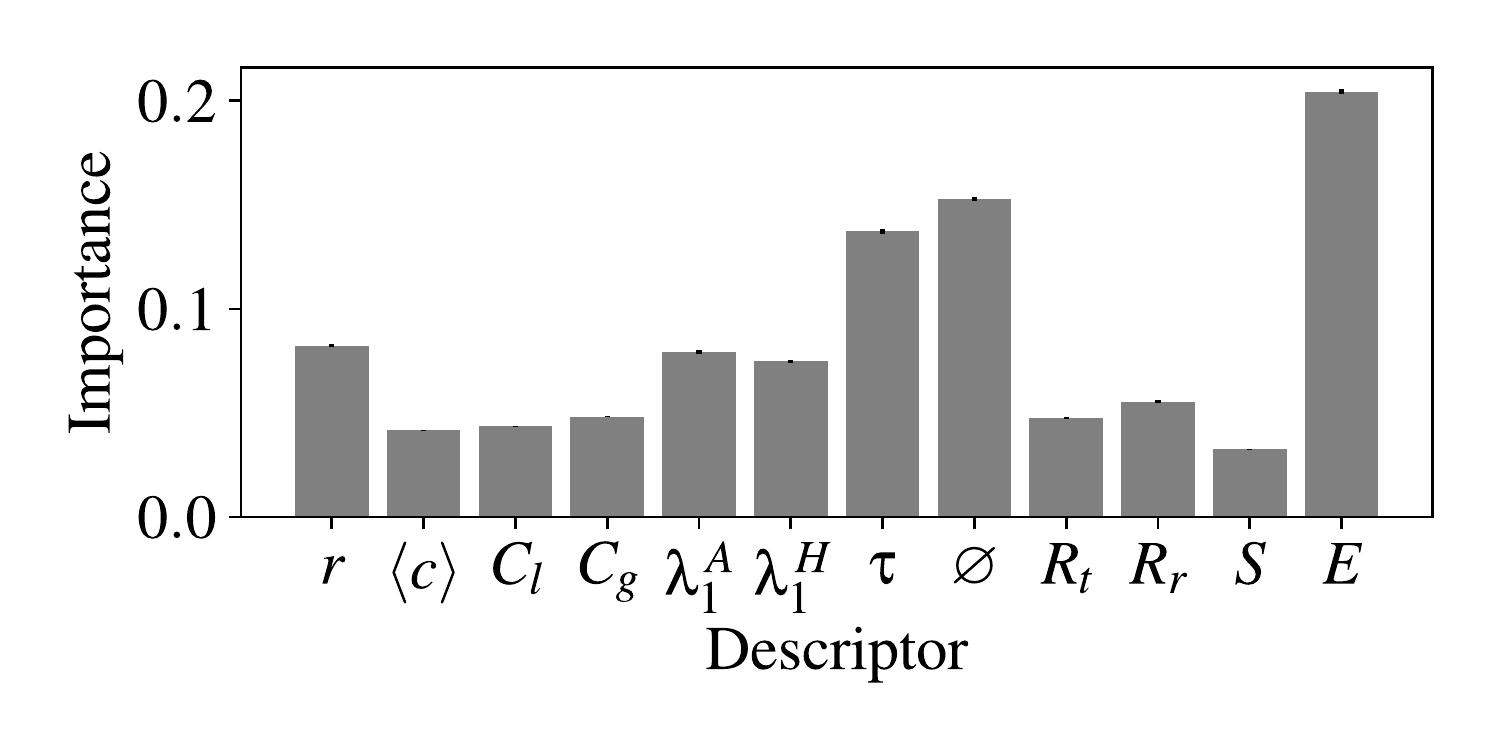}\\
    \includegraphicsl{(d)}{width=.5\columnwidth}{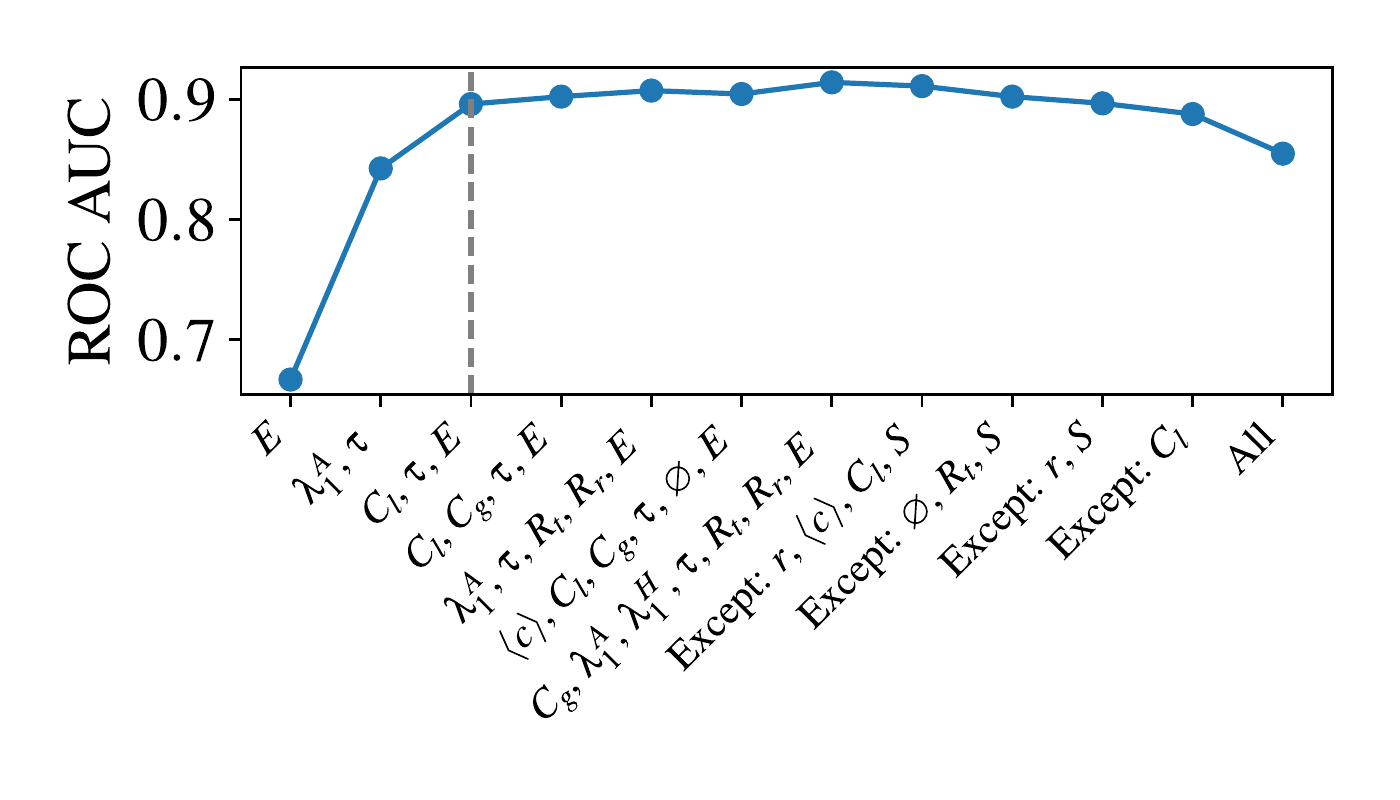} &
    \includegraphicsl{(e)}{width=.5\columnwidth}{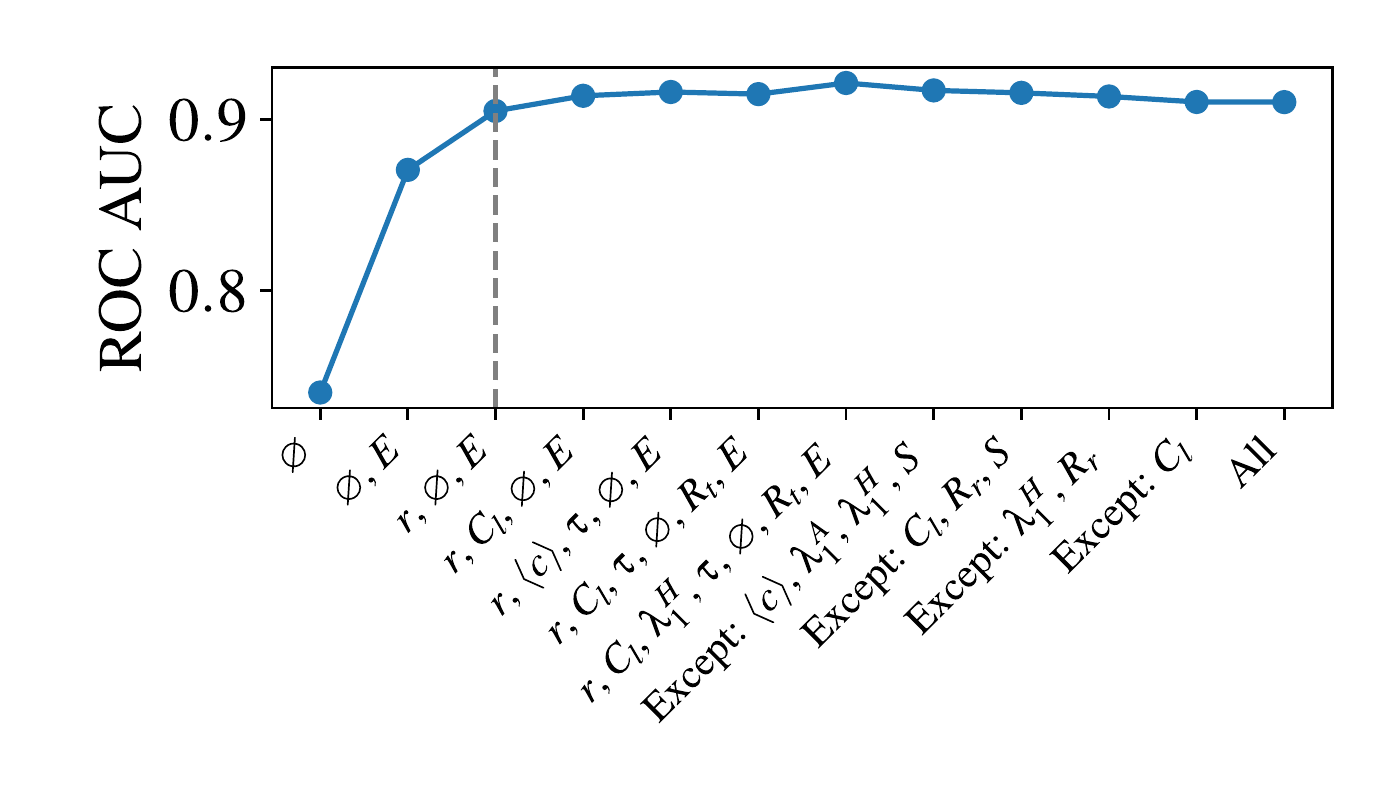}\\
    \includegraphicsl{(f)}{width=.5\columnwidth}{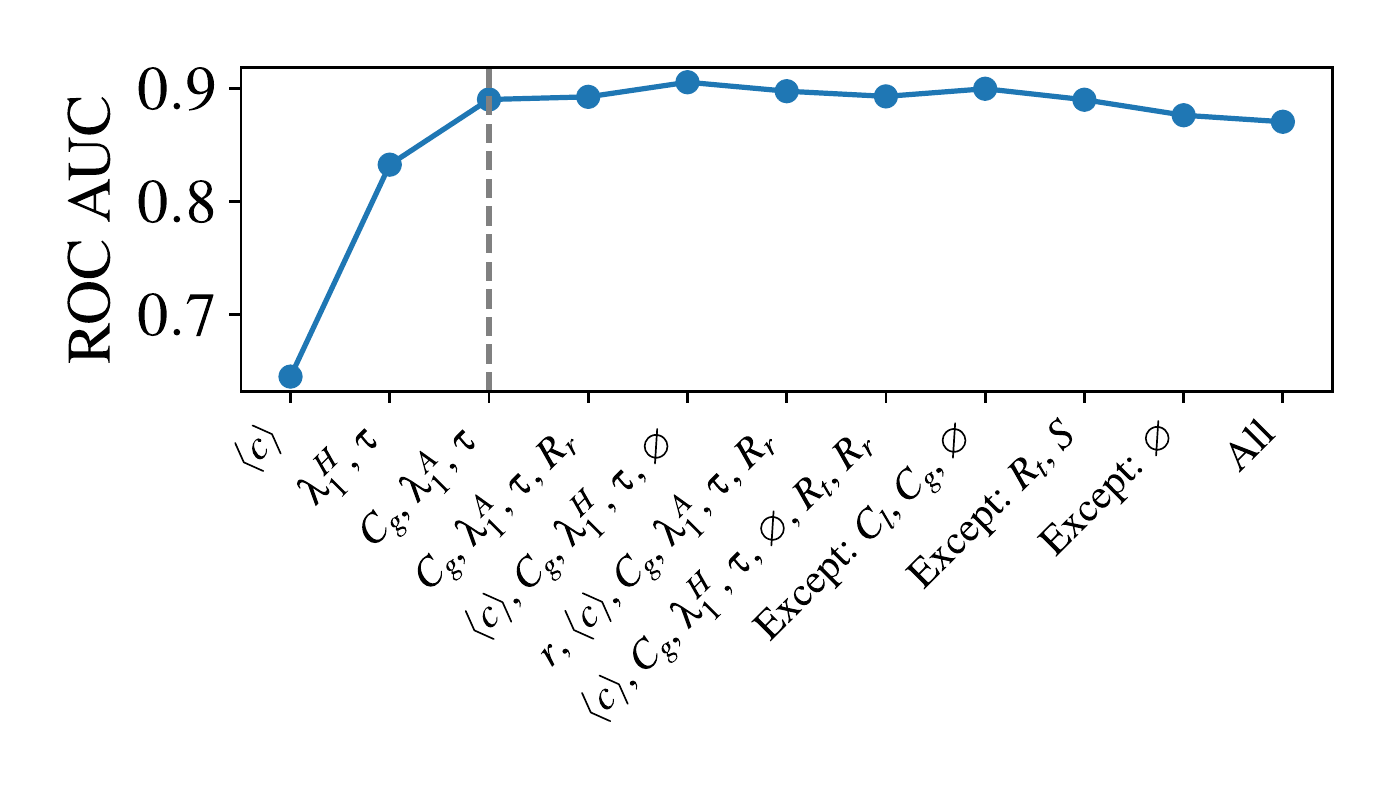} &
    \includegraphicsl{(g)}{width=.5\columnwidth}{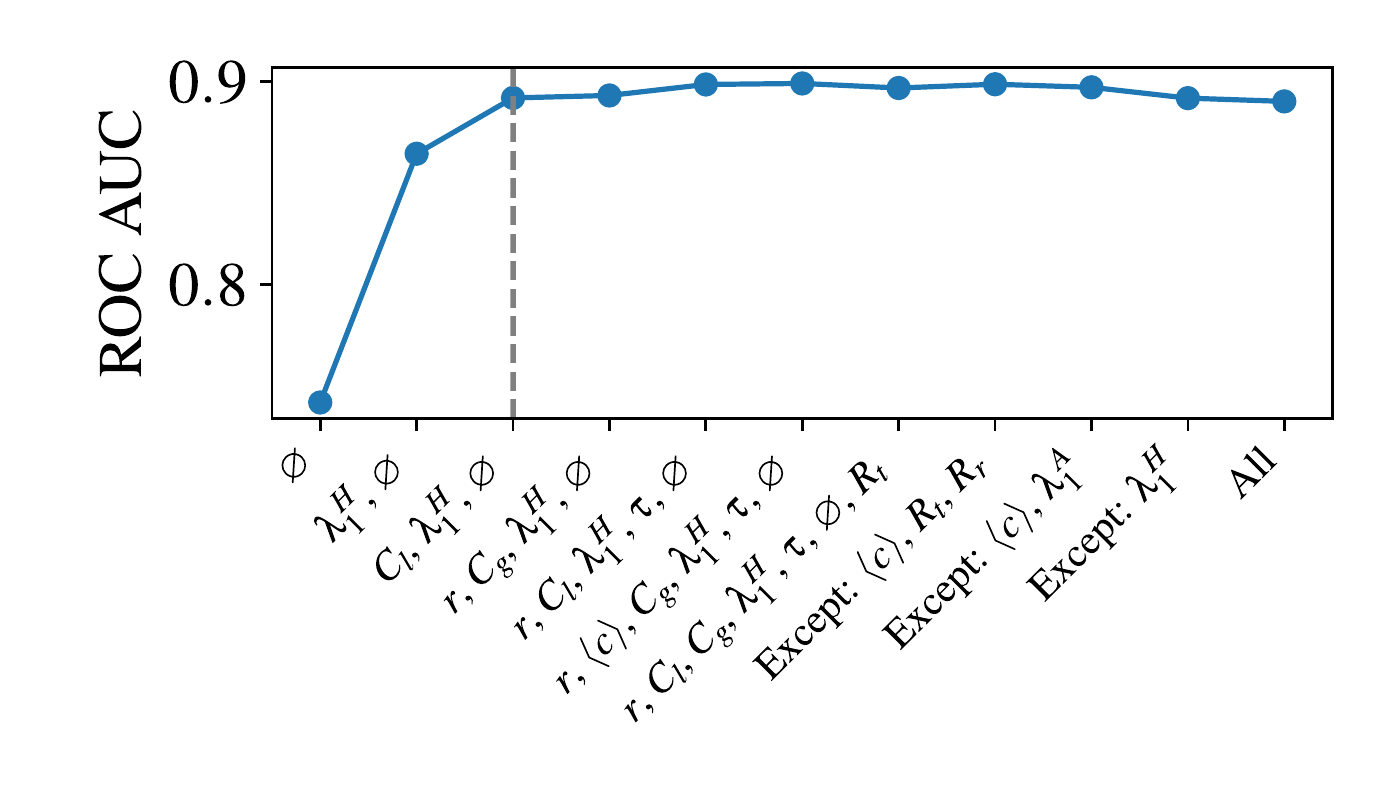}
  \end{tabular} \caption{Predictiveness of the quality of fit of the
  generative models considered, according to the empirical descriptor
  values, framed as a binary classification problem, as described in the
  text. (a) ROC curve for a leave-one-out random-forest classifier, (b)
  Gini feature importance for the configuration model, (c) same as (b)
  but for the DCSBM. Panels (d) and (e) show the best ROC AUC obtained
  for a set of descriptors of a given size, for the configuration model
  and DSCBM, respectively. Panels (f) and (g) show the same  as
  (d) and (e), respectively, but with the number of edges
  excluded from the analysis.\label{fig:prediction}}
\end{figure}

Now we address the question of whether it is possible to predict the
quality of fit of both models considered based solely on the empirical
values of the networks descriptors. If we can isolate the descriptors
which are most predictive, this would give us a general direction in
which more accurate models could be constructed.

In order to evaluate the predictability, we frame it as a binary
classification problem, where to each network $i$ is ascribed a binary
value $y_i = 0$ if we have simultaneously $|z_i|>3$ and
$|\Delta_i|>0.05$, or otherwise $y_i=1$. The feature vector for each
network is composed of the empirical values of the descriptors, $\bm x_i
= (r, \avg{c}, C_l, C_g, \lambda_1^A, \lambda_1^H, \tau, \varnothing,
R_r, R_t, S, E)$, with the addition of the number of edges $E$. For each
network $i$, we train a random forest classifier on the entire corpus
with that network removed, and evaluate the prediction score on the
held-out network. We then repeat this procedure for all networks in the
corpus, and evaluate how well the classifier is able to predict the
binary label. We present the results of this experiment in
Fig.~\ref{fig:prediction} (top) which shows the receiver operating
characteristic (ROC) curve, where the true positive rate and the false
positive rate are plotted for all threshold values used to reach a
classification. The area under the ROC curve (AUC), shown in the legend,
can be equivalently interpreted as the probability that a randomly
chosen true positive has a prediction score higher than a randomly
chosen true negative. For the DCSBM and configuration model, we obtain
an AUC value of 0.91 and 0.88, respectively. This indicates a fairly
high predictability, from which we can conclude that it is indeed often
possible to tell whether the models will provide a good or bad
agreement, based only on the descriptor values.

Further insight can be obtained by inspecting the importance of each
descriptor in the overall classification. We compute this via the
so-called Gini importance, defined as the total decrease in node
``impurity'' (i.e. how often a node in decision tree contributes to a
decision), weighted by the proportion of samples that reach that node,
averaged over all trees in the classifier.\footnote{We also computed
different a measure, called permutation importance, which leads to very
similar results (not shown).}  The results can be seen in
Fig.~\ref{fig:prediction} (b) and (c). In both cases, we see that the
number of edges is the most predictive descriptor, which is compatible
with what we had already seen in Fig.~\ref{fig:size}, namely that the
larger the networks are, the easier it becomes to reject a model
according to the $z$-score. Otherwise, as one would expect, the
importance of the remaining descriptors is largely compatible with their
reproducibility shown in Fig.~\ref{fig:desc-deviation}, where the
descriptors that agree the least with the inferred models tend to be the
most useful at predicting quality of fit beforehand.

This analysis allows us to emphasize two points: the characteristic time
of a random walk $\tau$ and the diameter $\varnothing$, both extremal
quantities of the network structure that are closely related, are the
most difficult descriptors to be captured by the DCSBM. Therefore, an
extension of the model that would cater for these properties would bring
the most benefit across all networks. However, beyond these two
descriptors, there is no substantial difference between the ones that
remain, indicating that there is no obvious direction that would bring a
systematic modelling improvement over all networks. On the other hand,
as we show in Appendix~\ref{app:descriptors}, the descriptor values and
their predictive posterior deviations show nontrivial correlations,
which means that if some of them are specifically targeted, it could
potentially improve the quality of fit of other descriptors.

In order to understand what is the minimal amount of information
required to predict the suitability of both models, and in this way
remove the redundancy provided by the different descriptors, we computed
the best ROC AUC obtained by a combination of descriptors of a given
size, as shown in Fig.~\ref{fig:prediction}(d) and (e). In both cases we
see that the predictability is saturated by only few
descriptors.\footnote{Since we optimized exhaustively for all descriptor
combinations of a given size, care should be taken to avoid overfitting,
despite the leave-one-out cross-validation, because the optimization was
performed the same set of networks. Because of this, we consider always
the smallest set of descriptors that reaches a ROC AUC close to the
optimum, not the actual optimum which is likely to be overfitting.} In
the case of the configuration model most of the predictability is
already achieved by a combination of $(C_l, \tau, E)$. For the DCSBM we
get instead $(r, \varnothing, E)$. If we remove the
number of edges from the set of features (since it is not informative on
the actual network structure), we obtain instead
$(C_g,\lambda_1^A,\tau)$ and $(C_l,\lambda^H_1,\varnothing)$, for the
configuration model and DCSBM, respectively. It should be emphasized
that if a descriptor does not appear in the minimal set this does not
mean it is not predictive of the quality of fit, only that it offers
largely redundant information in that regard. Thus, for both models if
we replace $\varnothing$ with $\tau$ or $\lambda^H_1$ with
$\lambda_1^A$, etc, we get similar results. This suggests that, besides
spatial embeddedness (which influence $\varnothing$ and $\tau$ the
most), the addition of explicit mechanisms for triangle formation (which
affects $C_g,C_l,\lambda^H_1, \lambda^A_1$ directly) might improve the
overall expressiveness of the DCSBM --- which in fact has been observed
in a more limited dataset~\cite{peixoto_disentangling_2021}.

\section{Conclusion}\label{sec:conclusion}

We performed a systematic analysis of posterior predictive checks of the
SBM on a diverse corpus of empirical networks, spanning a broad range of
sizes and domains. Using a variety of network descriptors, we observed
that the SBM is able to accurately capture the structure of the majority
of networks in the corpus. The types of networks that show the worst
agreement with DCSBM tend to possess a large diameter and a slow mixing
of random walks --- features that are commonly associated with a
low-dimensional spatial embedding, and a violation of the
``small-world'' property. For the other kinds of networks the agreement
tends to be fairly good, even for many networks with an abundance of
triangles, in contradiction to what is commonly assumed to be possible
with this class of models.

We have also identified the minimal set of network descriptors capable
of predicting the quality of fit of the SBM, which is composed of the
network diameter and characteristic time of a random walk as the most
important, followed by clustering as a secondary feature. This points to
the most productive directions in which this class of models could be
improved.

One of the limitations of our analysis is that it is conditioned on the
set of descriptors used, and thus shortcomings or successes of the model
with respect to other properties not analysed are not uncovered. A
natural extension of our work would be to consider an even broader set
of descriptors that could reveal more relevant dimensions for the
comparison. This kind of analysis is open ended, as there is no short
supply of possible network descriptors. We hope our work will motivate
further study in this direction, and with a larger variety of generative
models within or beyond the SBM family.

\section*{Aknowledgements}\label{sec:aknowledgements}
The computational results presented have been achieved using the Vienna
Scientific Cluster (VSC).

\bibliography{bib,data}

\appendix

\section{Posterior predictive sampling}\label{app:generation}
As described in the main text, we obtain samples from the posterior
predictive distribution of Eq.~\ref{eq:ppred} by first sampling from the
posterior distribution of Eq.~\ref{eq:posterior} using MCMC and then
generating new networks from the inferred models. More specifically, we
sample $(\A,\kk,\ee,\bb)$ from
\begin{equation}
  P(\A, \kk, \ee, \bb | \G) = \frac{P(\G|\A)P(\A | \kk, \ee, \bb)P(\kk, \ee, \bb)}{P(\G)},
\end{equation}
using the merge-split MCMC of Ref.~\cite{peixoto_merge-split_2020},
together with the agglomerative initialization heuristic of
Refs.~\cite{peixoto_efficient_2014,peixoto_hierarchical_2014}, and the
multigraph edge moves of Ref.~\cite{peixoto_latent_2020}. For networks
of size up to $E=10^5$ edges we observe good equilibration of the MCMC
runs, but for large networks it becomes too slow. For these large
networks we settle for a point estimate of the partition $\bb$ obtained
by several runs of the initialization algorithm and keeping the best
result, and then we equilibrate the chain according to $\A$ alone (which
affects $\kk$ and $\ee$), which tends to happen quickly. We have
verified that performing this calculation several times yields very
similar results. The only noticeable outcome of this shortcut for larger
networks is that it tends to reduce the variance of the posterior
predictive distributions, which can potentially contribute to the
elevated $z$-scores we obtained in our analysis. However, since the
relative deviation values we obtained did not seem to depend on the size
of the network, this gives us confidence that this approach does not
introduce significant biases.

Given a sample $(\A, \kk, \ee, \bb)$, we are interested only in
$(\kk,\ee,\bb)$ (and hence samples from their marginal distribution), so
we discard $\A$ and sample a new multigraph $\A'$ from the model of
Eq.~\ref{eq:dcsbm}. This can be done exactly with an efficient algorithm
that works similarly to what was proposed in
Refs.~\cite{parkkinen_block_2009,rohe_note_2018}, but is valid for the
microcanonical model: Given the parameters $(\kk,\ee,\bb)$ we proceed by
creating for each group $r$ a multiset of candidate nodes $\bm v_r$,
containing $k_i$ copies of each node $i$ with $b_i=r$. Then, for each
group pair $(r,s)$ with $r\le s$ and $e_{rs} > 0$, we repeat the
following three steps for an $e_{rs}$ number of times (or $e_{rs}/2$ if
$r=s$):
\begin{enumerate}
  \item We sample a node $i$ from the multiset $\bm v_r$ uniformly at
    random, and we remove it from the multiset.
  \item We sample a node $j$ from the multiset $\bm v_s$ uniformly at
    random, and we remove it from the multiset.
  \item We add an edge $(i,j)$ to $\A$ (i.e. increment $A_{ij}$ by one,
        or two if $i=j$).
\end{enumerate}
The resulting multigraph $\A$ is sampled exactly with a probability
given by Eq.\ref{eq:dcsbm}. Since the number of nonzero entries of $\bm
e$ cannot be larger than the total number of edges $E$, the whole
algorithm finishes in time $O(N+E)$, where $N$ is the number of
nodes.

Given a sample $\A$, we obtain a simple graph $\G$ simply by removing
all self-loops and truncating the edge multiplicities, i.e.
\begin{equation}
  G_{ij} =
  \begin{cases}
    1, \text{ if } A_{ij} > 0 \text{ and } i\ne j,\\
    0, \text{ otherwise}.
  \end{cases}
\end{equation}
Finally, given $\G$ we compute the network descriptor $f(\G)$ of interest.

A C++ implementation of every algorithm used in this analysis is freely
available as part of the \texttt{graph-tool}
library~\cite{peixoto_graph-tool_2014}.

\section{Network descriptors}\label{app:descriptors}
\begin{figure*}
  \includegraphicsl{(a)}{width=.49\textwidth}{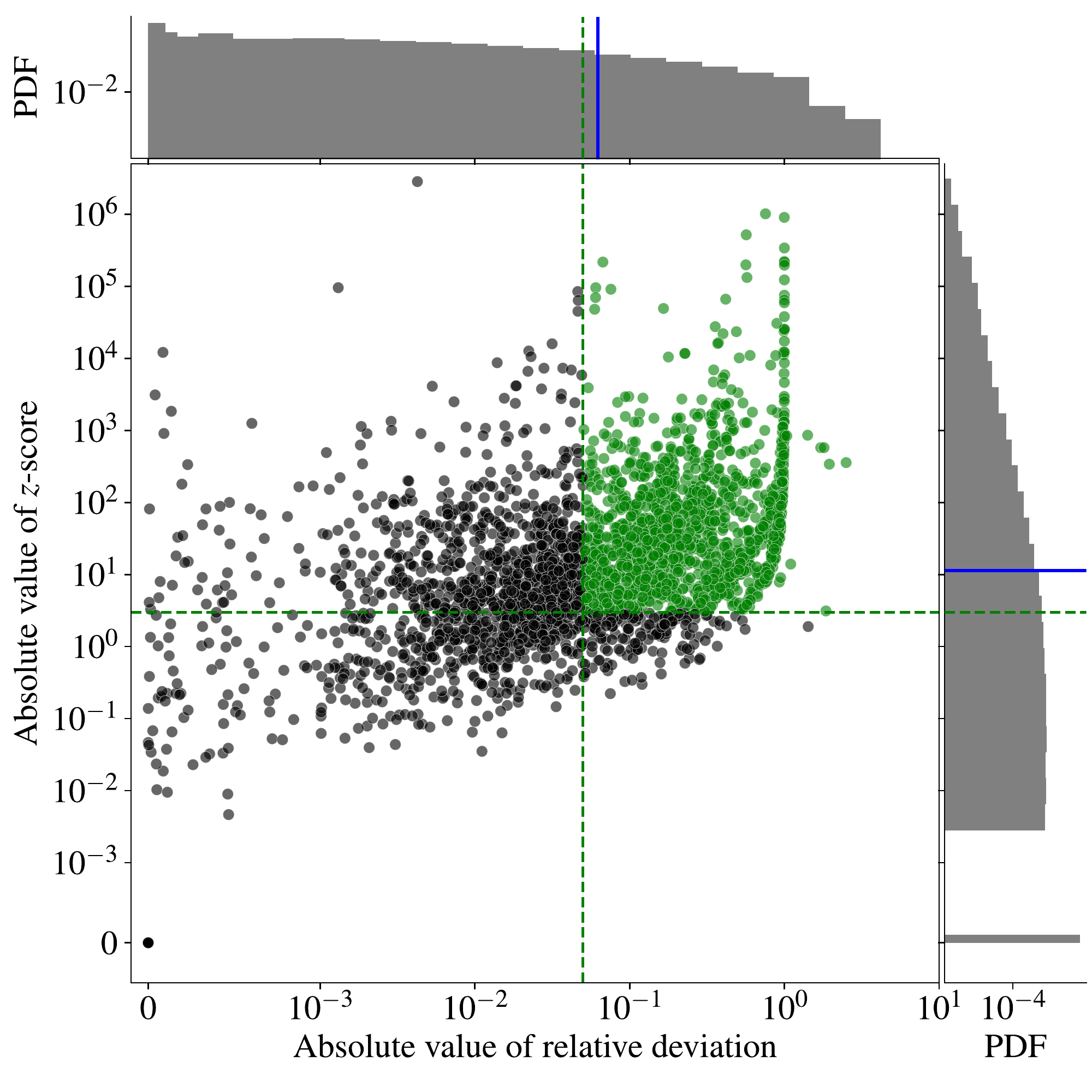}
  \includegraphicsl{(b)}{width=.49\textwidth}{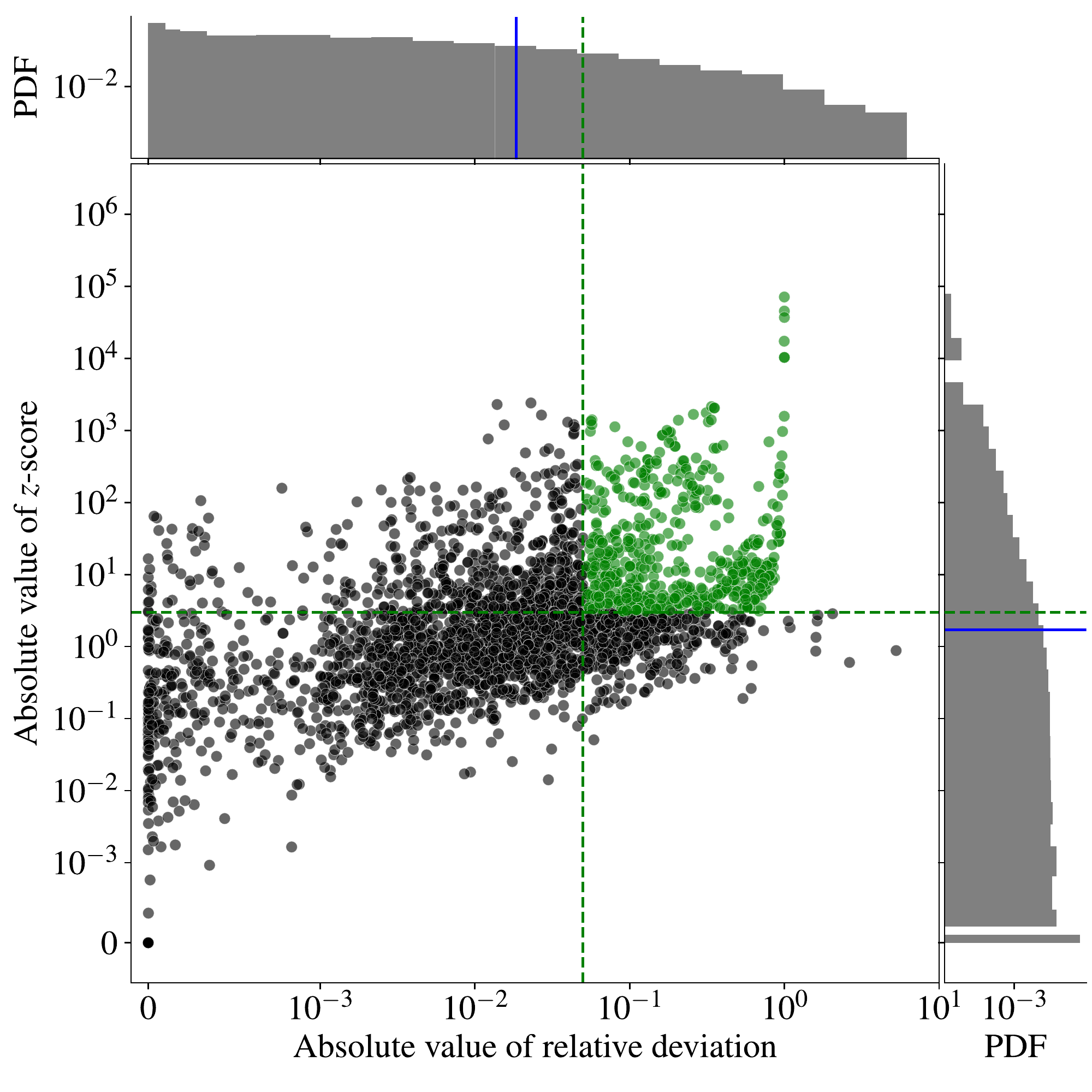}

  \caption{Absolute value of the $z$-score versus absolute value of
  relative deviation, for every descriptor value and network in the
  corpus, according to (a) the configuration model and (b) the
  DCSBM. The dashed lines mark the values $|z|=3$ and $|\Delta|=0.05$, and
  the histograms the marginal distributions. The solid blue lines mark
  the median values.\label{fig:dev-corr}}
\end{figure*}
\begin{figure*}
  \begin{tabular}{ccc}
    \includegraphicsl{(a)}{width=.33\textwidth}{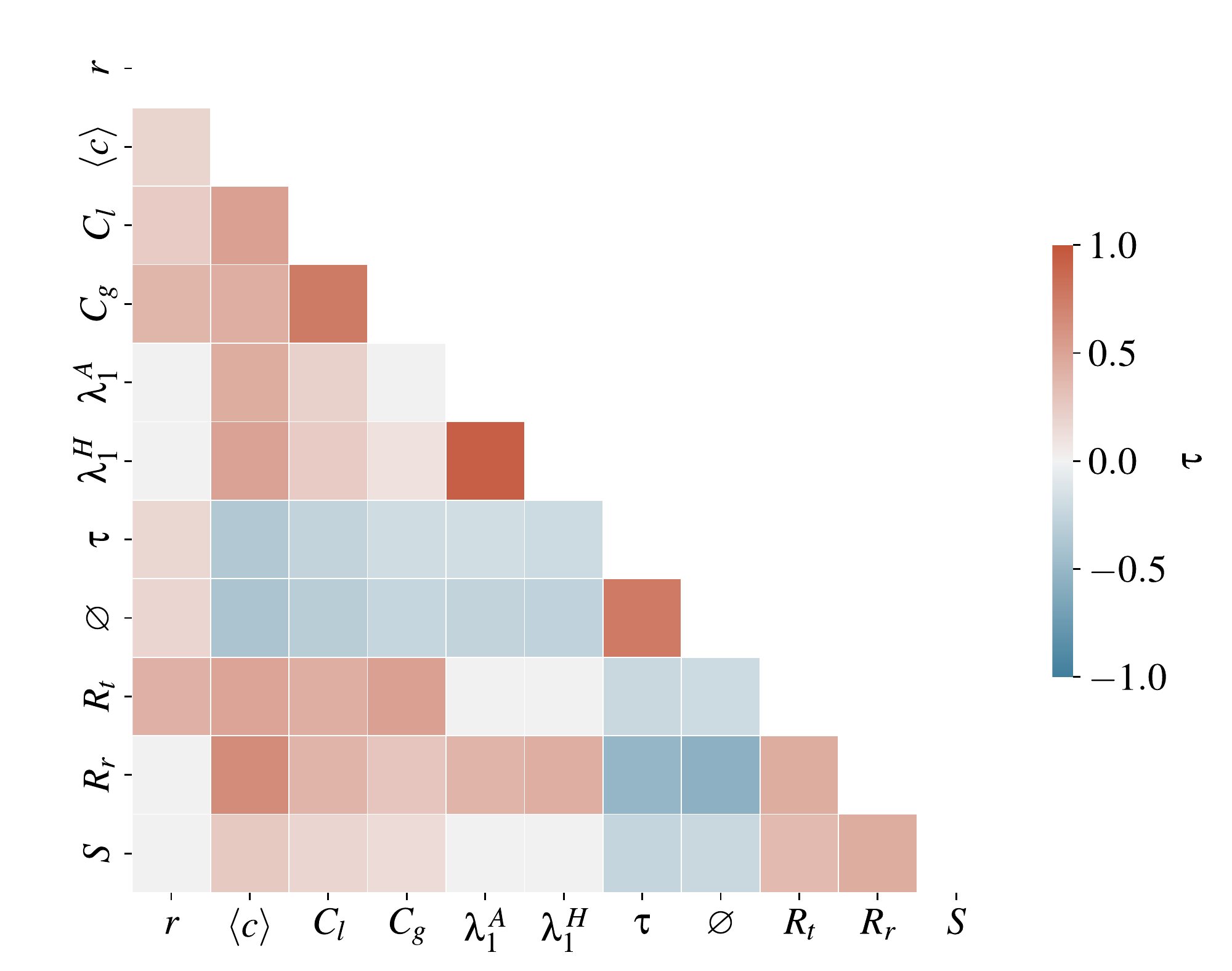} &
    \includegraphicsl{(b)}{width=.33\textwidth}{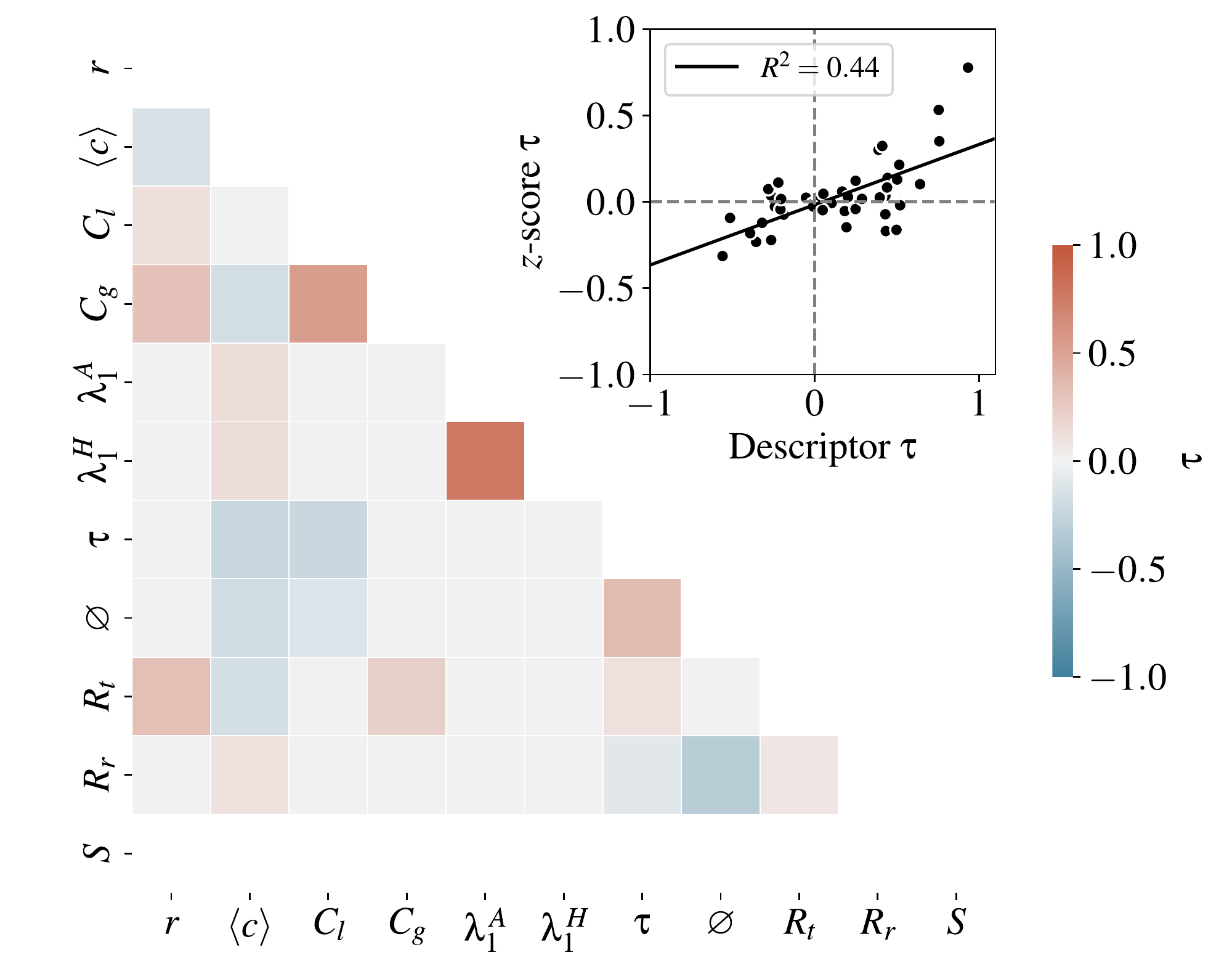} &
    \includegraphicsl{(c)}{width=.33\textwidth}{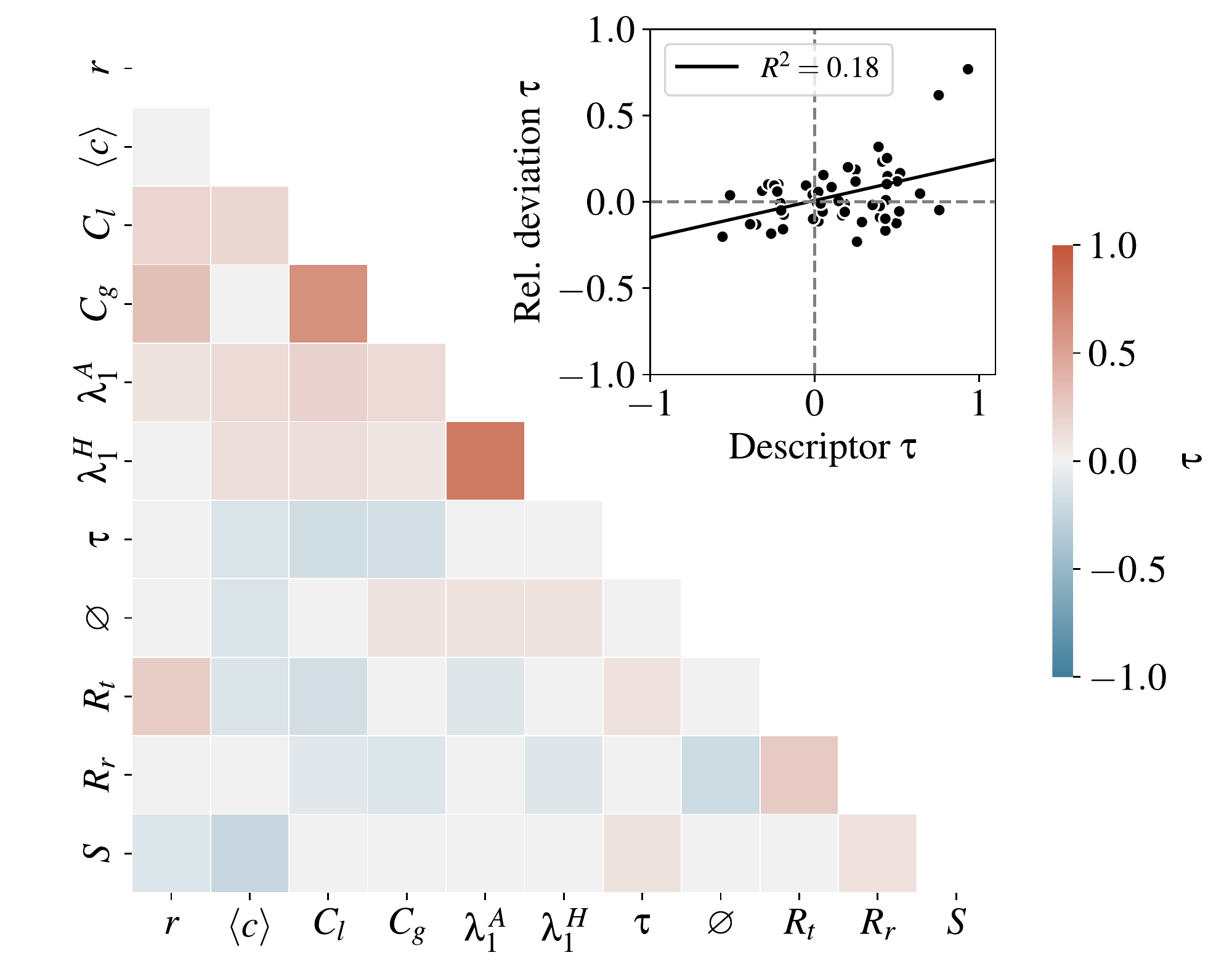}
  \end{tabular} \caption{(a) Kendall's correlation coefficient $\tau$
  between pairs of descriptor values across all networks in the
  corpus. Panels (b) and (c) show the same but for $z$-score and
  relative deviation values, respectively, according to the DCSBM. The
  insets show the correlation between coefficients from each respective
  panel and panel (a). \label{fig:desc-corr}}
\end{figure*}

Below are the definitions of the descriptors used in our analyses.

\begin{description}
\item[Degree assortativity, $r$] Defined as~\cite{newman_mixing_2003}
  \[
  r = \frac{\sum_{kk'}kk'(m_{kk'}-m_km_{k'})}{\sigma_k\sigma_{k'}},
  \]
  where $m_{kk'}$ is the fraction of edges with endpoints of degree $k$
  and $k'$, $m_k=\sum_{k'}m_{kk'}$, and $\sigma_k$ is the standard
  deviation of $m_k$.
\item[Mean $k$-core, $\avg{c}$] The $k$-core is a maximal set of
  vertices such that its induced subgraph only contains vertices with
  degree larger than or equal to $k$. The $k$-core value $c_i$ of node
  $i$ is the largest value of $k$ for which $i$ belongs to the
  $k$-core. The mean value is then
  \[\avg{c} = \frac{1}{N}\sum_ic_i.\]
  This can be computed in time $O(N+E)$ according to the algorithm of
  Ref.~\cite{batagelj_fast_2011}.

\item[Mean local clustering coefficient, $C_l$] The local clustering
  coefficient~\cite{watts_collective_1998} of node $i$ is given by
  \[
  C_i = \frac{\sum_{jk}G_{ij}G_{ki}G_{jk}}{k_i(k_i-1)}.
  \]
  It measures the fraction of pairs of neighbors that are also
  connected. The mean value is then just
  \[C_l = \frac{1}{N}\sum_iC_i.\]

\item[Global clustering coefficient, $C_g$] The global clustering
  coefficient of is given by
  \[
  C_g = \frac{\sum_{ijk}G_{ij}G_{ki}G_{jk}}{\sum_ik_i(k_i-1)}.
  \]
  It measures the fraction of connected triads that close to form a
  triangle.

\item[Leading eigenvalue of adjacency matrix, $\lambda_1^A$] The leading
  eigenvalue of the adjacency matrix is the largest value of $\lambda$
  which solves
  \[
  \G \bm x = \lambda \bm x,
  \]
  where $\bm x$ is the associated eigenvector.
\item[Leading eigenvalue of Hashimoto matrix, $\lambda_1^H$] The leading
  eigenvalue of the Hashimoto (a.k.a. non-backtracking)
  matrix~\cite{hashimoto_zeta_1989} is the largest value of $\lambda$
  which solves
  \[
  \bm H \bm x = \lambda \bm x,
  \]
  where $\bm x$ is the associated eigenvector, and $\bm H$ is an
  asymmetric $E\times E$ matrix with entries defined as
  \[
  H_{k\to l,i\to j} =
        \begin{cases}
            1 & \text{if } G_{kl} = G_{ij} = 1, l=i, k\ne j,\\
            0 & \text{otherwise}.
        \end{cases}
  \]
\item[Characteristic time of a random walk, $\tau$] The characteristic
  time of a random walk is obtained via the second largest eigenvalue $\lambda_2^T \in [0,1]$ of
  the transition matrix $\bm T$, with entries
  \[
  T_{ij} = \frac{G_{ij}}{k_j},
  \]
  where $k_i = \sum_j G_{ji}$. It is defined as
  \[
  \tau = -\ln \lambda_2^T.
  \]
  If the network is disconnected, we compute $\tau$ only on the largest
  component.
\item[Pseudo-diameter, $\varnothing$] The pseudo-diameter is an
  approximate graph diameter. It is obtained by starting from an
  arbitrary source node, and finding a target node that is farthest away
  from the source. This process is repeated by treating the target as
  the new starting node, and ends when the graph distance no longer
  increases. This graph distance is taken to be the pseudo-diameter. The
  algorithm runs in time $O(N+E)$.

  If the network is disconnected, $\varnothing$ is taken as the maximum
  of pseudo-diameters of the connected components.

\item[Node percolation profile (random removal), $R_r$] We chose a
  random node order, and remove nodes sequentially from the graph
  according to it. If $S_i$ is the fraction of nodes in the largest
  component after the $i$-th removal, then the profile value is
  \[
  R_r = \frac{1}{N}\sum_iS_i.
  \]
  The value is averaged over several node orderings.
\item[Node percolation profile (targeted removal), $R_t$] The computation
 is the same as $R_r$, but the nodes are always removed in decreasing
 order of the degree.

\item[Fraction of nodes in the largest component, $S$] A component is a
maximal set of nodes that are connected by a path. The largest component
is the component with the largest number of nodes, and $S$ is the
fraction of all nodes that belong to it.

\end{description}

In Fig.~\ref{fig:dev-corr} we show how the $z$-scores and relative
deviation values are related for every network descriptor, according to
both models used. In Fig.~\ref{fig:desc-corr} we show Kendall's $\tau$
correlation coefficient among the descriptor values themselves, as well
as their $z$-scores and relative deviations, according to the DCSBM. The
insets show how the correlations among the deviations are themselves
also correlated with the descriptor correlations.


\section{Dataset descriptions}\label{app:data}

Below are descriptions of the network datasets used in this work. The
codenames in the first row correspond to the respective entries in the
Netzschleuder repository~\cite{peixoto_netzschleuder_2020} where the
networks can be downloaded. Some of the descriptions were obtained from
the Colorado Index of Complex Networks~\cite{clauset_colorado_2016}.

For all networks, the versions considered in this work were transformed
into simple graphs, i.e. symmetrized versions of directed networks
and/or with parallel edges and self-loops removed.

\begingroup
\squeezetable
\begin{longtable*}{p{12em}p{35em}lll}
\hline
Name & Description & $N$ & $E$ &  Domain\\ \hline
\input{data-edited.tex}
\end{longtable*}
\endgroup

\end{document}

%% file: data-edited.tex
\href{https://networks.skewed.de/net/blumenau_drug}{\texttt{blumenau\_drug}} & A network of drug-drug interactions, extracted from 18 months of electronic health records (EHRs) from the city of Blumenau in Southern Brazil~\cite{Brattig_Correia_2019}. & 75 & 181 & Biological\\
\href{https://networks.skewed.de/net/budapest_connectome#male_20k}{\texttt{budapest\_connectome} (1)} & \multirow[t]{2}{35em}{Brain graphs derived from connectomes of 477 people, computed from the Human Connectome Project~\cite{Szalkai_2015}.} & $1015$ & $53586$ & Biological\\
\href{https://networks.skewed.de/net/budapest_connectome#female_20k}{\texttt{budapest\_connectome} (2)} &  & 1015 & 62552 & Biological\\
\href{https://networks.skewed.de/net/celegans_2019#male_chemical_synapse}{\texttt{celegans\_2019} (1)} & \multirow[t]{4}{35em}{Networks among neurons of both the adult male and adult hermaphrodite worms C. elegans, constructed from electron microscopy series, to include directed edges (chemical) and undirected (gap junction), and spanning including nodes for muscle and non-muscle end organs~\cite{Cook_2019}.} & $514$ & $2832$ & Biological\\
\href{https://networks.skewed.de/net/celegans_2019#male_chemical_corrected}{\texttt{celegans\_2019} (2)} &  & 575 & 4500 & Biological\\
\href{https://networks.skewed.de/net/celegans_2019#hermaphrodite_chemical_corrected}{\texttt{celegans\_2019} (3)} &  & 454 & 4172 & Biological\\
\href{https://networks.skewed.de/net/celegans_2019#hermaphrodite_gap_junction_corrected}{\texttt{celegans\_2019} (4)} &  & 469 & 1433 & Biological\\
\href{https://networks.skewed.de/net/celegans_interactomes#Interolog}{\texttt{celegans\_interactomes} (1)} & \multirow[t]{3}{35em}{Ten networks of protein-protein interactions in Caenorhabditis elegans (nematode), from yeast two-hybrid experiments, biological process maps, literature curation, orthologous interactions, and genetic interactions~\cite{Simonis_2008}.} & $2724$ & $13564$ & Biological\\
\href{https://networks.skewed.de/net/celegans_interactomes#Phenotypes}{\texttt{celegans\_interactomes} (2)} &  & 912 & 22738 & Biological\\
\href{https://networks.skewed.de/net/celegans_interactomes#BPmaps}{\texttt{celegans\_interactomes} (3)} &  & 537 & 517 & Biological\\
\href{https://networks.skewed.de/net/collins_yeast}{\texttt{collins\_yeast}} & Network of protein-protein interactions in Saccharomyces cerevisiae (budding yeast), measured by co-complex associations identified by high-throughput affinity purification and mass spectrometry (AP/MS)~\cite{Collins_2007}. & 1622 & 9070 & Biological\\
\href{https://networks.skewed.de/net/ecoli_transcription#v1.1}{\texttt{ecoli\_transcription} (1)} & \multirow[t]{1}{35em}{Network of operons and their pairwise interactions for E. coli~\cite{Shen_Orr_2002}.} & $423$ & $519$ & Biological\\
\href{https://networks.skewed.de/net/foodweb_baywet}{\texttt{foodweb\_baywet}} & Networks of carbon exchanges among species in the cypress wetlands of South Florida. One network covers the wet and the other the dry season~\cite{ulanowicz1999network}. & 128 & 2075 & Biological\\
\href{https://networks.skewed.de/net/foodweb_little_rock}{\texttt{foodweb\_little\_rock}} & A food web among the species found in Little Rock Lake in Wisconsin~\cite{Martinez_1991}. & 183 & 2434 & Biological\\
\href{https://networks.skewed.de/net/fresh_webs#SuttonSu}{\texttt{fresh\_webs} (1)} & \multirow[t]{2}{35em}{Trophic-level species interactions in streams in New Zealand, Maine and North Carolina~\cite{Thompson_2003}.} & $94$ & $424$ & Biological\\
\href{https://networks.skewed.de/net/fresh_webs#DempstersSu}{\texttt{fresh\_webs} (2)} &  & 107 & 965 & Biological\\
\href{https://networks.skewed.de/net/genetic_multiplex#Rattus}{\texttt{genetic\_multiplex} (1)} & \multirow[t]{4}{35em}{Multiplex networks representing different types of genetic interactions, for different organisms. Layers represent (i) physical, (ii) association, (iii) co-localization, (iv) direct, and (v) suppressive, (vi) additive or synthetic genetic interaction~\cite{2014}.} & $2640$ & $3677$ & Biological\\
\href{https://networks.skewed.de/net/genetic_multiplex#HumanHIV1}{\texttt{genetic\_multiplex} (2)} &  & 1005 & 1155 & Biological\\
\href{https://networks.skewed.de/net/genetic_multiplex#Gallus}{\texttt{genetic\_multiplex} (3)} &  & 313 & 325 & Biological\\
\href{https://networks.skewed.de/net/genetic_multiplex#Sacchcere}{\texttt{genetic\_multiplex} (4)} &  & 6570 & 223542 & Biological\\
\href{https://networks.skewed.de/net/human_brains#BNU1_0025875_1_DTI_DS01216}{\texttt{human\_brains} (1)} & \multirow[t]{15}{35em}{Networks of neural interactions extracted from human patients using the Magnetic Resonance One-Click Pipeline (MROCP), where nodes are voxels of neural tissue and edges represent connections by single fibers~\cite{Gray_Roncal_2013}.} & $1215$ & $13768$ & Biological\\
\href{https://networks.skewed.de/net/human_brains#NKIENH_0144702_1_DTI_CPAC200}{\texttt{human\_brains} (2)} &  & 200 & 1231 & Biological\\
\href{https://networks.skewed.de/net/human_brains#SWU4_0025850_1_DTI_DS00140}{\texttt{human\_brains} (3)} &  & 139 & 873 & Biological\\
\href{https://networks.skewed.de/net/human_brains#MRN1313_TEV_1_DTI_slab1068}{\texttt{human\_brains} (4)} &  & 1771 & 3645 & Biological\\
\href{https://networks.skewed.de/net/human_brains#SWU4_0025863_2_DTI_Talairach}{\texttt{human\_brains} (5)} &  & 1105 & 19543 & Biological\\
\href{https://networks.skewed.de/net/human_brains#SWU4_0025821_1_DTI_slab907}{\texttt{human\_brains} (6)} &  & 1527 & 3939 & Biological\\
\href{https://networks.skewed.de/net/human_brains#MRN114_M87143273_1_DTI_desikan}{\texttt{human\_brains} (7)} &  & 70 & 1219 & Biological\\
\href{https://networks.skewed.de/net/human_brains#Jung2015_M87126525_1_DTI_CPAC200}{\texttt{human\_brains} (8)} &  & 200 & 2808 & Biological\\
\href{https://networks.skewed.de/net/human_brains#HNU1_0025451_5_DTI_desikan}{\texttt{human\_brains} (9)} &  & 70 & 1301 & Biological\\
\href{https://networks.skewed.de/net/human_brains#HNU1_0025441_9_DTI_slab907}{\texttt{human\_brains} (10)} &  & 1632 & 5218 & Biological\\
\href{https://networks.skewed.de/net/human_brains#BNU1_0025879_1_DTI_DS16784}{\texttt{human\_brains} (11)} &  & 16783 & 430493 & Biological\\
\href{https://networks.skewed.de/net/human_brains#BNU1_0025881_1_DTI_DS72784}{\texttt{human\_brains} (12)} &  & 72783 & 2411659 & Biological\\
\href{https://networks.skewed.de/net/human_brains#HNU1_0025456_8_DTI_DS72784}{\texttt{human\_brains} (13)} &  & 72783 & 3720694 & Biological\\
\href{https://networks.skewed.de/net/human_brains#Jung2015_M87126525_1_DTI_DS72784}{\texttt{human\_brains} (14)} &  & 72783 & 4205222 & Biological\\
\href{https://networks.skewed.de/net/human_brains#NKIENH_0192736_1_DTI_DS72784}{\texttt{human\_brains} (15)} &  & 72783 & 7175769 & Biological\\
\href{https://networks.skewed.de/net/interactome_figeys}{\texttt{interactome\_figeys}} & A network of human proteins and their binding interactions~\cite{Ewing_2007}. & 2239 & 6432 & Biological\\
\href{https://networks.skewed.de/net/interactome_yeast}{\texttt{interactome\_yeast}} & A network of protein-protein binding interactions among yeast proteins~\cite{Coulomb_2005}. & 1870 & 2203 & Biological\\
\href{https://networks.skewed.de/net/kegg_metabolic#fnu}{\texttt{kegg\_metabolic} (1)} & \multirow[t]{3}{35em}{Metabolic networks of various species, as extracted from the Kyoto Encyclopedia of Genes and Genomes (KEGG) database in March 2006~\cite{Huss_2007}.} & $1031$ & $2485$ & Biological\\
\href{https://networks.skewed.de/net/kegg_metabolic#hsa}{\texttt{kegg\_metabolic} (2)} &  & 1917 & 5803 & Biological\\
\href{https://networks.skewed.de/net/kegg_metabolic#bbu}{\texttt{kegg\_metabolic} (3)} &  & 505 & 1144 & Biological\\
\href{https://networks.skewed.de/net/macaque_neural}{\texttt{macaque\_neural}} & A network of cortical regions in the Macaque cortex~\cite{1993}. & 47 & 313 & Biological\\
\href{https://networks.skewed.de/net/malaria_genes#HVR_5}{\texttt{malaria\_genes} (1)} & \multirow[t]{4}{35em}{Networks of recombinant antigen genes from the human malaria parasite P. falciparum~\cite{Larremore_2013}.} & $307$ & $2684$ & Biological\\
\href{https://networks.skewed.de/net/malaria_genes#HVR_8}{\texttt{malaria\_genes} (2)} &  & 307 & 3961 & Biological\\
\href{https://networks.skewed.de/net/malaria_genes#HVR_9}{\texttt{malaria\_genes} (3)} &  & 307 & 7579 & Biological\\
\href{https://networks.skewed.de/net/malaria_genes#HVR_1}{\texttt{malaria\_genes} (4)} &  & 307 & 2812 & Biological\\
\href{https://networks.skewed.de/net/messal_shale}{\texttt{messal\_shale}} & A network of feeding links among taxa based on the 48 million years old uppermost early Eocene Messel Shale~\cite{Dunne_2014}. & 700 & 6395 & Biological\\
\href{https://networks.skewed.de/net/nematode_mammal}{\texttt{nematode\_mammal}} & A global interaction web of interactions between nematodes and their host mammal species, extracted from the helminthR package and dataset~\cite{Dallas_2018}. & 30516 & 61597 & Biological\\
\href{https://networks.skewed.de/net/plant_pol_kato}{\texttt{plant\_pol\_kato}} & A bipartite network of plants and pollinators from Kyoto University Forest of Ashu, Japan, from 1984 to 1987~\cite{kato1990insect}. & 772 & 1206 & Biological\\
\href{https://networks.skewed.de/net/plant_pol_robertson}{\texttt{plant\_pol\_robertson}} & A bipartite network of plants and pollinators, from southwestern Illinois, USA~\cite{Robertson_1928}. & 1884 & 15255 & Biological\\
\href{https://networks.skewed.de/net/reactome}{\texttt{reactome}} & A network of human proteins and their binding interactions, extracted from Reactome project~\cite{Joshi_Tope_2004}. & 6327 & 146160 & Biological\\
\href{https://networks.skewed.de/net/yeast_transcription}{\texttt{yeast\_transcription}} & Network of operons and their pairwise interactions, via transcription factor-based regulation, within the yeast Saccharomyces cerevisiae~\cite{Milo_2002}. & 916 & 1081 & Biological\\
\href{https://networks.skewed.de/net/caida_as#20070917}{\texttt{caida\_as}} & \multirow[t]{1}{35em}{Autonomous System (AS) relationships on the Internet, from 2004-2007~\cite{caida_as}.} & $8020$ & $18203$ & Technological\\
\href{https://networks.skewed.de/net/gnutella#08}{\texttt{gnutella} (1)} & \multirow[t]{2}{35em}{Gnutella peer-to-peer file sharing network from 5-31 August 2002~\cite{Ripeanu_2002}.} & $6301$ & $20777$ & Technological\\
\href{https://networks.skewed.de/net/gnutella#25}{\texttt{gnutella} (2)} &  & 22687 & 54705 & Technological\\
\href{https://networks.skewed.de/net/internet_as}{\texttt{internet\_as}} & A symmetrized snapshot of the structure of the Internet at the level of Autonomous Systems (ASs), reconstructed from BGP tables posted by the University of Oregon Route Views Project~\cite{1405.0483v2}. & 22963 & 48436 & Technological\\
\href{https://networks.skewed.de/net/internet_top_pop#Tw}{\texttt{internet\_top\_pop} (1)} & \multirow[t]{5}{35em}{Assorted snapshots of internet graph at the Point of Presence (PoP) level (which lies between the IP and AS levels), collected from around the world and at various times. The earliest snapshots are for ARPANET (1969-1972), with a few more from pre-2000~\cite{Knight_2011}.} & $76$ & $115$ & Technological\\
\href{https://networks.skewed.de/net/internet_top_pop#TataNld}{\texttt{internet\_top\_pop} (2)} &  & 145 & 186 & Technological\\
\href{https://networks.skewed.de/net/internet_top_pop#Ntt}{\texttt{internet\_top\_pop} (3)} &  & 47 & 63 & Technological\\
\href{https://networks.skewed.de/net/internet_top_pop#Cogentco}{\texttt{internet\_top\_pop} (4)} &  & 197 & 243 & Technological\\
\href{https://networks.skewed.de/net/internet_top_pop#Kdl}{\texttt{internet\_top\_pop} (5)} &  & 754 & 895 & Technological\\
\href{https://networks.skewed.de/net/jdk}{\texttt{jdk}} & A network of class dependencies within the JDK (Java SE Development Kit) 1.6~\cite{Kunegis_2013}. & 6434 & 53658 & Technological\\
\href{https://networks.skewed.de/net/jung}{\texttt{jung}} & A network of software class dependency within the JUNG 2.0~\cite{_ubelj_2012}. & 6120 & 50290 & Technological\\
\href{https://networks.skewed.de/net/linux}{\texttt{linux}} & A network of Linux (v3.16) source code file inclusion~\cite{Kunegis_2013}. & 30837 & 213217 & Technological\\
\href{https://networks.skewed.de/net/power}{\texttt{power}} & A network representing the Western States Power Grid of the United States~\cite{Watts_1998}. & 4941 & 6594 & Technological\\
\href{https://networks.skewed.de/net/route_views#19990829}{\texttt{route\_views} (1)} & \multirow[t]{6}{35em}{733 daily network snapshots denoting BGP traffic among autonomous systems (ASs) on the Internet, from the Oregon Route Views Project, spanning 8 November 1997 to 2 January 2000. Data collected by NLANR/MOAT~\cite{route_views}.} & $103$ & $239$ & Technological\\
\href{https://networks.skewed.de/net/route_views#19981231}{\texttt{route\_views} (2)} &  & 512 & 1181 & Technological\\
\href{https://networks.skewed.de/net/route_views#19991208}{\texttt{route\_views} (3)} &  & 767 & 1734 & Technological\\
\href{https://networks.skewed.de/net/route_views#19991210}{\texttt{route\_views} (4)} &  & 1486 & 3172 & Technological\\
\href{https://networks.skewed.de/net/route_views#20000102}{\texttt{route\_views} (5)} &  & 6474 & 12572 & Technological\\
\href{https://networks.skewed.de/net/route_views#19991206}{\texttt{route\_views} (6)} &  & 6301 & 12226 & Technological\\
\href{https://networks.skewed.de/net/software_dependencies#flamingo}{\texttt{software\_dependencies} (1)} & \multirow[t]{6}{35em}{Several networks of software dependencies. Nodes represent libraries and a directed edge denotes a library dependency on another~\cite{_ubelj_2012,1202.3188v1}.} & $388$ & $514$ & Technological\\
\href{https://networks.skewed.de/net/software_dependencies#guava}{\texttt{software\_dependencies} (2)} &  & 838 & 1063 & Technological\\
\href{https://networks.skewed.de/net/software_dependencies#org}{\texttt{software\_dependencies} (3)} &  & 799 & 3579 & Technological\\
\href{https://networks.skewed.de/net/software_dependencies#colt}{\texttt{software\_dependencies} (4)} &  & 550 & 1153 & Technological\\
\href{https://networks.skewed.de/net/software_dependencies#jmail}{\texttt{software\_dependencies} (5)} &  & 282 & 505 & Technological\\
\href{https://networks.skewed.de/net/software_dependencies#weka}{\texttt{software\_dependencies} (6)} &  & 2124 & 4809 & Technological\\
\href{https://networks.skewed.de/net/topology}{\texttt{topology}} & An integrated snapshot of the structure of the Internet at the level of Autonomous Systems (ASs), reconstructed from multiple sources, including the RouteViews and RIPE BGP trace collectors, route servers, looking glasses, and the Internet Routing Registry databases~\cite{Zhang_2005}. & 34761 & 107720 & Technological\\
\href{https://networks.skewed.de/net/adjnoun}{\texttt{adjnoun}} & A network of word adjacencies of common adjectives and nouns in the novel ``David Copperfield'' by Charles Dickens~\cite{Newman_2006}. & 112 & 425 & Informational\\
\href{https://networks.skewed.de/net/bag_of_words#enron}{\texttt{bag\_of\_words}} & \multirow[t]{1}{35em}{Five text collections in the form of bags-of-words~\cite{newman_bag_of_words,bag_of_words}.} & $67963$ & $3710420$ & Informational\\
\href{https://networks.skewed.de/net/baidu}{\texttt{baidu}} & Four networks from Chinese online encyclopedias Baidu~\cite{Niu_2011}. & 2141300 & 17014946 & Informational\\
\href{https://networks.skewed.de/net/berkstan_web}{\texttt{berkstan\_web}} & The web graph of Berkeley and Stanford Universities~\cite{0810.1355v1}. & 685231 & 6649470 & Informational\\
\href{https://networks.skewed.de/net/bible_nouns}{\texttt{bible\_nouns}} & A network of noun phrases (places and names) in the King James Version of the Bible~\cite{bible_nouns}. & 1773 & 9131 & Informational\\
\href{https://networks.skewed.de/net/citeseer}{\texttt{citeseer}} & Citations among papers indexed by the CiteSeer digital library~\cite{Bollacker_1998}. & 384413 & 1736145 & Informational\\
\href{https://networks.skewed.de/net/cora}{\texttt{cora}} & Citations among papers indexed by CORA, from 1998, an early computer science research paper search engine~\cite{McCallum_2000}. & 23166 & 89157 & Informational\\
\href{https://networks.skewed.de/net/dblp_cite}{\texttt{dblp\_cite}} & Citations among papers contained in the DBLP computer science bibliography~\cite{Ley_2002}. & 12590 & 49636 & Informational\\
\href{https://networks.skewed.de/net/dbtropes_feature}{\texttt{dbtropes\_feature}} & A bipartite network of artistic works (movies, novels, etc.) and their tropes (stylistic conventions or devices), as extracted from tvtropes~\cite{Kunegis_2013}. & 152093 & 3232134 & Informational\\
\href{https://networks.skewed.de/net/discogs_label}{\texttt{discogs\_label}} & Two bipartite networks of the affiliations between musical labels and either musical genres or musical ``styles,'' as given in the discogs.com database~\cite{Kunegis_2013}. & 270786 & 481661 & Informational\\
\href{https://networks.skewed.de/net/edit_wikibooks#kk}{\texttt{edit\_wikibooks} (1)} & \multirow[t]{3}{35em}{Two bipartite user-page networks extracted from Wikipedia, about books~\cite{wikimedia}.} & $1162$ & $1213$ & Informational\\
\href{https://networks.skewed.de/net/edit_wikibooks#ur}{\texttt{edit\_wikibooks} (2)} &  & 1584 & 1748 & Informational\\
\href{https://networks.skewed.de/net/edit_wikibooks#sq}{\texttt{edit\_wikibooks} (3)} &  & 7177 & 7732 & Informational\\
\href{https://networks.skewed.de/net/edit_wikinews#sq}{\texttt{edit\_wikinews} (1)} & \multirow[t]{5}{35em}{Two bipartite user-page networks extracted from Wikipedia, about news events~\cite{wikimedia}.} & $2511$ & $4986$ & Informational\\
\href{https://networks.skewed.de/net/edit_wikinews#no}{\texttt{edit\_wikinews} (2)} &  & 4523 & 8891 & Informational\\
\href{https://networks.skewed.de/net/edit_wikinews#hu}{\texttt{edit\_wikinews} (3)} &  & 5541 & 10545 & Informational\\
\href{https://networks.skewed.de/net/edit_wikinews#sd}{\texttt{edit\_wikinews} (4)} &  & 2208 & 2753 & Informational\\
\href{https://networks.skewed.de/net/edit_wikinews#bg}{\texttt{edit\_wikinews} (5)} &  & 4457 & 5942 & Informational\\
\href{https://networks.skewed.de/net/edit_wikiquote#co}{\texttt{edit\_wikiquote} (1)} & \multirow[t]{5}{35em}{A bipartite user-page network extracted from Wikiquotes~\cite{wikimedia}.} & $270$ & $243$ & Informational\\
\href{https://networks.skewed.de/net/edit_wikiquote#mr}{\texttt{edit\_wikiquote} (2)} &  & 1041 & 1109 & Informational\\
\href{https://networks.skewed.de/net/edit_wikiquote#uz}{\texttt{edit\_wikiquote} (3)} &  & 704 & 800 & Informational\\
\href{https://networks.skewed.de/net/edit_wikiquote#ku}{\texttt{edit\_wikiquote} (4)} &  & 1333 & 2731 & Informational\\
\href{https://networks.skewed.de/net/edit_wikiquote#ang}{\texttt{edit\_wikiquote} (5)} &  & 625 & 823 & Informational\\
\href{https://networks.skewed.de/net/edit_wiktionary#mo}{\texttt{edit\_wiktionary} (1)} & \multirow[t]{5}{35em}{Three bipartite user-page networks extracted from Wiktionary,  for French, German, and English~\cite{wikimedia}.} & $271$ & $285$ & Informational\\
\href{https://networks.skewed.de/net/edit_wiktionary#ab}{\texttt{edit\_wiktionary} (2)} &  & 289 & 276 & Informational\\
\href{https://networks.skewed.de/net/edit_wiktionary#als}{\texttt{edit\_wiktionary} (3)} &  & 1271 & 1270 & Informational\\
\href{https://networks.skewed.de/net/edit_wiktionary#sm}{\texttt{edit\_wiktionary} (4)} &  & 8552 & 34589 & Informational\\
\href{https://networks.skewed.de/net/edit_wiktionary#si}{\texttt{edit\_wiktionary} (5)} &  & 3016 & 6263 & Informational\\
\href{https://networks.skewed.de/net/google_web}{\texttt{google\_web}} & A web graph representing a crawl of a portion of the general WWW, from a 2002 Google Programming contest~\cite{0810.1355v1}. & 916428 & 4322051 & Informational\\
\href{https://networks.skewed.de/net/movielens_100k}{\texttt{movielens\_100k}} & Three bipartite networks that make up the MovieLens 100K Dataset, a stable benchmark dataset of 100,000 ratings from 1000 users on 1700 movies~\cite{movielens_100k}. & 24129 & 71154 & Informational\\
\href{https://networks.skewed.de/net/polblogs}{\texttt{polblogs}} & A directed network of hyperlinks among a large set of U.S~\cite{Adamic_2005}. & 1490 & 16715 & Informational\\
\href{https://networks.skewed.de/net/polbooks}{\texttt{polbooks}} & A network of books about U.S~\cite{Pasternak_1970}. & 105 & 441 & Informational\\
\href{https://networks.skewed.de/net/scotus_majority#2008}{\texttt{scotus\_majority} (1)} & \multirow[t]{1}{35em}{Network of legal citations by the US Supreme Court (SCOTUS)~\cite{Fowler_2008,Fowler_2007}.} & $25417$ & $216456$ & Informational\\
\href{https://networks.skewed.de/net/trec_web}{\texttt{trec\_web}} & A web graph network originally constructed in 2003 as a testbed for information-retrieval techniques, including web search engines~\cite{Bailey_2003}. & 1601787 & 6679248 & Informational\\
\href{https://networks.skewed.de/net/unicodelang}{\texttt{unicodelang}} & A bipartite network of languages and the countries in which they are spoken, as estimated by Unicode~\cite{Kunegis_2013}. & 868 & 1255 & Informational\\
\href{https://networks.skewed.de/net/us_patents}{\texttt{us\_patents}} & Citations among patents in the United States, as found in the National Bureau of Economic Research (NBER) database, from 1975 to 1999~\cite{Hall_2001}. & 3774768 & 16518947 & Informational\\
\href{https://networks.skewed.de/net/webkb#webkb_texas_link1}{\texttt{webkb} (1)} & \multirow[t]{5}{35em}{Web graphs crawled from four Computer Science departments in 1998, with each page manually classified into one of 7 categories: course, department, faculty, project, staff, student, or other~\cite{Slattery_1998}.} & $286$ & $493$ & Informational\\
\href{https://networks.skewed.de/net/webkb#webkb_washington_link1}{\texttt{webkb} (2)} &  & 433 & 954 & Informational\\
\href{https://networks.skewed.de/net/webkb#webkb_wisconsin_link1}{\texttt{webkb} (3)} &  & 300 & 565 & Informational\\
\href{https://networks.skewed.de/net/webkb#webkb_cornell_link1}{\texttt{webkb} (4)} &  & 349 & 696 & Informational\\
\href{https://networks.skewed.de/net/webkb#webkb_wisconsin_cocite}{\texttt{webkb} (5)} &  & 348 & 16625 & Informational\\
\href{https://networks.skewed.de/net/wiki_science}{\texttt{wiki\_science}} & A network of scientific fields, extracted from the English Wikipedia in early 2020~\cite{Calderone2020}. & 687 & 6523 & Informational\\
\href{https://networks.skewed.de/net/word_adjacency#french}{\texttt{word\_adjacency} (1)} & \multirow[t]{2}{35em}{Networks of word adjacency in texts of several languages including English, French, Spanish and Japanese~\cite{Milo_2004}.} & $8325$ & $23841$ & Informational\\
\href{https://networks.skewed.de/net/word_adjacency#japanese}{\texttt{word\_adjacency} (2)} &  & 2704 & 7998 & Informational\\
\href{https://networks.skewed.de/net/word_assoc}{\texttt{word\_assoc}} & A network of word associations showing the count of such associations as collected from subjects, from the Edinburgh Associative Thesaurus (EAT)~\cite{kiss1973associative}. & 23132 & 297094 & Informational\\
\href{https://networks.skewed.de/net/wordnet}{\texttt{wordnet}} & A network of English words from the WordNet, denoting relationships between words (synonymy, hyperonymy, meronymy, etc.)~\cite{Fellbaum_2010} & 146005 & 656999 & Informational\\
\href{https://networks.skewed.de/net/yahoo_ads}{\texttt{yahoo\_ads}} & A network of words extracted from phrases on which advertisers bid, in Yahoo! advertisements~\cite{Kunegis_2013}. & 653260 & 2931698 & Informational\\
\href{https://networks.skewed.de/net/chicago_road}{\texttt{chicago\_road}} & A transportation network of Chicago, USA, from an unknown date (probably late 20th century)~\cite{chicago_road}. & 12982 & 20627 & Transportation\\
\href{https://networks.skewed.de/net/contiguous_usa}{\texttt{contiguous\_usa}} & A network of contiguous states in the USA, in which each state is a node and two nodes are connected if they share a land-based geographic border~\cite{knuth1993stanford}. & 49 & 107 & Transportation\\
\href{https://networks.skewed.de/net/eu_airlines}{\texttt{eu\_airlines}} & A multiplex network of airline routes among European airports, where each of the 37 edge types represents routes by a different airline~\cite{2013}. & 450 & 2953 & Transportation\\
\href{https://networks.skewed.de/net/euroroad}{\texttt{euroroad}} & A network of international ``E-roads,'' mostly in Europe~\cite{_ubelj_2011}. & 1174 & 1417 & Transportation\\
\href{https://networks.skewed.de/net/faa_routes}{\texttt{faa\_routes}} & A network of air traffic routes, from the FAA (Federal Aviation Administration) National Flight Data Center (NFDC) preferred routes database~\cite{faa_routes}. & 1226 & 2408 & Transportation\\
\href{https://networks.skewed.de/net/london_transport}{\texttt{london\_transport}} & Multiplex network with 3 edge types representing links within the three layers of London train stations: Underground, Overground and DLR~\cite{2014}. & 369 & 430 & Transportation\\
\href{https://networks.skewed.de/net/openflights}{\texttt{openflights}} & A network of regularly occurring flights among airports worldwide, extracted from the openflights.org dataset~\cite{openflights}. & 3214 & 18858 & Transportation\\
\href{https://networks.skewed.de/net/openstreetmap#01-AL-cities-street_networks:0100124_Abbeville}{\texttt{openstreetmap} (1)} & \multirow[t]{14}{35em}{The road network for the entire United States, as extracted from the OpenStreetMap project~\cite{Boeing_2019}.} & $351$ & $434$ & Transportation\\
\href{https://networks.skewed.de/net/openstreetmap#01-AL-cities-street_networks:0102320_Argo}{\texttt{openstreetmap} (2)} &  & 354 & 350 & Transportation\\
\href{https://networks.skewed.de/net/openstreetmap#01-AL-cities-street_networks:0143120_Lincoln}{\texttt{openstreetmap} (3)} &  & 831 & 923 & Transportation\\
\href{https://networks.skewed.de/net/openstreetmap#12-FL-tracts-street_networks:12095017103}{\texttt{openstreetmap} (4)} &  & 1603 & 2188 & Transportation\\
\href{https://networks.skewed.de/net/openstreetmap#01-AL-counties-street_networks:01079_Lawrence_County}{\texttt{openstreetmap} (5)} &  & 4240 & 5102 & Transportation\\
\href{https://networks.skewed.de/net/openstreetmap#01-AL-counties-street_networks:01127_Walker_County}{\texttt{openstreetmap} (6)} &  & 8904 & 10549 & Transportation\\
\href{https://networks.skewed.de/net/openstreetmap#37-NC-cities-street_networks:3718320_Dunn}{\texttt{openstreetmap} (7)} &  & 724 & 1048 & Transportation\\
\href{https://networks.skewed.de/net/openstreetmap#37-NC-cities-street_networks:3709060_Burlington}{\texttt{openstreetmap} (8)} &  & 2371 & 3295 & Transportation\\
\href{https://networks.skewed.de/net/openstreetmap#48-TX-cities-street_networks:4865726_Santa_Fe}{\texttt{openstreetmap} (9)} &  & 684 & 823 & Transportation\\
\href{https://networks.skewed.de/net/openstreetmap#48-TX-cities-street_networks:4862600_Robstown}{\texttt{openstreetmap} (10)} &  & 500 & 780 & Transportation\\
\href{https://networks.skewed.de/net/openstreetmap#08-CO-cities-street_networks:0807850_Boulder}{\texttt{openstreetmap} (11)} &  & 3377 & 4698 & Transportation\\
\href{https://networks.skewed.de/net/openstreetmap#06-CA-cities-street_networks:0670224_Santee}{\texttt{openstreetmap} (12)} &  & 1609 & 1972 & Transportation\\
\href{https://networks.skewed.de/net/openstreetmap#06-CA-cities-street_networks:0649187_Moraga}{\texttt{openstreetmap} (13)} &  & 612 & 688 & Transportation\\
\href{https://networks.skewed.de/net/openstreetmap#17-IL-urbanized_areas-street_networks:16264_Chicago_IL--IN_Urbanized_Area}{\texttt{openstreetmap} (14)} &  & 209734 & 297196 & Transportation\\
\href{https://networks.skewed.de/net/roadnet#CA}{\texttt{roadnet} (1)} & \multirow[t]{3}{35em}{Road networks from three US states (CA, PA, TX), in which edges are stretches of road and vertices are intersections of roads~\cite{0810.1355v1}.} & $1971281$ & $2766607$ & Transportation\\
\href{https://networks.skewed.de/net/roadnet#PA}{\texttt{roadnet} (2)} &  & 1090920 & 1541898 & Transportation\\
\href{https://networks.skewed.de/net/roadnet#TX}{\texttt{roadnet} (3)} &  & 1393383 & 1921660 & Transportation\\
\href{https://networks.skewed.de/net/urban_streets#brasilia}{\texttt{urban\_streets} (1)} & \multirow[t]{19}{35em}{Urban street networks, corresponding to 1-square-mile maps of 20 cities around the world~\cite{2006,2017}.} & $179$ & $230$ & Transportation\\
\href{https://networks.skewed.de/net/urban_streets#los-angeles}{\texttt{urban\_streets} (2)} &  & 240 & 339 & Transportation\\
\href{https://networks.skewed.de/net/urban_streets#vienna}{\texttt{urban\_streets} (3)} &  & 467 & 691 & Transportation\\
\href{https://networks.skewed.de/net/urban_streets#new-york}{\texttt{urban\_streets} (4)} &  & 248 & 418 & Transportation\\
\href{https://networks.skewed.de/net/urban_streets#richmond}{\texttt{urban\_streets} (5)} &  & 697 & 1084 & Transportation\\
\href{https://networks.skewed.de/net/urban_streets#san-francisco}{\texttt{urban\_streets} (6)} &  & 169 & 271 & Transportation\\
\href{https://networks.skewed.de/net/urban_streets#venice}{\texttt{urban\_streets} (7)} &  & 1840 & 2397 & Transportation\\
\href{https://networks.skewed.de/net/urban_streets#cairo}{\texttt{urban\_streets} (8)} &  & 1496 & 2252 & Transportation\\
\href{https://networks.skewed.de/net/urban_streets#savannah}{\texttt{urban\_streets} (9)} &  & 584 & 958 & Transportation\\
\href{https://networks.skewed.de/net/urban_streets#irvine2}{\texttt{urban\_streets} (10)} &  & 217 & 222 & Transportation\\
\href{https://networks.skewed.de/net/urban_streets#bologna}{\texttt{urban\_streets} (11)} &  & 541 & 771 & Transportation\\
\href{https://networks.skewed.de/net/urban_streets#new-delhi}{\texttt{urban\_streets} (12)} &  & 252 & 328 & Transportation\\
\href{https://networks.skewed.de/net/urban_streets#seoul}{\texttt{urban\_streets} (13)} &  & 869 & 1307 & Transportation\\
\href{https://networks.skewed.de/net/urban_streets#washington}{\texttt{urban\_streets} (14)} &  & 192 & 302 & Transportation\\
\href{https://networks.skewed.de/net/urban_streets#barcelona}{\texttt{urban\_streets} (15)} &  & 210 & 323 & Transportation\\
\href{https://networks.skewed.de/net/urban_streets#ahmedabad}{\texttt{urban\_streets} (16)} &  & 2870 & 4375 & Transportation\\
\href{https://networks.skewed.de/net/urban_streets#paris}{\texttt{urban\_streets} (17)} &  & 335 & 494 & Transportation\\
\href{https://networks.skewed.de/net/urban_streets#london}{\texttt{urban\_streets} (18)} &  & 488 & 729 & Transportation\\
\href{https://networks.skewed.de/net/urban_streets#walnut-creek}{\texttt{urban\_streets} (19)} &  & 169 & 196 & Transportation\\
\href{https://networks.skewed.de/net/us_air_traffic}{\texttt{us\_air\_traffic}} & Yearly snapshots of flights among all commercial airports in the United States from 1990 to today~\cite{us_air_traffic}. & 2278 & 58228 & Transportation\\
\href{https://networks.skewed.de/net/us_roads#DC}{\texttt{us\_roads} (1)} & \multirow[t]{5}{35em}{The road networks of the 50 US States and the District of Columbia based on UA Census 2000 TIGER/Line Files~\cite{us_roads}.} & $9559$ & $14841$ & Transportation\\
\href{https://networks.skewed.de/net/us_roads#ME}{\texttt{us\_roads} (2)} &  & 194505 & 212345 & Transportation\\
\href{https://networks.skewed.de/net/us_roads#NJ}{\texttt{us\_roads} (3)} &  & 330386 & 431398 & Transportation\\
\href{https://networks.skewed.de/net/us_roads#VA}{\texttt{us\_roads} (4)} &  & 630639 & 705083 & Transportation\\
\href{https://networks.skewed.de/net/us_roads#NY}{\texttt{us\_roads} (5)} &  & 716215 & 886897 & Transportation\\
\href{https://networks.skewed.de/net/7th_graders}{\texttt{7th\_graders}} & A network of friendships among 29 seventh grade students in Victoria, Australia~\cite{1976}. & 29 & 250 & Social\\
\href{https://networks.skewed.de/net/academia_edu}{\texttt{academia\_edu}} & Snapshot of the follower relationships among users of academia.edu, a platform for academics to share research papers, scraped in 2011~\cite{Fire_2013}. & 200169 & 1022441 & Social\\
\href{https://networks.skewed.de/net/add_health#comm31}{\texttt{add\_health} (1)} & \multirow[t]{7}{35em}{A directed network of friendships obtained through a social survey of high school students in 1994. The ADD HEALTH data are constructed from the in-school questionnaire; 90,118 students representing 84 communities took this survey in 1994-95~\cite{Moody_2001}.} & $900$ & $1648$ & Social\\
\href{https://networks.skewed.de/net/add_health#comm49}{\texttt{add\_health} (2)} &  & 1929 & 7035 & Social\\
\href{https://networks.skewed.de/net/add_health#comm57}{\texttt{add\_health} (3)} &  & 1282 & 3487 & Social\\
\href{https://networks.skewed.de/net/add_health#comm6}{\texttt{add\_health} (4)} &  & 111 & 378 & Social\\
\href{https://networks.skewed.de/net/add_health#comm71}{\texttt{add\_health} (5)} &  & 74 & 358 & Social\\
\href{https://networks.skewed.de/net/add_health#comm80}{\texttt{add\_health} (6)} &  & 624 & 1745 & Social\\
\href{https://networks.skewed.de/net/add_health#comm84}{\texttt{add\_health} (7)} &  & 1755 & 4017 & Social\\
\href{https://networks.skewed.de/net/arxiv_authors#GrQc}{\texttt{arxiv\_authors} (1)} & \multirow[t]{1}{35em}{Scientific collaborations between authors of papers submitted to arxiv.org~\cite{Leskovec_evolution_2007}.} & $26197$ & $14484$ & Social\\
\href{https://networks.skewed.de/net/arxiv_collab#hep-th-1999}{\texttt{arxiv\_collab}} & \multirow[t]{1}{35em}{Collaboration graphs for scientists, extracted from the arXiv (physics)~\cite{Newman_2001}.} & $8361$ & $15751$ & Social\\
\href{https://networks.skewed.de/net/bitcoin_alpha}{\texttt{bitcoin\_alpha}} & A network of who-trusts-whom relationships among users of the Bitcoin Alpha platform~\cite{Kumar_2016}. & 3783 & 14124 & Social\\
\href{https://networks.skewed.de/net/ceo_club}{\texttt{ceo\_club}} & A bipartite network of the memberships of chief executive officers and the social organizations (clubs) to which they belong, from the Minneapolis-St. Paul area~\cite{Faust_1997}. & 40 & 95 & Social\\
\href{https://networks.skewed.de/net/chess}{\texttt{chess}} & A network among chess players (nodes) giving the chess match outcomes (edges), for game-by-game results among the world’s top chess players~\cite{chess}. & 7301 & 55899 & Social\\
\href{https://networks.skewed.de/net/copenhagen#calls}{\texttt{copenhagen} (1)} & \multirow[t]{3}{35em}{A network of social interactions among university students within the Copenhagen Networks Study, over a period of four weeks, sampled every 5 minutes~\cite{Sapiezynski_2019}.} & $536$ & $621$ & Social\\
\href{https://networks.skewed.de/net/copenhagen#fb_friends}{\texttt{copenhagen} (2)} &  & 800 & 6418 & Social\\
\href{https://networks.skewed.de/net/copenhagen#sms}{\texttt{copenhagen} (3)} &  & 568 & 697 & Social\\
\href{https://networks.skewed.de/net/crime}{\texttt{crime}} & A network of associations among suspects, victims, and/or witnesses involved in crimes in St. Louis in the 1990s~\cite{decker1991st}. & 1380 & 1476 & Social\\
\href{https://networks.skewed.de/net/cs_department}{\texttt{cs\_department}} & Multiplex network consisting of 5 edge types corresponding to online and offline relationships (Facebook, leisure, work, co-authorship, lunch) between employees of the Computer Science department at Aarhus~\cite{1303.4986v1}. & 61 & 353 & Social\\
\href{https://networks.skewed.de/net/dbpedia_country}{\texttt{dbpedia\_country}} & A bipartite network of the affiliations between notable people and countries of the world, as extracted from Wikipedia via the DBpedia project~\cite{Auer_2007}. & 592414 & 624402 & Social\\
\href{https://networks.skewed.de/net/dbpedia_occupation}{\texttt{dbpedia\_occupation}} & A bipartite network of the affiliations between notable people and occupations, as extracted from Wikipedia by the DBpedia project~\cite{Auer_2007}. & 229307 & 250945 & Social\\
\href{https://networks.skewed.de/net/dnc}{\texttt{dnc}} & A network representing the exchange of emails among members of the Democratic National Committee, in the email data leak released by WikiLeaks in 2016~\cite{Kunegis_2013}. & 2029 & 10429 & Social\\
\href{https://networks.skewed.de/net/dolphins}{\texttt{dolphins}} & An undirected social network of frequent associations observed among 62 dolphins (Tursiops) in a community living off Doubtful Sound, New Zealand, from 1994-2001~\cite{Lusseau_2003}. & 62 & 159 & Social\\
\href{https://networks.skewed.de/net/ego_social#facebook_414}{\texttt{ego\_social} (1)} & \multirow[t]{3}{35em}{Ego networks associated with a set of accounts of three social media platforms (Facebook, Google+, and Twitter)~\cite{1210.8182v3}.} & $150$ & $1693$ & Social\\
\href{https://networks.skewed.de/net/ego_social#facebook_1912}{\texttt{ego\_social} (2)} &  & 747 & 30025 & Social\\
\href{https://networks.skewed.de/net/ego_social#gplus_104105354262797387583}{\texttt{ego\_social} (3)} &  & 452 & 12513 & Social\\
\href{https://networks.skewed.de/net/email_company}{\texttt{email\_company}} & A network of emails among employee email addresses at a mid-sized manufacturing company~\cite{Michalski_2011}. & 167 & 3250 & Social\\
\href{https://networks.skewed.de/net/email_enron}{\texttt{email\_enron}} & The Enron email corpus, containing all the email communication from the Enron corporation, which was made public as a result of legal action~\cite{Klimt_2004}. & 36692 & 183831 & Social\\
\href{https://networks.skewed.de/net/escorts}{\texttt{escorts}} & A bipartite network of escort and individuals who buy sex from them in Brazil, extracted from a Brazilian online community for such ratings~\cite{Rocha_2011}. & 16730 & 39044 & Social\\
\href{https://networks.skewed.de/net/facebook_friends}{\texttt{facebook\_friends}} & A small anonymized Facebook ego network, from April 2014. Nodes are Facebook profiles, and an edge exists if the two profiles are ``friends'' on Facebook~\cite{1706.02356v2}. & 362 & 1988 & Social\\
\href{https://networks.skewed.de/net/facebook_organizations#S1}{\texttt{facebook\_organizations} (1)} & \multirow[t]{6}{35em}{Six networks of friendships among users on Facebook who indicated employment at one of the target corporation~\cite{1303.3741v2}.} & $320$ & $2369$ & Social\\
\href{https://networks.skewed.de/net/facebook_organizations#S2}{\texttt{facebook\_organizations} (2)} &  & 165 & 726 & Social\\
\href{https://networks.skewed.de/net/facebook_organizations#M1}{\texttt{facebook\_organizations} (3)} &  & 1429 & 19357 & Social\\
\href{https://networks.skewed.de/net/facebook_organizations#M2}{\texttt{facebook\_organizations} (4)} &  & 3862 & 87324 & Social\\
\href{https://networks.skewed.de/net/facebook_organizations#L1}{\texttt{facebook\_organizations} (5)} &  & 5793 & 30753 & Social\\
\href{https://networks.skewed.de/net/facebook_organizations#L2}{\texttt{facebook\_organizations} (6)} &  & 5524 & 94218 & Social\\
\href{https://networks.skewed.de/net/facebook_wall}{\texttt{facebook\_wall}} & Friendship relationships and interactions (wall posts) for a subset of the Facebook social network in 2009, recorded over a 2 year period~\cite{Viswanath_2009}. & 46952 & 183412 & Social\\
\href{https://networks.skewed.de/net/fediverse}{\texttt{fediverse}} & An early snapshot of the federation network among web publishers using the ActivityPub protocol~\cite{rochko_2018}. & 4860 & 426351 & Social\\
\href{https://networks.skewed.de/net/flickr_groups}{\texttt{flickr\_groups}} & Bipartite networks of the affiliations between users and groups on several online social network sites, including Flickr, YouTube, LiveJournal, and Orkut, extracted in 2007~\cite{Mislove_2007}. & 499610 & 8545307 & Social\\
\href{https://networks.skewed.de/net/football_tsevans}{\texttt{football\_tsevans}} & A network of American football games between Division IA colleges during regular season Fall 2000~\cite{Girvan_2002,Evans_2010}. & 115 & 613 & Social\\
\href{https://networks.skewed.de/net/foursquare#NYC_restaurant_tips}{\texttt{foursquare} (1)} & \multirow[t]{2}{35em}{Two bipartite networks of users and restaurant locations in New York City on Foursquare, from 24 October 2011 to 20 February 2012~\cite{Yang_2013}.} & $6410$ & $9472$ & Social\\
\href{https://networks.skewed.de/net/foursquare#NYC_restaurant_checkin}{\texttt{foursquare} (2)} &  & 4936 & 13472 & Social\\
\href{https://networks.skewed.de/net/highschool}{\texttt{highschool}} & A network of friendships among male students in a small high school in Illinois from 1958~\cite{coleman1964introduction}. & 70 & 274 & Social\\
\href{https://networks.skewed.de/net/hiv_transmission}{\texttt{hiv\_transmission}} & A set of networks of HIV transmissions between people through sexual, needle-sharing, or social connections, based on combining 8 datasets collected from 1988 to 2001~\cite{Morris_2011}. & 35229 & 48889 & Social\\
\href{https://networks.skewed.de/net/hyves}{\texttt{hyves}} & A network of friendships among users of Hyves, an online social networking site in the Netherlands (comparable to Facebook at the time)~\cite{hyves}. & 1402673 & 2777419 & Social\\
\href{https://networks.skewed.de/net/jazz_collab}{\texttt{jazz\_collab}} & The network of collaborations among jazz musicians, and among jazz bands, extracted from The Red Hot Jazz Archive digital database, covering bands that performed between 1912 and 1940~\cite{GLEISER_2003}. & 198 & 2742 & Social\\
\href{https://networks.skewed.de/net/karate#78}{\texttt{karate} (1)} & \multirow[t]{1}{35em}{Network of friendships among members of a university karate club~\cite{Zachary_1977}.} & $34$ & $78$ & Social\\
\href{https://networks.skewed.de/net/kidnappings}{\texttt{kidnappings}} & Bipartite network of members of the Abu Sayyaf Group in the Philippines, and the kidnapping events they were involved in~\cite{Gerdes_2014}. & 351 & 402 & Social\\
\href{https://networks.skewed.de/net/lastfm#band}{\texttt{lastfm} (1)} & \multirow[t]{1}{35em}{User-band networks from the music website \texttt{last.fm}~\cite{Celma_2010}.} & $175069$ & $898062$ & Social\\
\href{https://networks.skewed.de/net/lesmis}{\texttt{lesmis}} & The network of scene coappearances of characters in Victor Hugo's novel "Les Miserables". Edge weights denote the number of such occurrences~\cite{knuth1993stanford}. & 77 & 254 & Social\\
\href{https://networks.skewed.de/net/libimseti}{\texttt{libimseti}} & A network of ratings given between users at Libimseti.cz, a Czech online dating website~\cite{Kunegis_2012}. & 220970 & 17233144 & Social\\
\href{https://networks.skewed.de/net/mislove_osn#youtube}{\texttt{mislove\_osn} (1)} & \multirow[t]{1}{35em}{Network structure for four large online social networks~\cite{Mislove_2007}.} & $1138499$ & $2990443$ & Social\\
\href{https://networks.skewed.de/net/netscience}{\texttt{netscience}} & A coauthorship network among scientists working on network science, from 2006. This network is a one-mode projection from the bipartite graph of authors and their scientific publications~\cite{Newman_2006}. & 1589 & 2742 & Social\\
\href{https://networks.skewed.de/net/new_zealand_collab}{\texttt{new\_zealand\_collab}} & A network of scientific collaborations among institutions in New Zealand~\cite{Aref_2018}. & 1511 & 4273 & Social\\
\href{https://networks.skewed.de/net/petster}{\texttt{petster}} & A network of friendships among users on \texttt{catster.com} and \texttt{dogster.com}~\cite{Kunegis_2013}. & 623766 & 15695166 & Social\\
\href{https://networks.skewed.de/net/physician_trust}{\texttt{physician\_trust}} & A network of trust relationships among physicians in four midwestern (USA) cities in 1966~\cite{Coleman_1957}. & 241 & 923 & Social\\
\href{https://networks.skewed.de/net/physics_collab#pierreAuger}{\texttt{physics\_collab}} & \multirow[t]{1}{35em}{Coauthorships among  the Pierre Auger Collaboration of physicists~\cite{1408.2925v1}.} & $514$ & $6482$ & Social\\
\href{https://networks.skewed.de/net/reality_mining}{\texttt{reality\_mining}} & A network of human proximities among students at Massachusetts Institute of Technology (MIT), as measured by personal mobile phones~\cite{Eagle_2005}. & 96 & 2539 & Social\\
\href{https://networks.skewed.de/net/residence_hall}{\texttt{residence\_hall}} & A network of friendships among students living in a residence hall at Australian National University~\cite{Freeman_1998}. & 217 & 1839 & Social\\
\href{https://networks.skewed.de/net/sp_high_school#diaries}{\texttt{sp\_high\_school} (1)} & \multirow[t]{2}{35em}{Contacts and friendship relations between students in a high school in Marseilles, France, in December 2013~\cite{mastrandrea_2015}.} & $329$ & $348$ & Social\\
\href{https://networks.skewed.de/net/sp_high_school#survey}{\texttt{sp\_high\_school} (2)} &  & 329 & 406 & Social\\
\href{https://networks.skewed.de/net/sp_high_school_new#2011}{\texttt{sp\_high\_school\_new}} & \multirow[t]{1}{35em}{Network of contacts between students in a high school in Marseilles, France~\cite{2014}.} & $126$ & $1709$ & Social\\
\href{https://networks.skewed.de/net/sp_hypertext#intervals}{\texttt{sp\_hypertext}} & Network of contacts among attendees of the ACM Hypertext 2009 conference~\cite{1006.1260v3}. & $113$ & $2196$ & Social\\
\href{https://networks.skewed.de/net/sp_office}{\texttt{sp\_office}} & A temporal network of contacts between individuals, measured in an office building in France, from June 24 to July 3, 2013~\cite{G_NOIS_2015}. & 92 & 755 & Social\\
\href{https://networks.skewed.de/net/sp_primary_school}{\texttt{sp\_primary\_school}} & Network of contacts among students and teachers at a primary school in Lyon, France, on consecutive days of in October 2009~\cite{2011}. & 242 & 8317 & Social\\
\href{https://networks.skewed.de/net/student_cooperation}{\texttt{student\_cooperation}} & Network of cooperation among students in the ``Computer and Network Security'' course at Ben-Gurion University, in 2012~\cite{Fire_2012}. & 185 & 311 & Social\\
\href{https://networks.skewed.de/net/swingers}{\texttt{swingers}} & A bipartite sexual affiliation network representing ``swing unit'' couples (one node per couple) and the parties they attended~\cite{Niekamp_2013}. & 96 & 232 & Social\\
\href{https://networks.skewed.de/net/twitter}{\texttt{twitter}} & A network of following relationships from Twitter, from a snowball sample crawl across ``quality'' users in 2009~\cite{De_Choudhury_2010}. & 465017 & 833540 & Social\\
\href{https://networks.skewed.de/net/twitter_15m}{\texttt{twitter\_15m}} & A network representing follower-following relations among Twitter users associated with the 15-M Movement or Anti-austerity movement in Spain, in the period April-May 2011~\cite{Gonz_lez_Bail_n_2011}. & 87569 & 4708274 & Social\\
\href{https://networks.skewed.de/net/twitter_higgs#reply}{\texttt{twitter\_higgs}} & \multirow[t]{1}{35em}{Tweet reply network related to the discovery of the Higgs boson~\cite{De_Domenico_2013}.} & $38918$ & $29552$ & Social\\
\href{https://networks.skewed.de/net/ugandan_village#health-advice_3}{\texttt{ugandan\_village}} & Complete friendship and health advice social networks among households in 17 rural villages bordering Lake Victoria in Mayuge District, Uganda in 2013~\cite{Chami_2017}. & $185$ & $638$ & Social\\
\href{https://networks.skewed.de/net/us_agencies#michigan}{\texttt{us\_agencies} (1)} & \multirow[t]{2}{35em}{Web-based links between U.S. government agencies websites~\cite{Kosack_2018}.} & $1796$ & $47686$ & Social\\
\href{https://networks.skewed.de/net/us_agencies#hawaii}{\texttt{us\_agencies} (2)} &  & 234 & 515 & Social\\
\href{https://networks.skewed.de/net/us_congress#S112}{\texttt{us\_congress}} & Networks of bill co-sponsorship tendencies among US Congresspeople, from 1973 (93rd Congress) to 2016 (114th Congress)~\cite{Neal_2020,Neal_2014}. & $101$ & $3914$ & Social\\
\href{https://networks.skewed.de/net/wiki_talk#br}{\texttt{wiki\_talk} (1)} & \multirow[t]{2}{35em}{Interactions among users of 10 language-specific Wikipedias~\cite{Kunegis_2013}.} & $1181$ & $2330$ & Social\\
\href{https://networks.skewed.de/net/wiki_talk#oc}{\texttt{wiki\_talk} (2)} &  & 3144 & 4098 & Social\\
\href{https://networks.skewed.de/net/wikipedia-en-talk}{\texttt{wikipedia-en-talk}} & Nodes in the network represent (English) Wikipedia users and a directed edge from node $i$ to node $j$ represents that user $i$ at least once edited a talk page of user $j$~\cite{Leskovec_2010}. & 2394385 & 4659565 & Social\\
\href{https://networks.skewed.de/net/wikitree}{\texttt{wikitree}} & A multigraph network representing child-parent connections among family members, collected in 2012 from WikiTree, an online genealogical website with 13+ million profiles~\cite{1311.4276v2}. & 1382751 & 4810045 & Social\\
\href{https://networks.skewed.de/net/windsurfers}{\texttt{windsurfers}} & A network of interpersonal contacts among windsurfers in southern California during the Fall of 1986. The edge weights indicate the perception of social affiliations majored by the tasks in which each individual was asked to sort cards with other surfer’s name in the order of closeness~\cite{FREEMAN_1988}. & 43 & 336 & Social\\
\href{https://networks.skewed.de/net/amazon_copurchases#601}{\texttt{amazon\_copurchases}} & Network of items for sale on amazon.com and the items they ``recommend''~\cite{Leskovec_2007}. & $403394$ & $2443408$ & Economic\\
\href{https://networks.skewed.de/net/amazon_ratings}{\texttt{amazon\_ratings}} & A bipartite network of users and products on Amazon.com~\cite{Lim_2010}. & 3376972 & 5743258 & Economic\\
\href{https://networks.skewed.de/net/bookcrossing}{\texttt{bookcrossing}} & Bipartite network representing people and the books they have interacted with, from the BookCrossing website~\cite{Ziegler_2005}. & 445801 & 1149739 & Economic\\
\href{https://networks.skewed.de/net/corporate_directors}{\texttt{corporate\_directors}} & Bipartite network of directors and the companies on whose boards they sit, spanning 54 countries worldwide, constructed from data collected by the Financial Times~\cite{1910.07441v2}. & 356638 & 376918 & Economic\\
\href{https://networks.skewed.de/net/dbpedia_starring}{\texttt{dbpedia\_starring}} & A bipartite network of movies and the actors that played in them, as extracted from Wikipedia by the DBpedia project~\cite{Auer_2007}. & 157184 & 281396 & Economic\\
\href{https://networks.skewed.de/net/dbpedia_team}{\texttt{dbpedia\_team}} & Bipartite network of the affiliations (employment relations) between professional athletes and their teams, as extracted from Wikipedia by the DBpedia project~\cite{Auer_2007}. & 935627 & 1366466 & Economic\\
\href{https://networks.skewed.de/net/discogs_affiliation}{\texttt{discogs\_affiliation}} & A large bipartite network of the affiliations (contractual relations) among musical artists and record labels, as given in the discogs.com database~\cite{Kunegis_2013}. & 2025594 & 5302276 & Economic\\
\href{https://networks.skewed.de/net/epinions}{\texttt{epinions}} & A bipartite network of users and the products they rated on the website Epinions.com~\cite{massa2005controversial}. & 876252 & 13668320 & Economic\\
\href{https://networks.skewed.de/net/eu_procurements}{\texttt{eu\_procurements}} & A bipartite network of public EU procurement contracts, from 2008 to 2016, between issuing buyers (public institutions such as a ministry or city hall) and supplying winners (a private firm)~\cite{Wachs_2020}. & 839824 & 1841009 & Economic\\
\href{https://networks.skewed.de/net/eu_procurements_alt#EE_2009}{\texttt{eu\_procurements\_alt} (1)} & \multirow[t]{9}{35em}{Networks representing the annual national public procurement markets of 26 European countries from 2008-2016, inclusive~\cite{Wachs_2020}.} & $552$ & $588$ & Economic\\
\href{https://networks.skewed.de/net/eu_procurements_alt#CY_2011}{\texttt{eu\_procurements\_alt} (2)} &  & 585 & 588 & Economic\\
\href{https://networks.skewed.de/net/eu_procurements_alt#PT_2008}{\texttt{eu\_procurements\_alt} (3)} &  & 1038 & 1009 & Economic\\
\href{https://networks.skewed.de/net/eu_procurements_alt#SK_2011}{\texttt{eu\_procurements\_alt} (4)} &  & 1098 & 1118 & Economic\\
\href{https://networks.skewed.de/net/eu_procurements_alt#AT_2013}{\texttt{eu\_procurements\_alt} (5)} &  & 2189 & 2320 & Economic\\
\href{https://networks.skewed.de/net/eu_procurements_alt#LT_2015}{\texttt{eu\_procurements\_alt} (6)} &  & 1656 & 3132 & Economic\\
\href{https://networks.skewed.de/net/eu_procurements_alt#LV_2009}{\texttt{eu\_procurements\_alt} (7)} &  & 2097 & 2518 & Economic\\
\href{https://networks.skewed.de/net/eu_procurements_alt#IT_2016}{\texttt{eu\_procurements\_alt} (8)} &  & 9877 & 11185 & Economic\\
\href{https://networks.skewed.de/net/eu_procurements_alt#DE_2012}{\texttt{eu\_procurements\_alt} (9)} &  & 19438 & 23191 & Economic\\
\href{https://networks.skewed.de/net/fao_trade}{\texttt{fao\_trade}} & Multiplex network representing trade relationships between countries from the Food and Agricultural Organization of the United Nations~\cite{2015}. & 214 & 9420 & Economic\\
\href{https://networks.skewed.de/net/github}{\texttt{github}} & The bipartite project-user membership network of the software development hosting site GitHub~\cite{Chacon_2014}. & 177386 & 440237 & Economic\\
\href{https://networks.skewed.de/net/jester}{\texttt{jester}} & Two bipartite networks of users and jokes, extracted from the online joke recommender system Jester~\cite{Goldberg_2001}. & 73521 & 4136360 & Economic\\
\href{https://networks.skewed.de/net/stackoverflow}{\texttt{stackoverflow}} & A bipartite network of users and the posts they have favorited, from the online Q\&A site Stack Overflow~\cite{Kunegis_2013}. & 641876 & 1301942 & Economic\\
\href{https://networks.skewed.de/net/digg_votes}{\texttt{digg\_votes}} & A bipartite network between users and stories on digg.com from 2009~\cite{Hogg_2012}. & 142962 & 3010898 & Economic\\